\documentclass[12pt]{iopart} 
\expandafter\let\csname equation*\endcsname=\relax 
\expandafter\let\csname endequation*\endcsname=\relax 
\usepackage{amsmath,amssymb,amsthm} 
\usepackage{physics}
\usepackage{graphicx} 
\graphicspath{{Materials/}} 
\usepackage{xfrac}
\usepackage[hidelinks]{hyperref}
\usepackage{cleveref}
\usepackage{dsfont}
\usepackage{listings}
\usepackage{xcolor}
\usepackage[section]{placeins}
\usepackage{appendix}
\usepackage[font={small,rm}]{subcaption}
\usepackage{tikz}
\begin{document}
\title[Tunneling and Thermal Transitions in a Double Well Potential]{Quantum Tunnelling and Thermally Driven Transitions in a Double Well Potential at Finite Temperature}
\author{Robson Christie and Jessica Eastman}
\address{Department of Mathematics, Imperial College London, London SW7 2AZ,
United Kingdom}
\ead{robson.christie13@imperial.ac.uk, jessica.eastman@anu.edu.au}
\vspace{10pt}
\begin{indented}
\item[]\today
\end{indented}

\begin{abstract}
We explore dissipative quantum tunnelling, a phenomenon central to various physical and chemical processes, using a double-well potential model. This paper aims to bridge gaps in understanding the crossover from thermal activation to quantum tunnelling, a domain still shrouded in mystery despite extensive research. We study a Caldeira-Leggett-derived model of quantum Brownian motion and investigate the Lindblad and stochastic Schr\"{o}dinger dynamics numerically, seeking to offer new insights into the transition states in the crossover region. Our study has implications for quantum computing and understanding fundamental natural processes, highlighting the significance of quantum effects on transition rates and temperature influences on tunnelling.

Additionally, we introduce a new model for quantum Brownian motion which takes Lindblad form and is formulated as a modification of the widely known model found in Breuer and Petruccione. In our approach, we remove the zero-temperature singularity resulting in a better description of low-temperature quantum Brownian motion near a potential minima.

\end{abstract}

\section{Introduction}
Dissipative quantum tunnelling is a key phenomenon in understanding various physical and chemical processes, marked by the intriguing ability of quantum mechanical objects to pass through barriers. This concept, which has long captivated and challenged physicists, becomes even more complex when fluctuations and dissipation are involved. This complexity opens up numerous avenues for research.

One area that remains particularly elusive is the crossover from thermal activation, or barrier hopping, to quantum tunnelling. Despite considerable efforts to decipher the underlying mechanisms, this crossover continues to hold mysteries. This paper aims to delve into this crossover region, contributing to the extensive research in this field. The fundamental understanding of tunnelling behaviour in the presence of dissipation is not only academically intriguing but also crucial for numerous practical applications. This knowledge is particularly vital in advancing the field of quantum computing, where control over quantum tunnelling could revolutionize computational capabilities. Additionally, it provides deeper insights into natural phenomena such as enzyme catalysis and photosynthesis, processes essential to life yet not fully understood from a quantum mechanical perspective. A pertinent example illustrating the practical implications of this study is the diffusion of light elements such as hydrogen in metals \cite{Munakata, Kaneko}. This example demonstrates how a transition from thermal hopping to quantum tunnelling occurs at low temperatures, emphasizing the need to comprehend its effects on solid-state systems. More recently, there has been interest in dissipative quantum tunnelling in the context of proton tunnelling in DNA \cite{Alkhalili}. 

The literature on dissipative quantum tunnelling is extensive, one of the usual approaches in the field is to draw inspiration from the seminal work of Caldeira and Leggett (CL), who provided an explicit model and dynamical equation for the density matrix of a quantum system in contact with a thermal environment \cite{caldeira1983path}. Building upon this foundational work, Hänggi and his colleagues have made significant contributions to the theory of transitions in dissipative quantum systems, as evidenced by a series of papers \cite{hanggi1984tunnel, grabert1984crossover, hanggi1986escape, hanggi1987dissipative}, and summarized in a comprehensive review \cite{hanggi1990reaction}. These works have introduced quantum corrections to Kramers' rate theory, providing useful approximate expressions for transition rates that are highly accurate in the tunnelling regime (low temperature) and the thermal hopping regime (high temperature). It is important to note that other approaches have also been taken concerning specific examples, ie. Hydrogen diffusion in solids \cite{Munakata, Kaneko}.

The process of quantum tunnelling can be modelled simply but effectively with a single particle in a double well potential, a toy model widely used by many to understand the fundamental physics at play. In a cold atoms set-up, a double well may be created by a magneto-optical trap and a laser \cite{Oberthaler}.

To model the dissipative component of the dynamics we employ a model of quantum Brownian motion found in \cite{petruccione} which emerges from a minimally invasive modification of the Caldeira Leggett (CL) master equation. This modification ensures the complete positivity of the dynamics and facilitates the unravelling of density operator dynamics into stochastic Schrödinger pure state trajectories, each representing the state conditioned on an individual noise history. We use a primarily numerical approach to simulate individual stochastic realizations of dissipative quantum systems. This strategy offers insights into the transition states occurring in the crossover region without the encumbrances associated with deriving analytic expressions.

In \cref{sec:classical}, we revisit results related to classical Langevin dynamics \cite{pavliotis2014stochastic}. Our investigation focuses on the trajectories of Langevin dynamics and examines the influence of temperature and coupling strength on the transition rate to set the stage for comparison with quantum dynamics.

In \cref{sec:quantum}, we delve into the features and rates of both zero-temperature tunnelling and finite-temperature hybrid tunnelling-hopping regimes. Our analysis covers the qualitative features of the quantum trajectories and quantifies the impacts on the transition rate. We take note of the predominant factors of dissipation and decoherence in many cases \cite{schlosshauer2007decoherence}. Additionally, our study reveals an enhancement in interference effects and transition rates at low temperatures in the minimally invasive CL model, indicative of an anti-Zeno effect \cite{AntiZeno1,AntiZeno2}.

The final section, \cref{sec:modified}, addresses whether these anti-Zeno effects accurately describe quantum Brownian motion or are artefacts resulting from extending the CL master equation beyond its intended scope. We propose an alternative modification to transit the CL master equation into Lindblad form, which offers distinct advantages, especially at low temperatures. We also explore two types of double-well potentials to capture generic behaviours and modify the Petruccione model to yield better correspondence with the expected asymptotic thermal state.

\section{The classical model} \label{sec:classical}
In the present paper, we consider a particle in a double-well potential of the form 
\begin{equation}
   V(x)=\frac{1}{2}  x^2 +A e^{-x^2/2 \sigma^2},
\end{equation}
which corresponds to a harmonic potential superimposed with a Gaussian bump with standard deviation $\sigma$ and amplitude $A$. This is a similar potential to the one used in cold atom experiments \cite{Oberthaler}, with easy access to a magneto-optical trap and a laser beam.

Classically, the influence of temperature is most commonly modelled heuristically by introducing noise and friction into the dynamics. This is  described by a Langevin equation of the form
\begin{equation} \label{eq:Langevin_tilde}
\begin{pmatrix}
dx\\ dp
\end{pmatrix}=\left(\frac{1}{m}\begin{pmatrix} p\\0 \end{pmatrix}-\begin{pmatrix}0 \\ V'(x)\end{pmatrix}-2\gamma \begin{pmatrix} 0 \\ p \end{pmatrix} \right) dt+\begin{pmatrix}
0\\2 \sqrt{ \gamma m k_B T }
\end{pmatrix}d \xi,
\end{equation}
where $d \xi$ is a real It\^{o} process \cite{bernt}, and $x$ and $p$ are the particle's position and momentum, respectively.  That is, a classical particle under the influence of temperature is subject to stochastic energy fluctuations, as well as damping. The stochastic description of individual trajectories is dual to the deterministic Fokker-Planck master equation for the probability density in phase space \cite{pavliotis2014stochastic} which is of the form 
\begin{multline} \label{eq:Fokker}
	\frac{\partial}{\partial t}\rho(x,p)=-\frac{p}{m} \frac{\partial}{\partial x}\rho(x,p)+V'(x)\frac{\partial}{\partial p}\rho(x,p)\\ +  2 \gamma \frac{\partial}{\partial p}\left(p \rho(x,p) \right) + 2\sqrt{m \gamma k_B T}\frac{\partial^2}{\partial p^2} \rho(x,p).
\end{multline}
Langevin equations can be derived from microscopic Hamiltonian dynamics describing a particle and its thermal environment, using various levels of approximations. A well-known example of this approach is Zwanzig's derivation of effective Langevin dynamics for a particle linearly coupled to a bath of oscillators in thermal equilibrium \cite{zwanzig1973nonlinear}. Remarkably, the resulting Langevin dynamics are often applicable over a wider range of situations, going beyond the assumptions made during their derivation.

\begin{figure}[t]
\centering
\begin{tabular}{c c c c}
\begin{subfigure}[c]{.23\textwidth} 
	  \includegraphics[width=\textwidth]{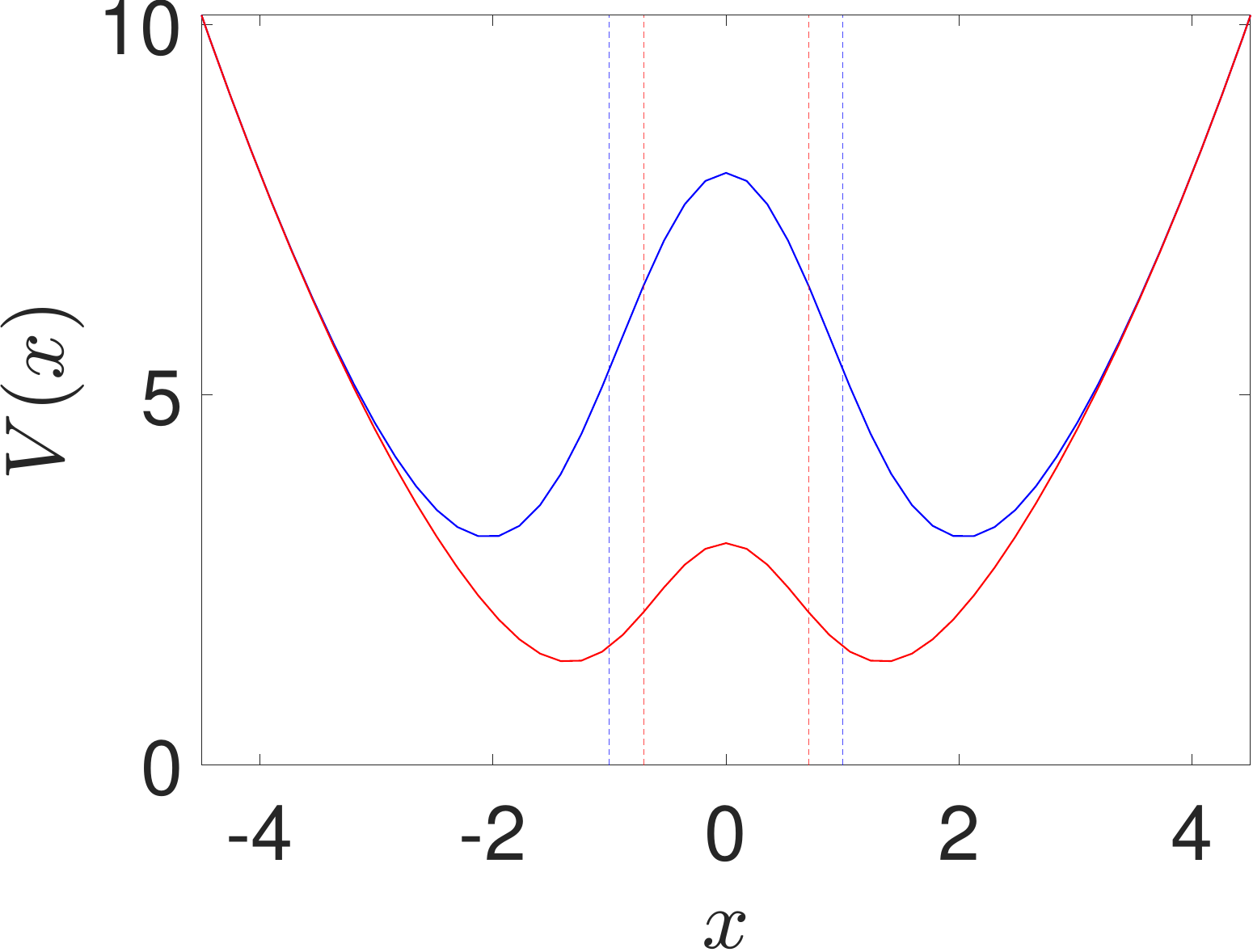}
 	 \caption{}\label{fig:GWellCompare}
\end{subfigure}
\begin{subfigure}[c]{.23\textwidth} 
	  \includegraphics[width=\textwidth]{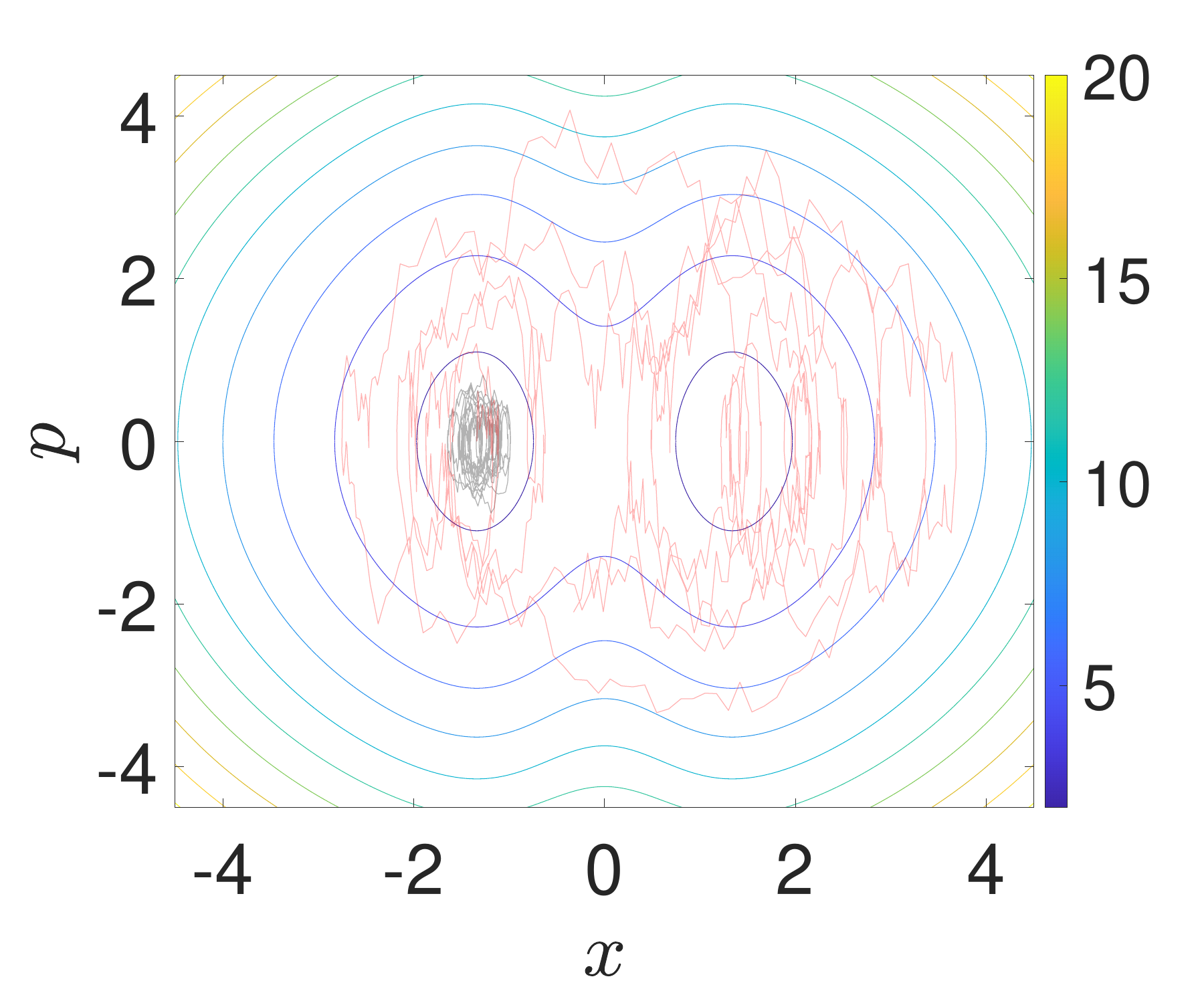}
 	 \caption{}\label{fig:HamPhase}
\end{subfigure}&
\begin{subfigure}[c]{.23\textwidth} 
	  \includegraphics[width=\textwidth]{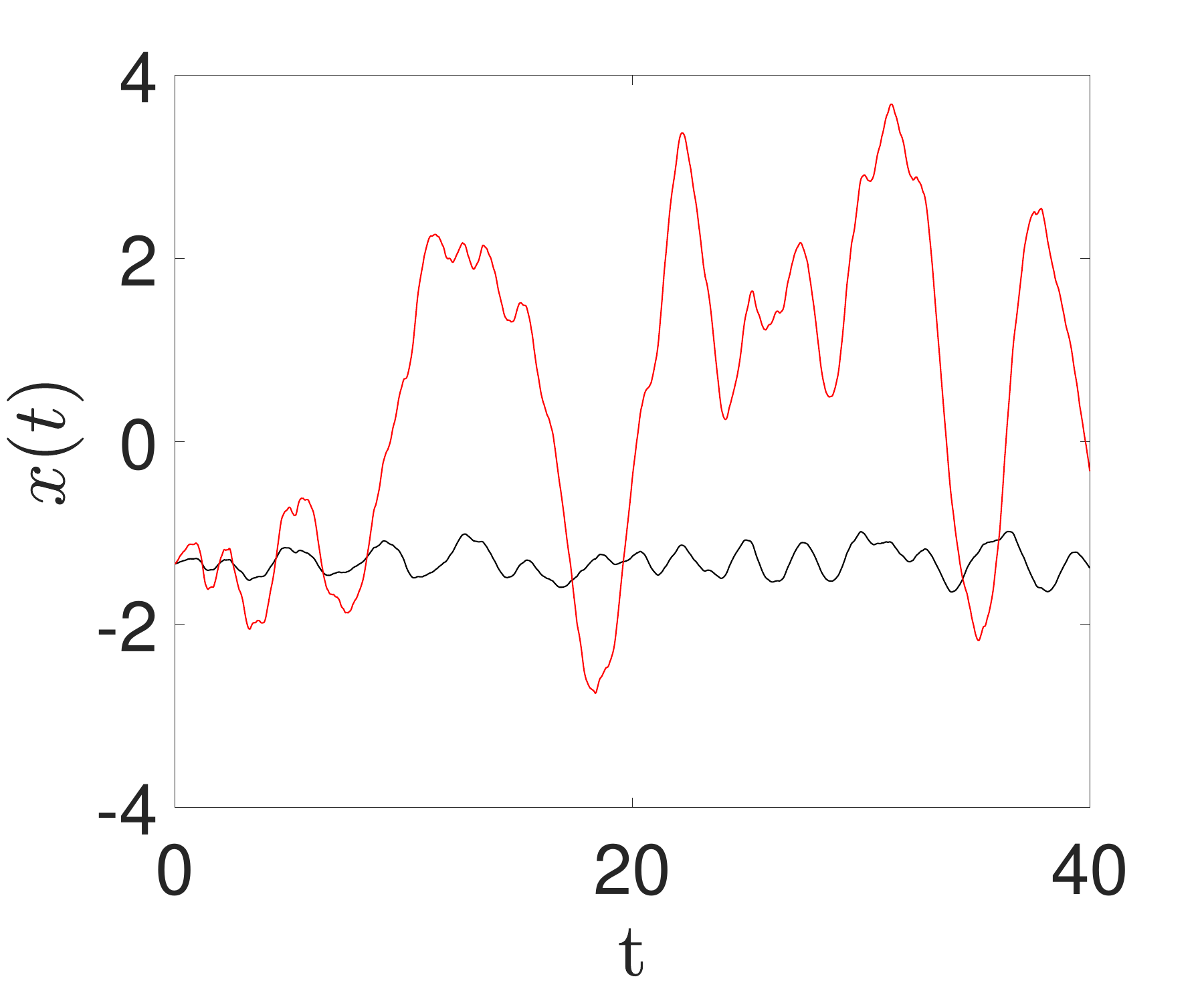}
	\caption{ }\label{fig:XSingleLangevin}
\end{subfigure}&
\begin{subfigure}[c]{.23\textwidth} 
	  \includegraphics[width=\textwidth]{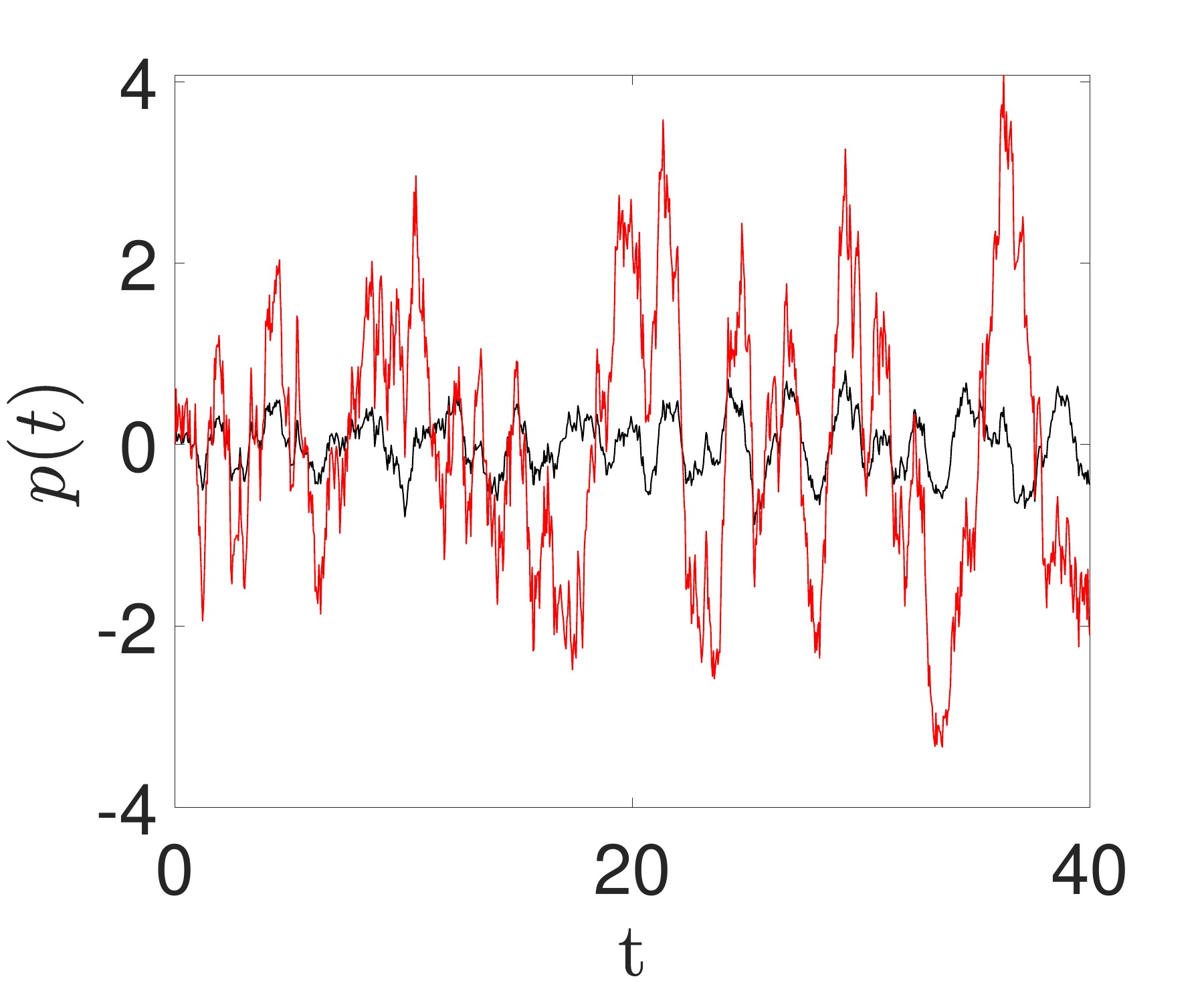}
 \caption{}\label{fig:PSingleLangevin}
\end{subfigure}
\end{tabular}
 \caption{Potential and dividing surfaces (a), and typical trajectories of the Langevin equation \eqref{eq:Langevin_tilde}, for temperatures $T=3$ (red) and $T=0.2$ (black) (b-d), in all plots $\gamma=0.2$. (a) shallow barrier ($A=3$ and $\sigma=\frac{1}{\sqrt{2}}$) in red and high barrier ($A=8$ and $\sigma=1$) in blue. (b-d): Two typical trajectories for the shallow case; (b) trajectories on top of a phase-space contour plot of the classical Hamiltonian function, (c) and (d) position and momentum as functions of time. \label{fig:classical_trajectories}} 
\end{figure}

Here we are interested in temperature-induced transitions between the two wells of our double-well potential. We consider a particle initially located in one of the minima. If at some point the energy of the particle exceeds the height of the barrier between the two energy wells, the particle can transition to the other well. Here we consider the simple case, where we initialise the particle in the left minimum with zero momentum. To define a transition time, one needs to choose a transition surface (in one dimension this reduces to a transition coordinate $x^*$) that separates the two wells, the first crossing of which defines the transition time \cite{ref:chandlerbook}. To avoid counting fast recrossings, we consider a transition coordinate slightly to the right of the barrier top.  Here we choose the dividing coordinate as
\begin{equation}
    x^{*}=\sigma,
\end{equation}
where $\sigma$ is the standard deviation of the barrier. The potential and the dividing coordinate are depicted in the left panel of figure \ref{fig:classical_trajectories} for two different sets of parameters, belonging to a shallow and a higher barrier, respectively.
In the remaining panels of figure \cref{fig:classical_trajectories}, we depict two individual trajectories for \( A = 3 \) and \( \sigma = \frac{1}{\sqrt{2}} \) for two different temperatures 
(a low temperature with $k_B T<<V_{\text{barrier}}$ in black, and a higher temperature with ($k_B T>V_{\text{barrier}}$) in red). (b) Depicts the phase space trajectories on top of a contour plot of the Hamiltonian function. (c,d) show the corresponding position and momentum evolution. We observe the expected behaviour with the particle trapped in the initial well for low temperature, and sporadic thermal hopping between the wells for the higher temperature.

\begin{figure}[t]
\centering
\begin{tabular}{c c c c}
\begin{subfigure}[c]{.23\textwidth} 
	  \includegraphics[width=\textwidth]{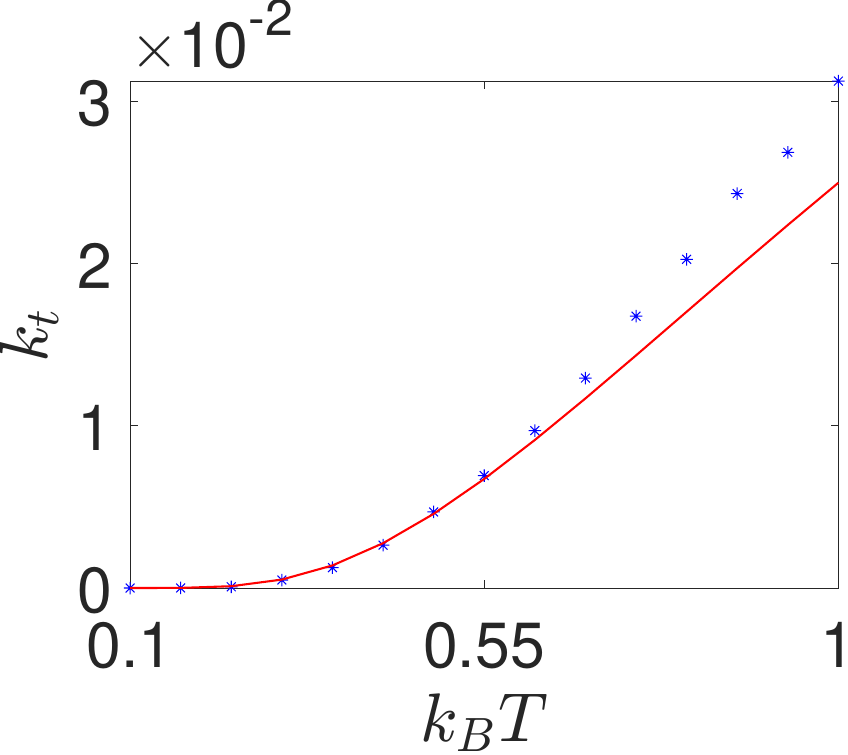}
 	 \caption{\label{fig:G_Shall_ArrheniusPlotHigh}}
\end{subfigure}&
\begin{subfigure}[c]{.23\textwidth} 
	  \includegraphics[width=\textwidth]{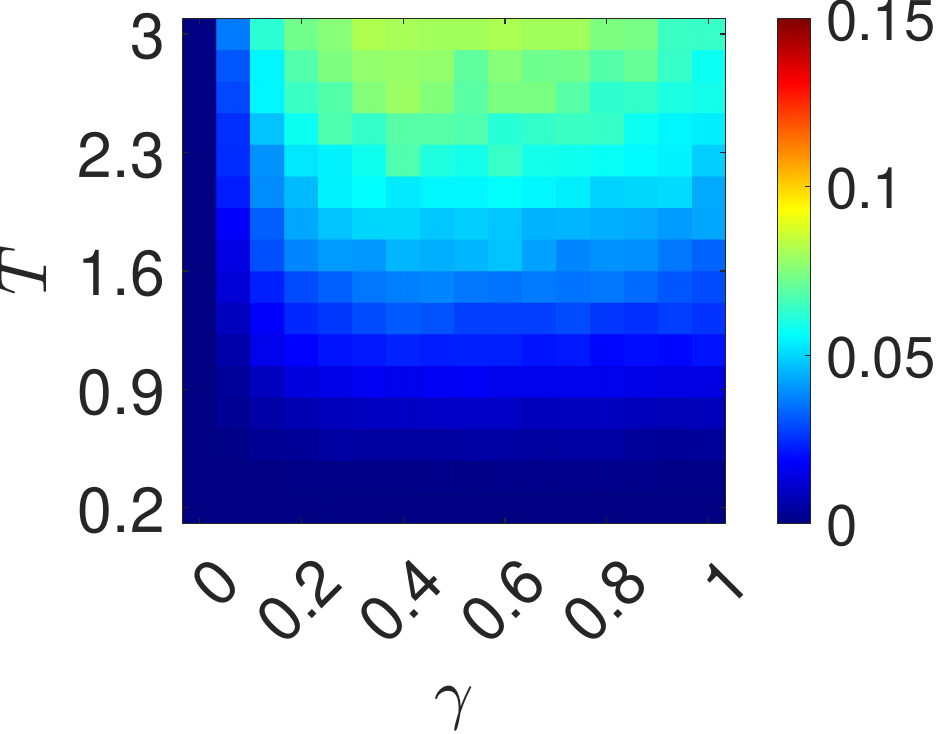}
 	 \caption{}\label{fig:G_Shall_RateLangevinCombined}
\end{subfigure}&
\begin{subfigure}[c]{.23\textwidth} 
	  \includegraphics[width=\textwidth]{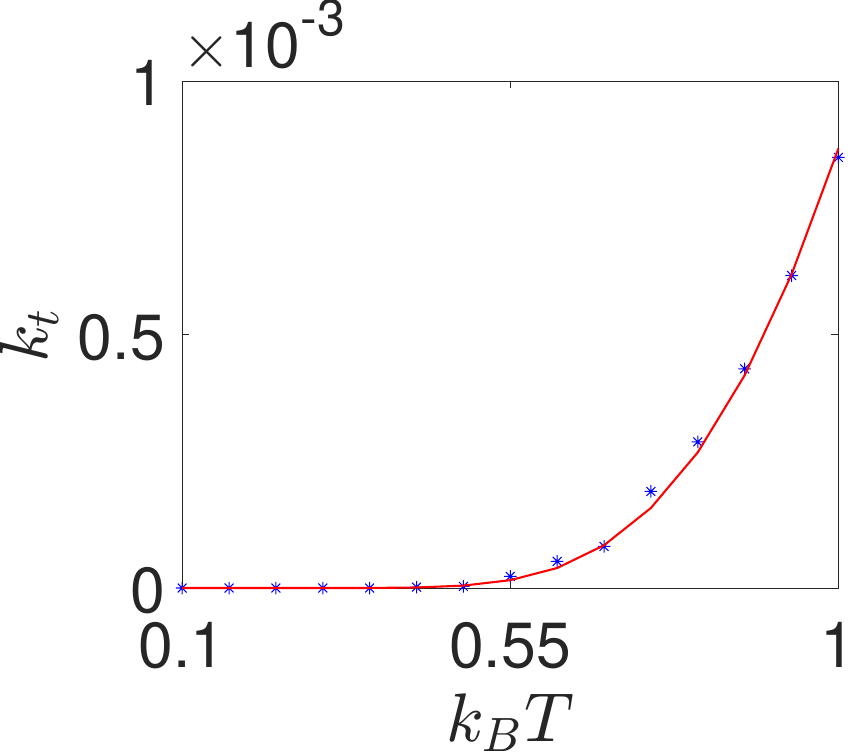}
 	 \caption{\label{fig:G_Deep_ArrheniusPlotHigh}}
\end{subfigure}&
\begin{subfigure}[c]{.23\textwidth} 
	  \includegraphics[width=\textwidth]{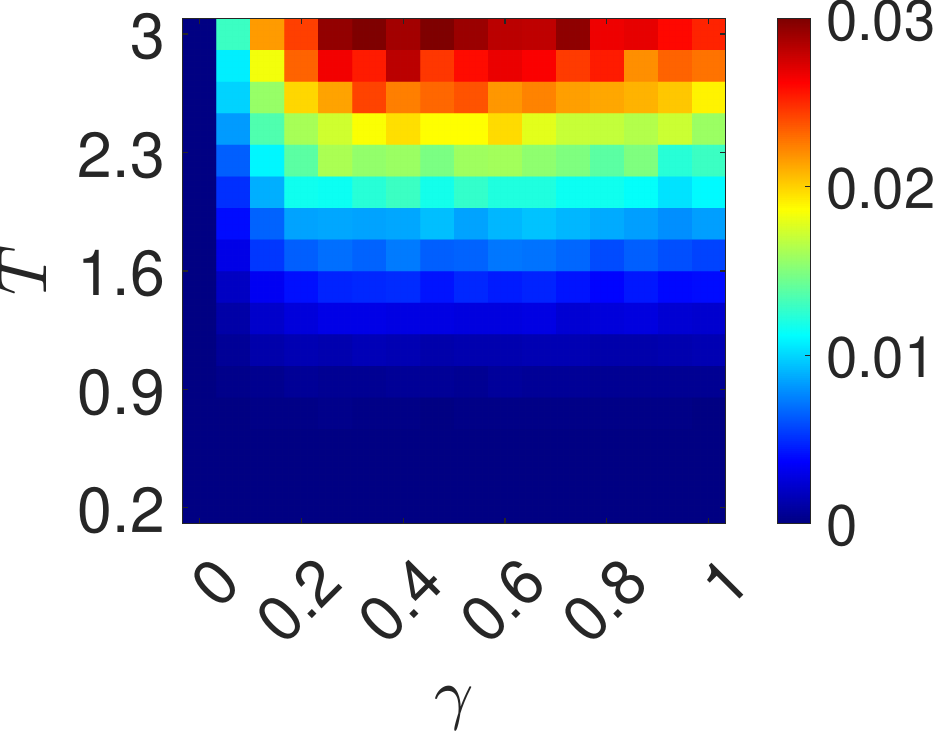}
 	 \caption{}\label{fig:G_Deep_RateLangevinCombined}
\end{subfigure}
\end{tabular}
 \caption{(a) and (c) show the transition rate (blue) and Arrhenius prediction (red) against temperature. (b) and (d) show the dependence of the transition rate on temperature and coupling strength. The left two plots correspond to a shallow double-well potential with $A=3$ and $\sigma=1/\sqrt{2}$ the right two plots correspond to a deep Gaussian well with $A=8$ and $\sigma=1$. \label{fig:Arrhenius}} 
\end{figure}

Building on the individual transition times, one can define an average transition rate as the inverse of the mean value of the transition times, when averaged over a suitable ensemble \cite{ref:chandlerbook}
\begin{equation}\label{eq:rate}
    k_t=\frac{1}{\mathbb{E}\left\{t_{\text{crossing}}\right\}}.
\end{equation}
In figure \ref{fig:Arrhenius} we show the resulting transition times in dependence on the temperature for the same parameters as in figure \ref{fig:classical_trajectories}, obtained from 5000 trajectories starting from the left minimum with zero momentum. The right plot shows a false colour plot of the transition time as a function of both temperature and the coupling parameter $\gamma$. As expected for a given temperature the transition time only has a weak dependence on the coupling parameter $\gamma$ in a relatively wide range of values for $\gamma$. The transition rate grows with the temperature, as is heuristically expected from the Arrhenius law \cite{arrhenius1889,hanggi1990reaction}, which estimates the rate as
\begin{equation} \label{eq:Arrhenius}
k_t\propto \exp\left(-\frac{V_B}{k_B T}\right)
\end{equation}
where $V_B$ is the barrier height as measured from the relevant minimum of the double well. For the double well, the barrier height is given by the expression
\begin{equation}
    V_{B}=A - \sigma^2\left(1 - \log \left( \frac{A}{\sigma^2} \right)\right)
\end{equation}
which takes values $V_B=1.6\quad\text{and}\quad 4.9$ for the shallow and deep wells respectively, the average kinetic energy of the particle in one dimension is given by $k_B T/2$. The Arrhenius estimate (solid red curve) is plotted in comparison to the numerical results arising from the Langevin dynamics (blue dots) in \cref{fig:Arrhenius}. The proportionality constant for the Arrhenius law is found by least squares fitting. We observe that the Arrhenius law is a good approximation of the classical transition times when $k_B T$ is much less than the barrier height, for larger temperatures we observe significant deviations from this simple approximation. 

\section{The quantum model} \label{sec:quantum}

To investigate the influence of temperature on the quantum dynamics in analogy to the classical description, we initially follow the approach of Caldeira and Leggett \cite{caldeira1983path}, who, building on Zwanzig's classical derivation of the Fokker-Planck equation, derived a dynamical equation for the reduced density matrix of a quantum system under the influence of a thermal bath as 
\begin{equation} \label{eq:Caldeira-Master}
\frac{d}{dt} \hat \rho = -\frac{i}{\hbar}[\hat H, \hat \rho]-\frac{i \gamma}{\hbar}[\hat X \{\hat P, \hat \rho\}]-\frac{2 m \gamma k_B T}{\hbar^2}[\hat X, [\hat X, \hat \rho]],\quad\text{with}\quad
\hat H=\frac{\hat P^{2}}{2 m}+V(\hat X).
\end{equation}
Implicit in the derivation of the above is the condition:
\begin{equation} \label{eq:TempCondition}
\hbar \gamma \ll 2 \pi k_B T.
\end{equation}
It is well known that the master equation \eqref{eq:Caldeira-Master} does not preserve the positivity of the density matrix at low temperatures, preventing its unravelling in terms of stochastic trajectories representing individual experimental runs. The most widely used modification of the Caldeira Leggett master equation is that of Petruccione and Breuer \cite{petruccione}, in which a ``minimally invasive" term of the form
\begin{equation} \label{eq:minimal}
-\frac{\gamma}{8 m k_B T}[\hat P,[\hat P,\hat \rho]]
\end{equation}
is added to the Caldeira-Leggett equation \eqref{eq:Caldeira-Master}, to bring it to Lindblad form. The additional term (\ref{eq:minimal}) is insignificant in the high-temperature regime. The resulting equation including the extra term is now of Lindblad form 
\cite{lindblad1976generators,gorini1976completely}:
\begin{equation} \label{eq:lindblad}
\frac{d}{d t}\hat{\rho}=-\frac{i}{\hbar}[\hat{H},\hat{\rho}] + \frac{1}{\hbar} \left(\hat{L} \hat{\rho} \hat{L}^{\dagger} -\frac{1}{2} \hat{L}^{\dagger} \hat{L}\hat{\rho}-\frac{1}{2}\hat{\rho} \hat{L}^{\dagger} \hat{L}\right),
\end{equation}
with
\begin{equation} \label{eq:heatbath}
\begin{aligned}
\hat{H}=\frac{\hat{P}^2}{2m}+V(\hat{X})+\frac{\gamma}{2}(\hat{X}\hat{P}+\hat{P}\hat{X}),\quad {\rm and}\quad \hat{L}= \sqrt{\frac{4 \gamma m k_B T}{\hbar}}\hat{X}+i\sqrt{\frac{\gamma \hbar}{4mk_B T}}\hat{P}.
\end{aligned}
\end{equation}
Similar to the unravelling of the Fokker-Planck equation by Langevin trajectories in the classical case, the Lindblad equation can be unravelled as an average over stochastic Schrödinger equations \cite{percival,belavkin1989nondemolition} of the form
\begin{multline} \label{eq:SSE}
\ket{d \psi } = \frac{1}{\hbar}(-{i}\hat{H} -\frac{1}{2 }\hat{L}^{\dagger}\hat{L}+\ev*{\hat{L}^{\dagger}}\hat{L} - \frac{1}{2}\ev*{\hat{L}^{\dagger}}\ev*{\hat{L}})\ket {\psi} dt \
+ \frac{1}{\sqrt{2 \hbar}}\left(\hat{L}-\ev*{\hat{L}}\right)\ket{\psi} (d\xi_R +i d\xi_I).
\end{multline}

To understand the connection between the stochastic Schrödinger equations (SSE) and the classical Langevin dynamics, we need to explore the classical limit of these equations. It is well-known that the ensemble averages (Lindblad expectation values) of position and momentum expectation values approach the dynamics of the classical Fokker-Planck dynamics, as described by Petruccione and others. This reveals a clear connection between the quantum dynamics of the Lindblad equation and the classical Fokker-Planck dynamics.

Furthermore, it is shown in \cite{schlosshauer2007decoherence} that in a Gaussian approximation, the Lindblad equation specified by \eqref{eq:heatbath} approaches the Fokker Planck \eqref{eq:Fokker} dynamics for up to quadratic potentials. However, it is important to note that the general case for potentials beyond the quadratic form is still an open question. This discrepancy requires further discussion and investigation to fully understand the relationship between the quantum and classical dynamics in different limits.

We will use the quantum dynamics generated by \eqref{eq:heatbath} to analyse the transition rates in the quantum case, for the same potential as in the classical system, given by  
\begin{equation} \label{eq:gausswell}
	 \hat{V}=\frac{m \omega^2}{2} \hat X ^2 +A e^{-\hat X^2/2 \sigma^2}.
\end{equation}
We consider the scenario of an initial Gaussian wavepacket with zero momentum expectation value, placed in one of the minima of the double-well potential, and define as transition time the time at which the expectation value $\ev*{\hat X}$ crosses the dividing surface $x^*$; we then take the average of this first crossing time over many trials and take the inverse to calculate the transition rate $k_c$ as in the classical case (\ref{eq:rate}) \cite{ref:chiara}.

\subsection{Quantum dynamics for vanishing temperature} 

\begin{figure}[t]
\centering
\begin{tabular}{c c c c}
\begin{subfigure}[c]{.23\textwidth} 
	  \includegraphics[width=\textwidth]{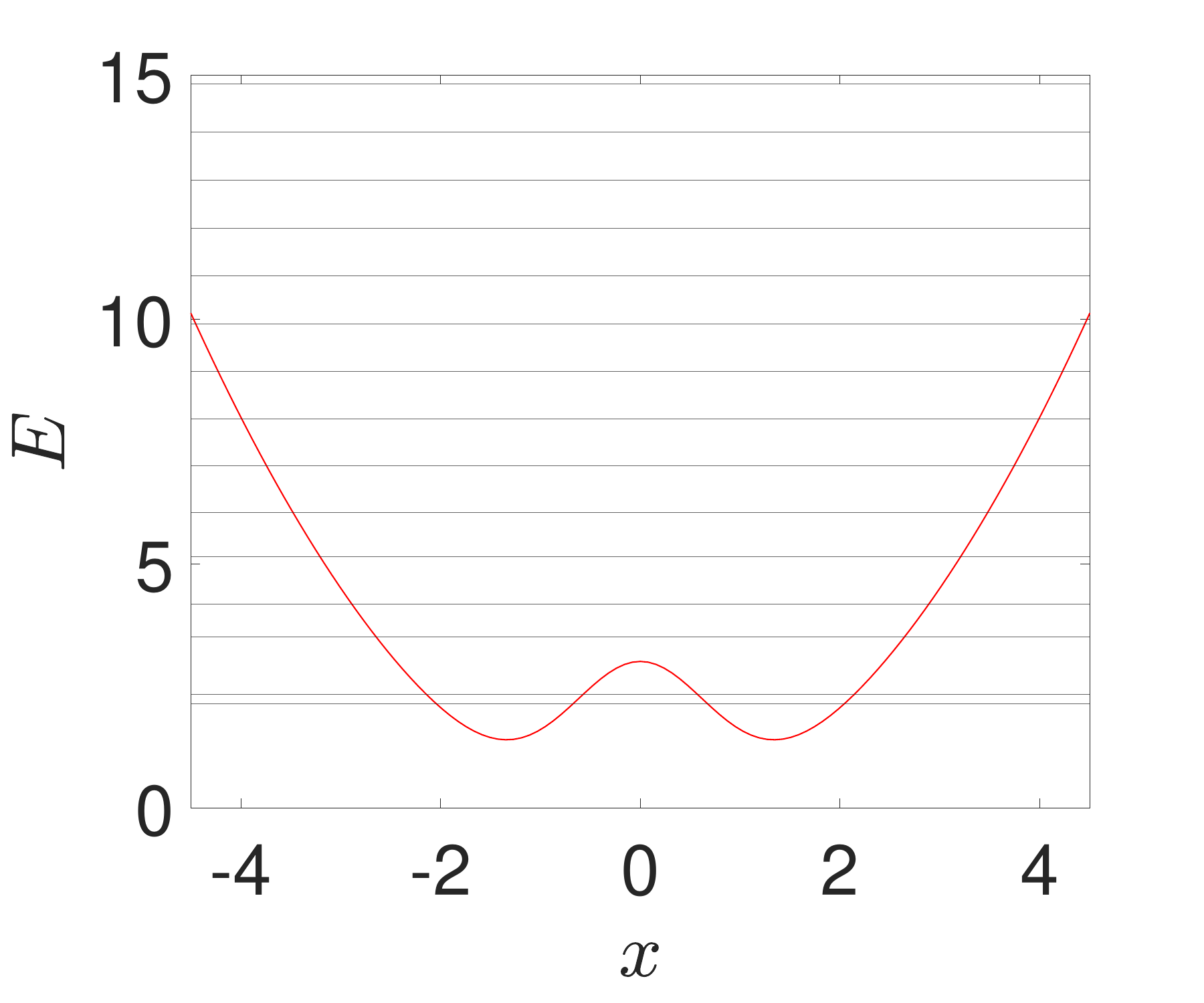}
 	 \caption{\label{fig:GaussWellEnergyLevels}}
\end{subfigure}&
\begin{subfigure}[c]{.23\textwidth} 
	  \includegraphics[width=\textwidth]{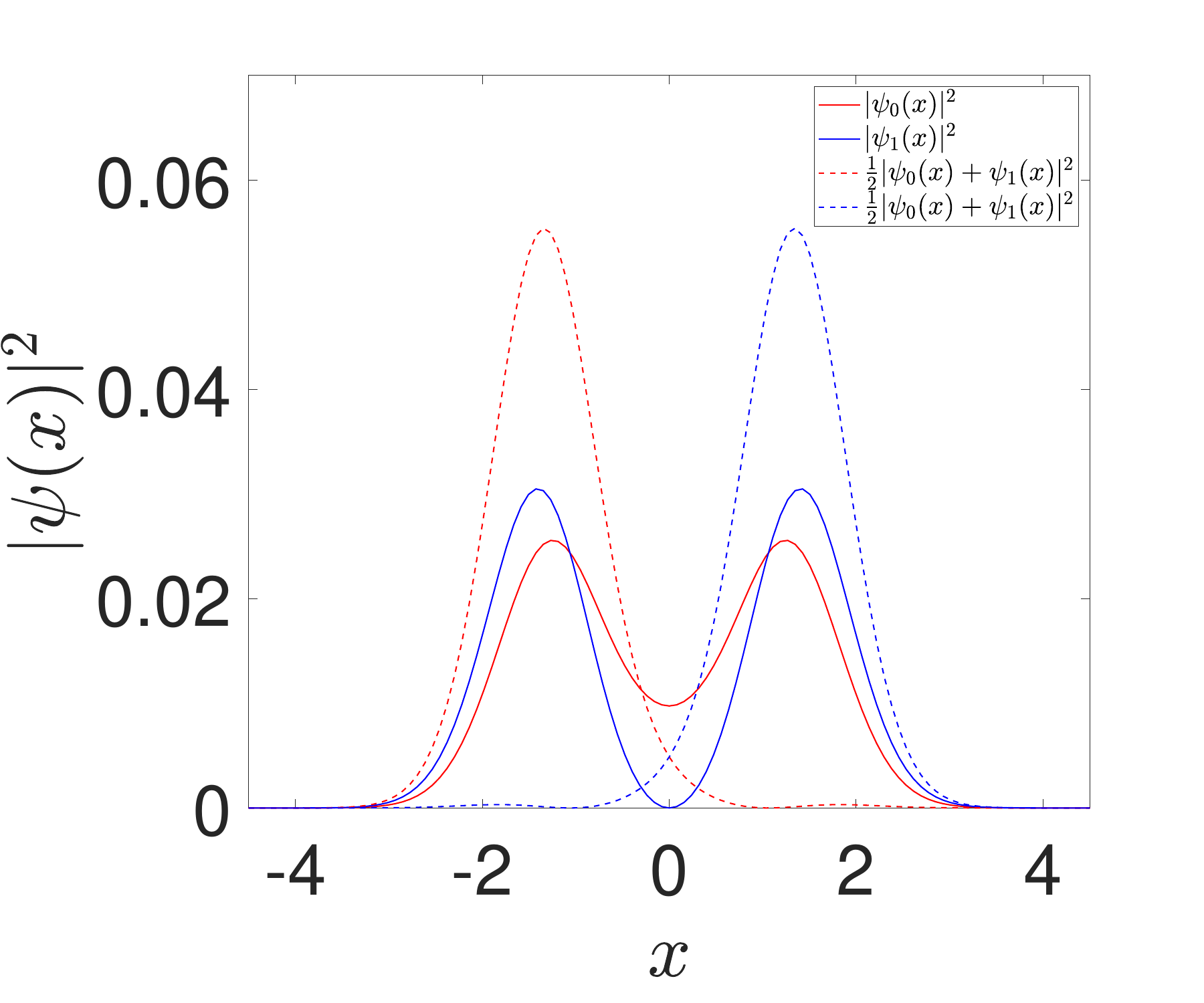}
 	 \caption{\label{fig:Eigenfunctions}}
\end{subfigure}&
\begin{subfigure}[c]{.23\textwidth} 
	  \includegraphics[width=\textwidth]{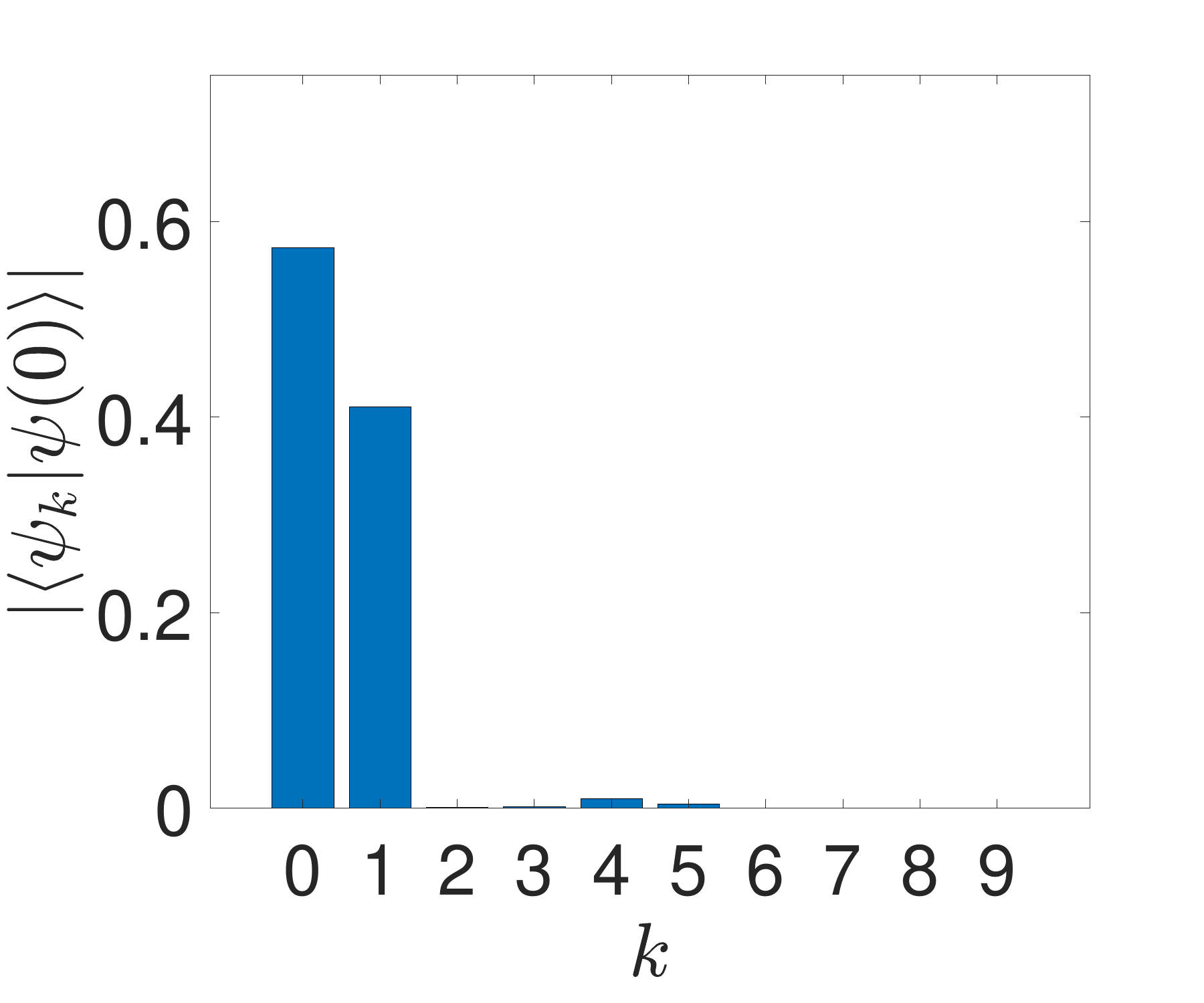}
 	 \caption{\label{fig:HistogramHamiltonian}}
\end{subfigure}&
\begin{subfigure}[c]{.22\textwidth} 
	  \includegraphics[width=\textwidth]{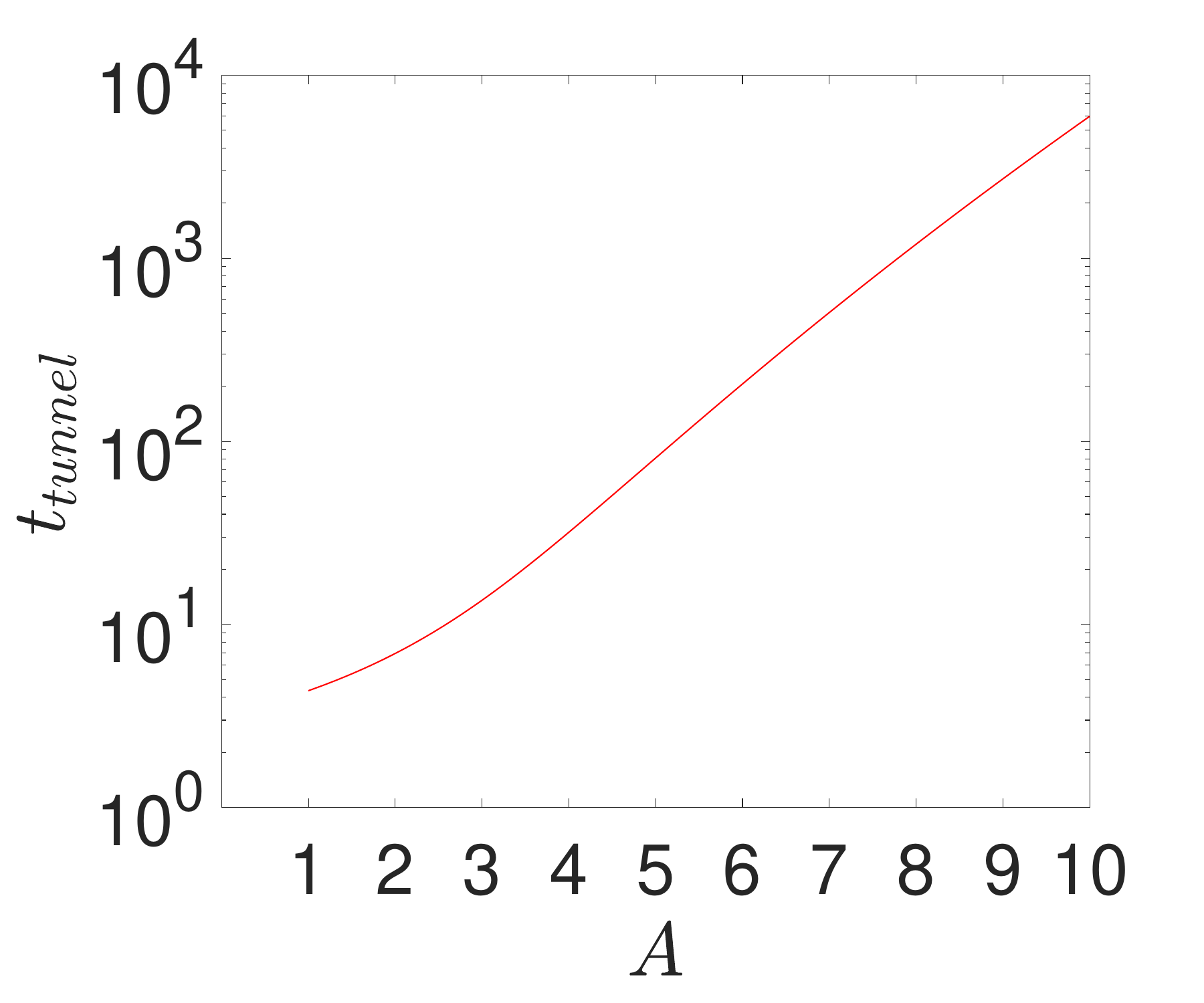}
 	 \caption{\label{fig:AmpTunnellPlot}}
\end{subfigure}
\end{tabular}
\caption{Eigenstates in the quadratic well with Gaussian barrier \eqref{eq:gausswell} with parameters $A=3$,$\gamma=0$ and $\sigma=1/\sqrt2$. (a) Potential $V(x)$ and eigenenergies; (b) Position probability densities of the ground (solid red) and 1st excited (solid blue) state and their symmetric and antisymmetric superpositions. (c) Coefficients of a coherent state at the minimum of the left well expanded in the eigenbasis. (d) The tunneling time (log scale) vs the amplitude of the central Gaussian barrier $A$. \label{fig:TunnelPlots}} 
\end{figure}

In the quantum case, at zero temperature, the initial wave packet tunnels coherently between the two minima, with a period that is approximately given by 
\begin{equation} \label{eq:t-tunnel}
    t_{\text{tunnel}}=\frac{\pi \hbar}{E_1-E_0},
\end{equation}
where $E_0$ and $E_1$ denote the ground state and first excited state energies, respectively. This periodic behaviour is often referred to as \textit{dynamical tunnelling} and stems from the fact that a symmetric superposition of the ground and first excited state is localised in one of the sides of the double well, and an initial Gaussian wave packet localised in one of the minima has little contribution from higher energy states. This is illustrated for a particular example in figure \ref{fig:TunnelPlots}, the left panel of which depicts the energy levels in a relatively shallow double well potential. \Cref{fig:Eigenfunctions} depicts the two lowest Hamiltonian eigenstates as well as their symmetric and antisymmetric superposition. Here the symmetric superposition $\frac{1}{\sqrt2}(\ket{\psi_0}+\ket{\psi_1})$ is localized the left well; whilst the antisymmetric superposition is localized the right well. \Cref{fig:HistogramHamiltonian} shows a histogram of the coefficients of a coherent state at the minimum of the left well in the basis of the eigenstates.
In general, increasing the amplitude $A$ of the central Gaussian raises the energy levels of the even eigenstates to a greater degree than the odd eigenstates; the lower energy eigenstates are also affected to a greater degree than the higher energy states. This phenomenon in a double well is well known and explicitly calculated for the model considered here in \cite{fassari2020two} using perturbation theory. The right panel in figure \ref{fig:TunnelPlots} depicts the tunnelling time estimated from the numerically obtained energy gap between the ground state and first excited state as a function of the parameter $A$, which controls the barrier height. As intuitively obvious, the tunnelling time increases (approximately exponentially) with increasing barrier height. 

\begin{figure}
\centering 
\begin{tabular}{c c c c}
\begin{subfigure}[c]{.23\textwidth} 
	  \includegraphics[width=\textwidth]{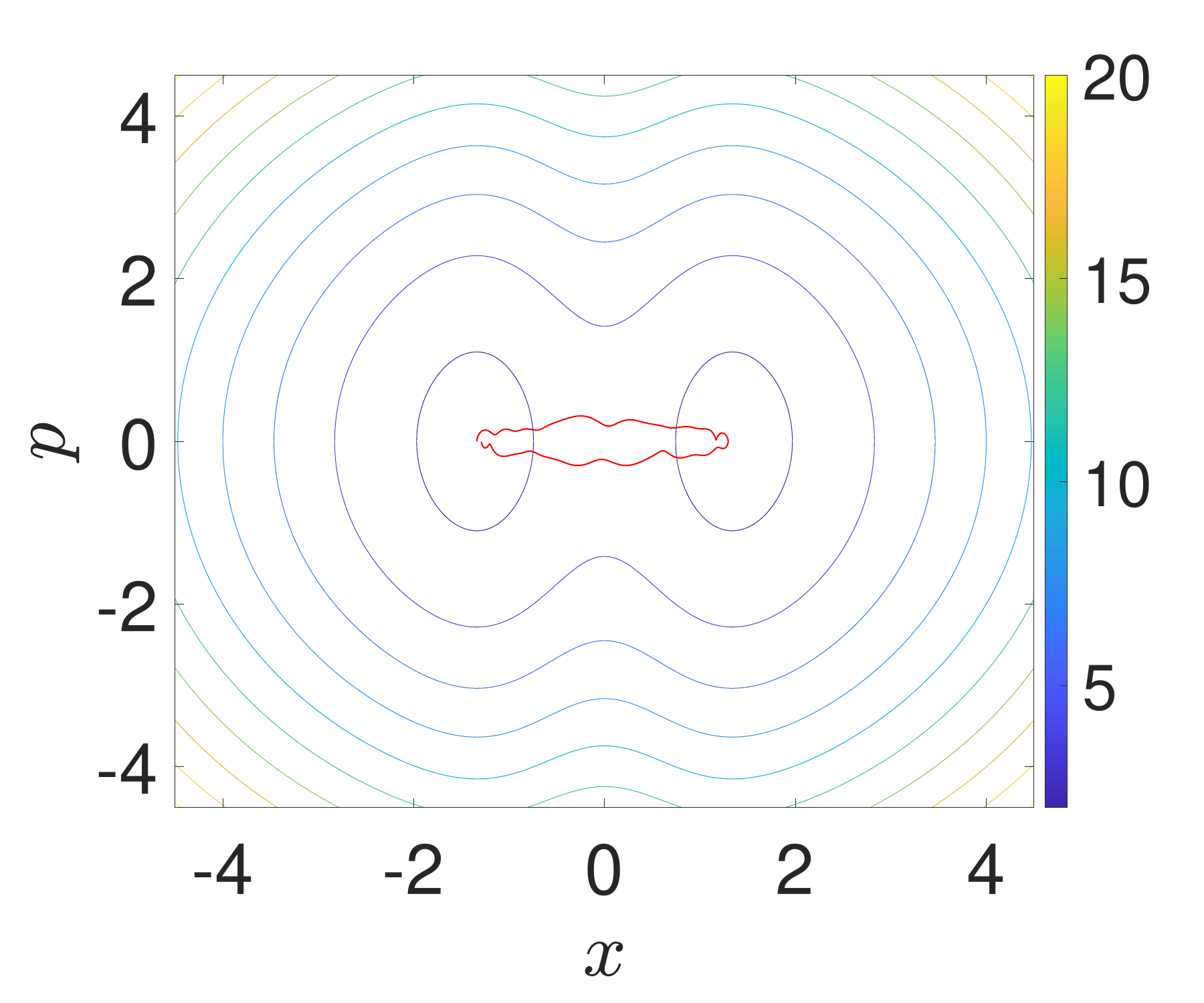}
 	 \caption{}\label{fig:HGaussContour_tunnel}
\end{subfigure}&
\begin{subfigure}[c]{.23\textwidth} 
	  \includegraphics[width=\textwidth]{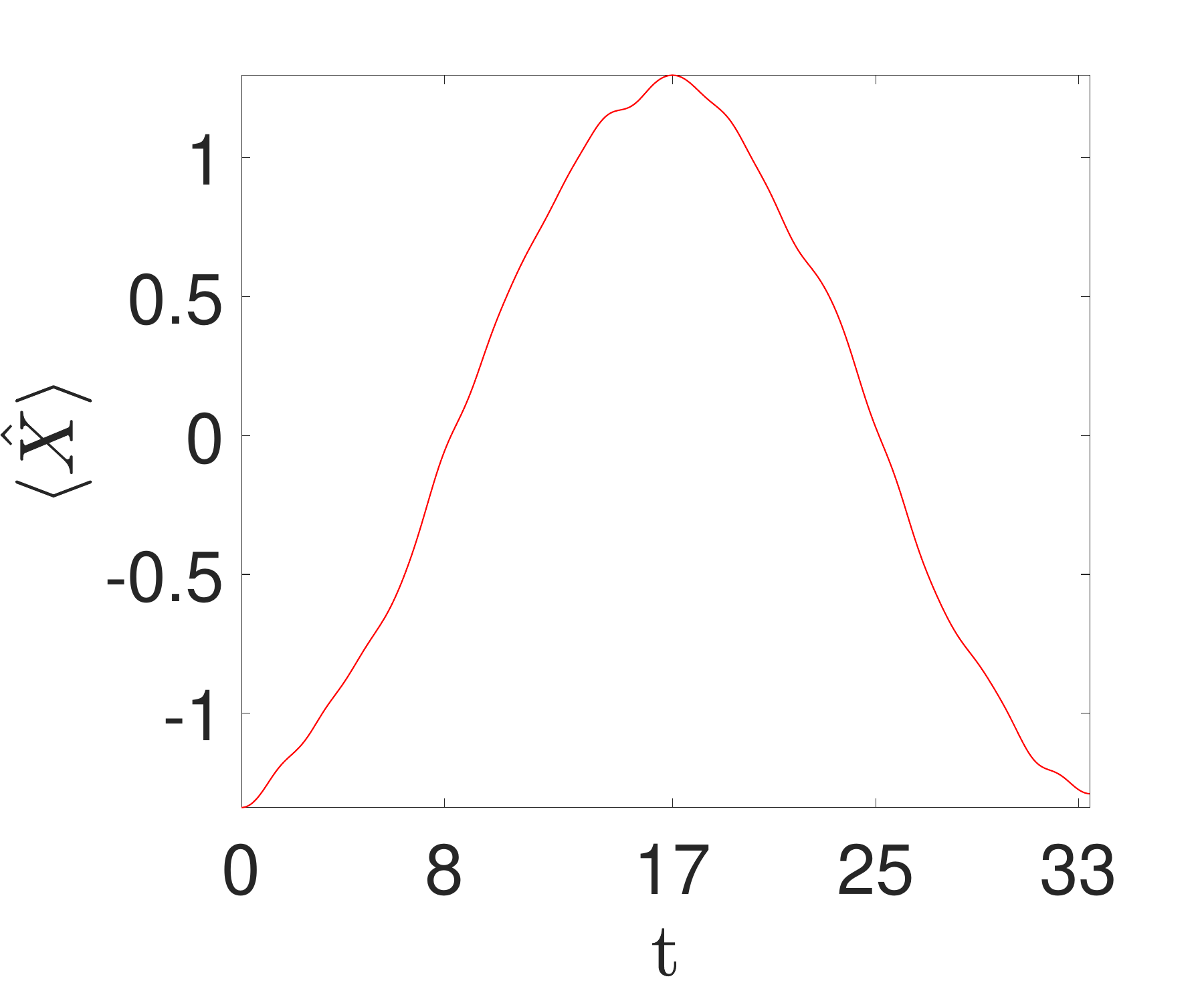}
 	 \caption{}\label{fig:XDoubleHam}
\end{subfigure}&
\begin{subfigure}[c]{.23\textwidth} 
	  \includegraphics[width=\textwidth]{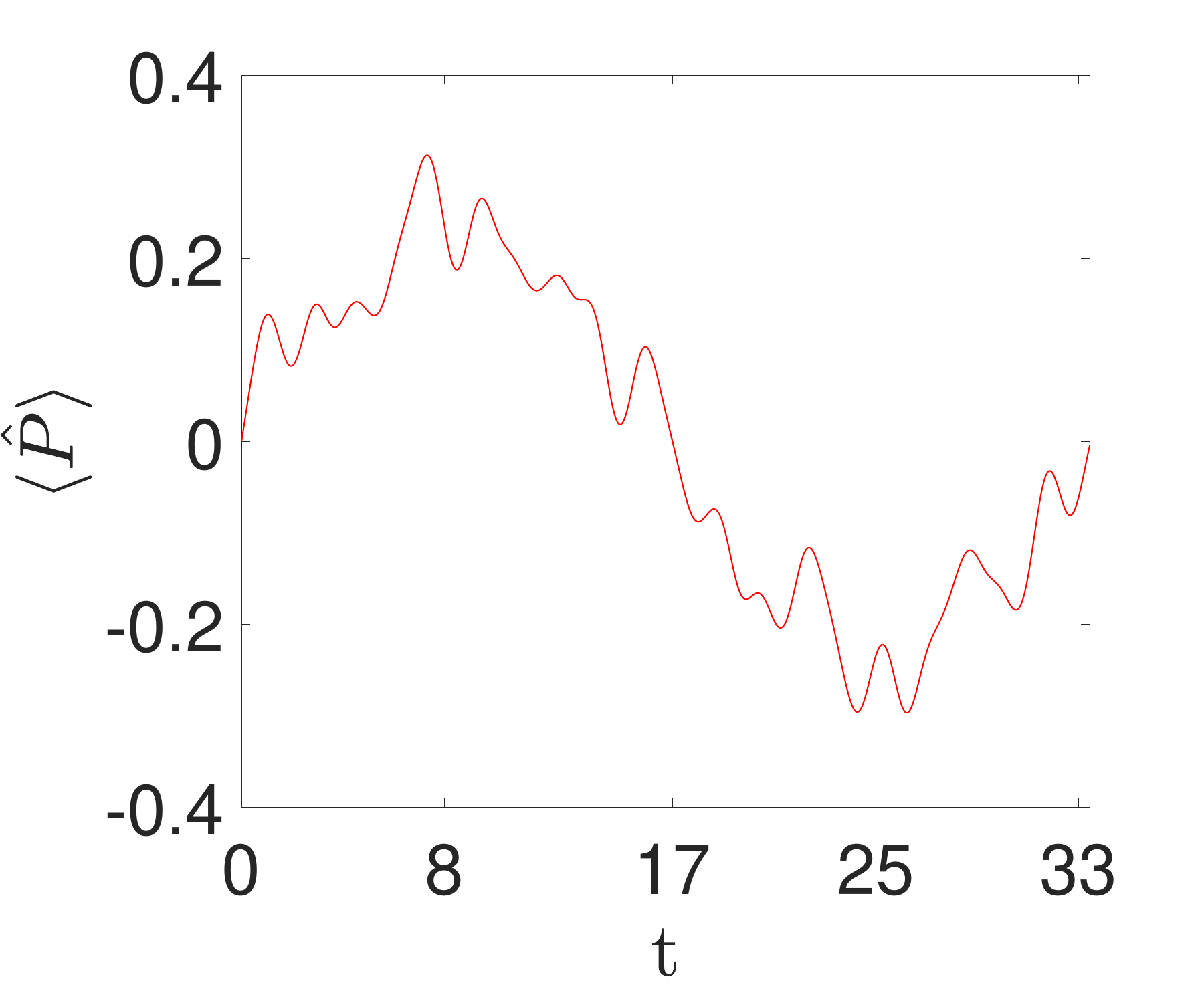}
 	 \caption{}\label{fig:PDoubleHam}
\end{subfigure}&
\begin{subfigure}[c]{.23\textwidth} 
	  \includegraphics[width=\textwidth]{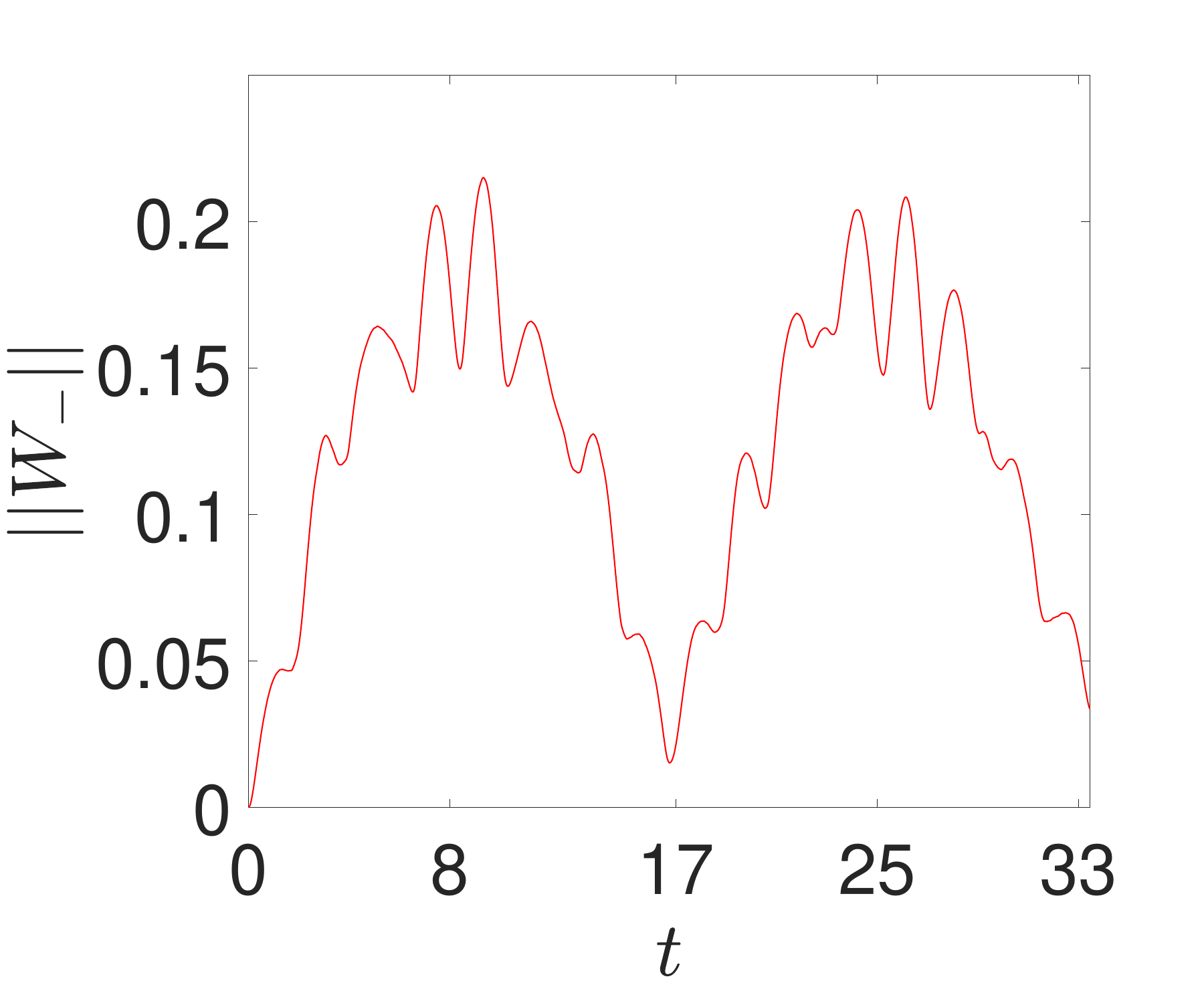}
 	 \caption{}\label{fig:WigNegDoubleHam}
\end{subfigure}\\
\begin{subfigure}[c]{.22\textwidth} \label{fig:WigStartDoubleHam}
	  \includegraphics[width=\textwidth]{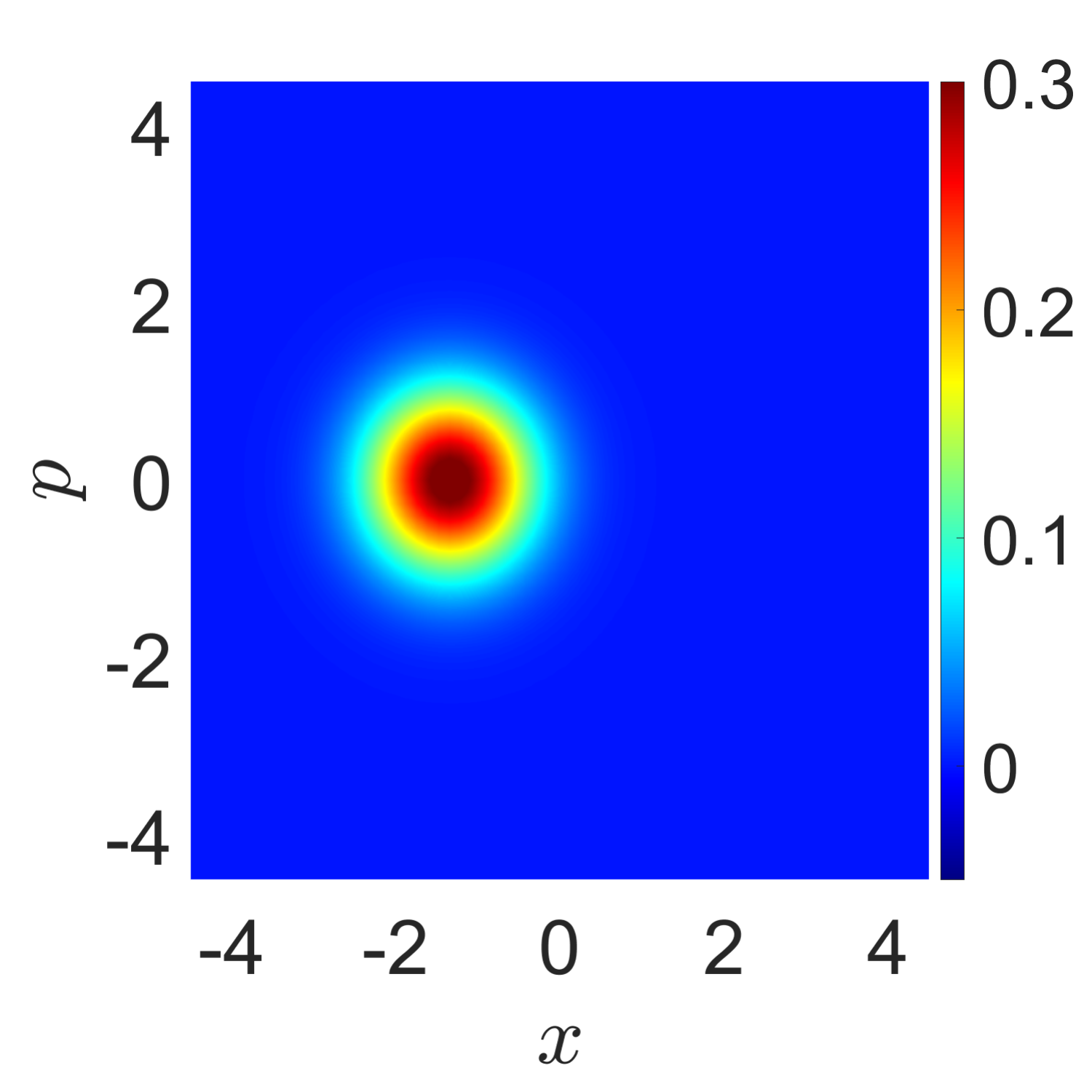}
 	 \caption{: t=0}
\end{subfigure}&
\begin{subfigure}[c]{.22\textwidth} \label{fig:WigMidDoubleHam}
	  \includegraphics[width=\textwidth]{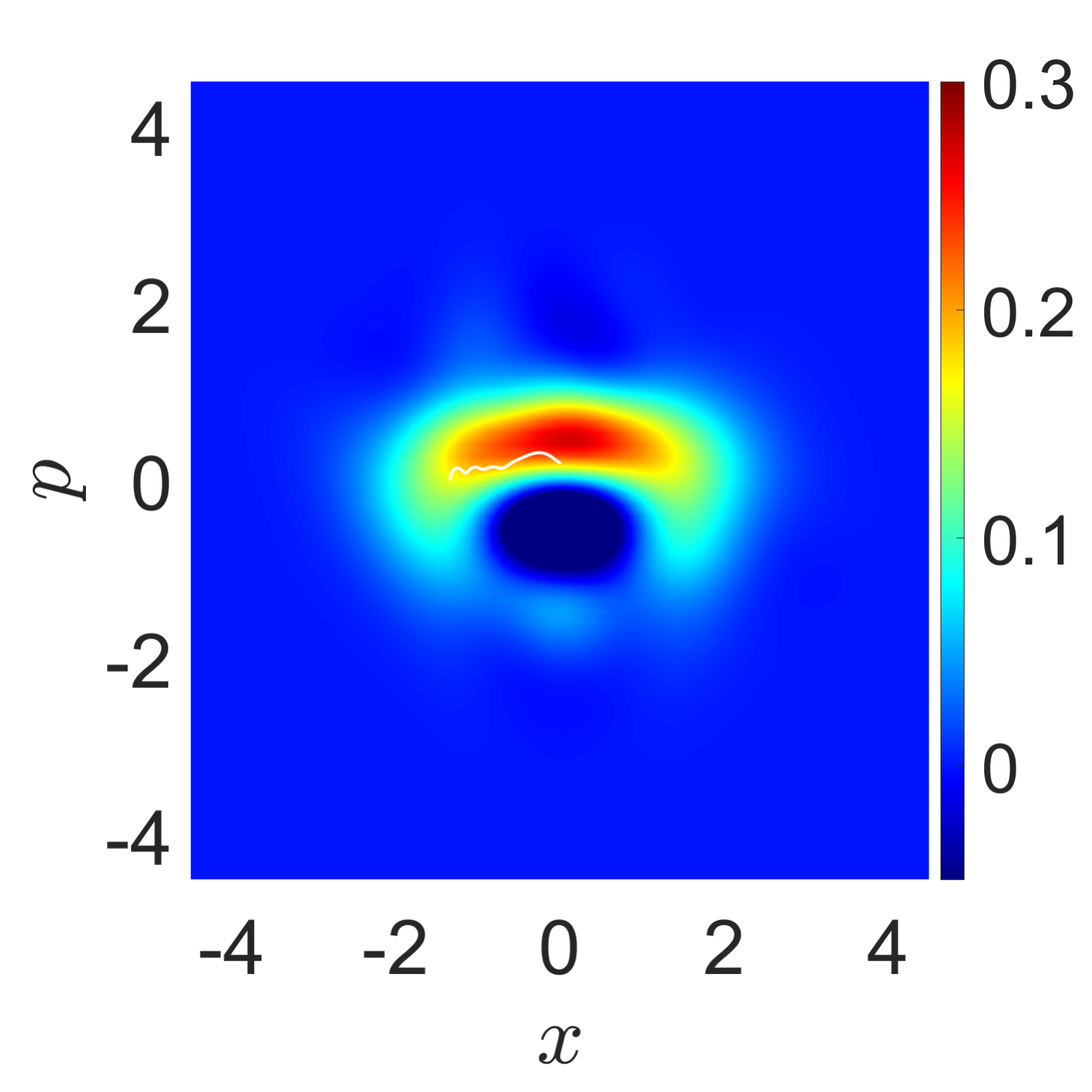}
 	 \caption{: t=8.35}
\end{subfigure}&
\begin{subfigure}[c]{.22\textwidth} \label{fig:WigEndDoubleHam}
	  \includegraphics[width=\textwidth]{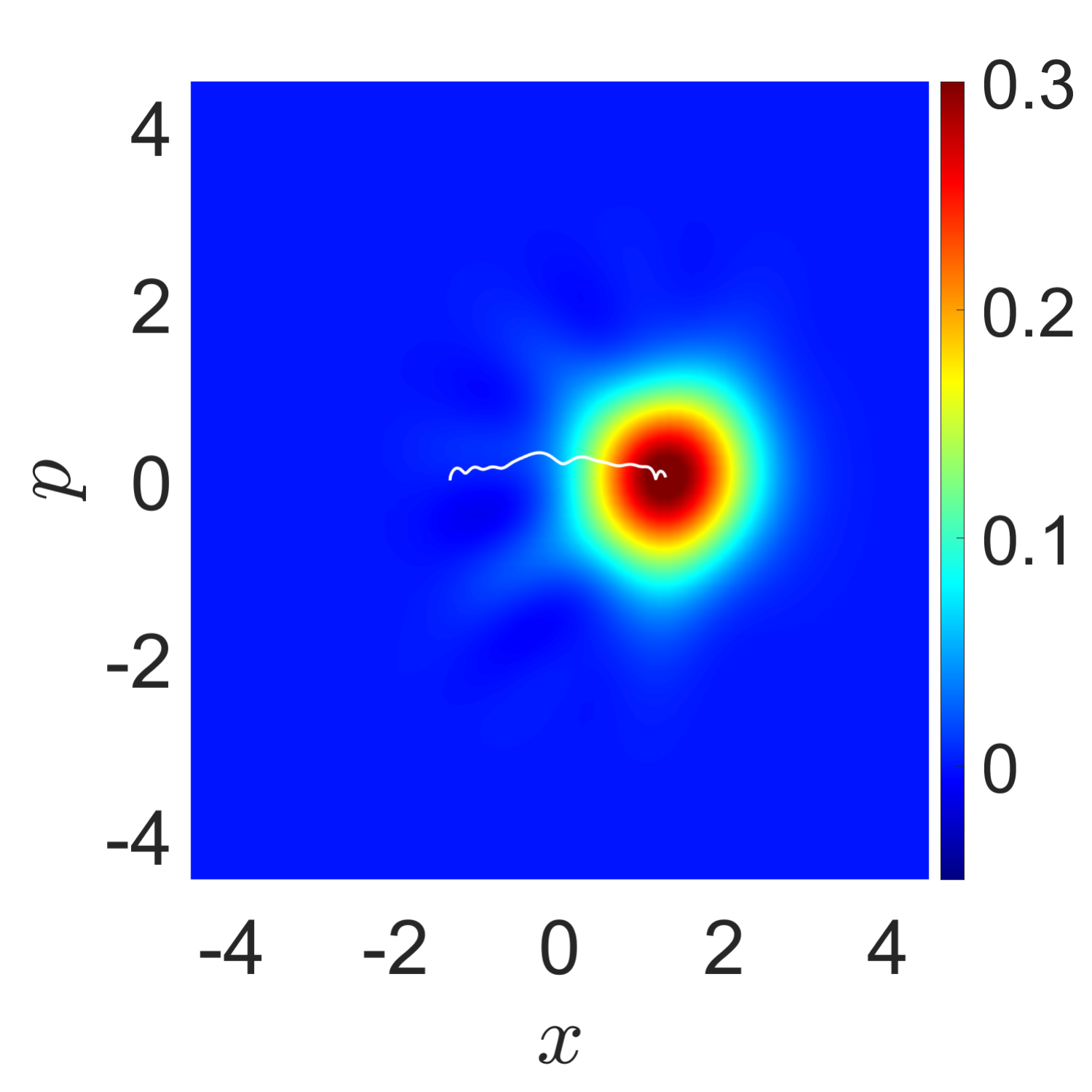}
 	 \caption{: t=16.72}
\end{subfigure}&
\begin{subfigure}[c]{.22\textwidth} \label{fig:WigEnd2DoubleHam}
	  \includegraphics[width=\textwidth]{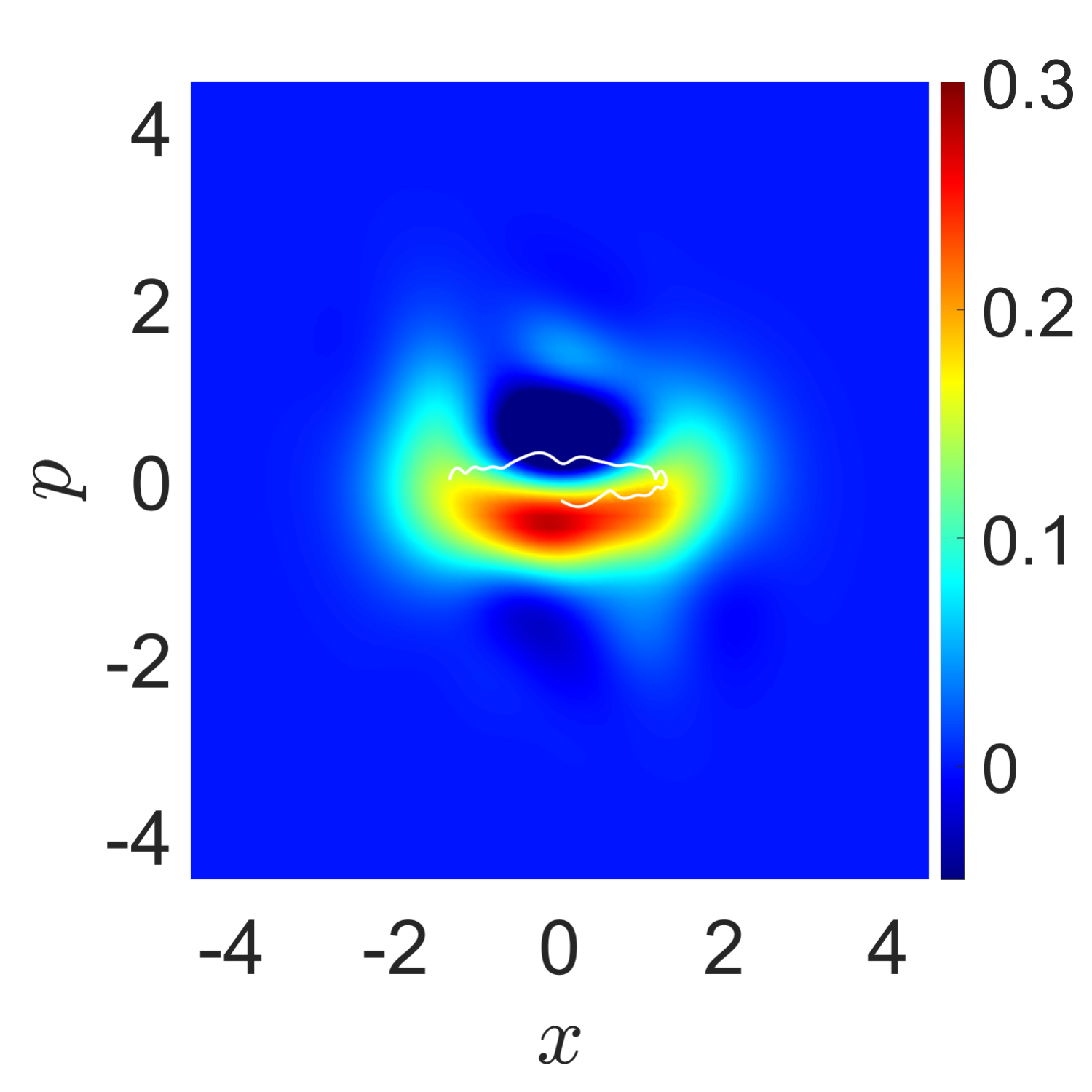}
 	 \caption{: t=25.08}
\end{subfigure}
\end{tabular}
\caption{Dynamical tunnelling in the shallow quadratic well with Gaussian barrier $V_2$ \cref{eq:gausswell} with $A=3$ and $\sigma=\tfrac{1}{\sqrt2}$, for an initial coherent state placed at the minimum of the left well. Top row: (a) Expectation values of position ($\langle \hat X\rangle$) and momentum ($\langle \hat P\rangle$) in phase space (on top of the contour plot of the energy), (b) time dependence of $\langle\hat X\rangle$, (c) time-dependence of $\langle \hat P\rangle$, (d) time-dependence of the negativity of the Wigner function.
Bottom row: Corresponding Wigner functions at selected times,  (e): $t=0$, (f): $t=\frac12 t_{\text{tunnel}}$, (g): $t=t_{\text{tunnel}}$, and (h): $t=\frac32 t_{\text{tunnel}}$, where the tunnel time is given by \eqref{eq:t-tunnel} as $t_{tunnel}\approx 33.44$.\label{fig:TunnelPlots2}}
\end{figure}

In \cref{fig:TunnelPlots2} we depict expectation values of position and momentum, as well as the negativity of the Wigner function for the time evolution of an initial coherent state centred at the minimum of the left well. The corresponding classical system would not experience any dynamical change at all, as the initial configuration corresponds to a stable fixed point. In the quantum system, we observe the expected tunnelling behaviour. The position expectation value oscillates between the left and the right minimum with a period, which is obtained from \eqref{eq:t-tunnel} and numerically calculated eigenenergies as  $t_{tunnel}\approx 33.44$. While the oscillation in the position expectation value is accompanied by an oscillation of the momentum expectation, as intuitively expected from a classical transition, the value of the momentum remains small at all times (and in particular $\frac{\langle \hat P\rangle^2}{2}$ remains lower than the value of the potential barrier). Thus, the transition differs crucially from a classical one. This is most obvious when considering the Wigner function, which is depicted in the lower panel of the figure at four selected times. While the phase-space center moves between the two minima, the Wigner function elongates and stretches over both minima, while developing negative areas related to interference effects, and clearly indicating the non-classicality of the process \cite{Wignegnegativity}. The negativity of the Wigner function, defined as
\begin{equation} \label{eq:WigNeg}
    \norm{W_{-}}=\int \frac{dx dp} {4 \pi \hbar}\bigg(W(x,p)-\abs{W(x,p)}\bigg),
\end{equation}
is depicted in the right panel of the top row. We observe an oscillating behaviour with maxima of the negativity corresponding to the crossings through the barrier by the position expectation value. 

\begin{figure}[t]
\centering
\begin{tabular}{c c c c}
\begin{subfigure}[c]{.23\textwidth} 
	  \includegraphics[width=\textwidth]{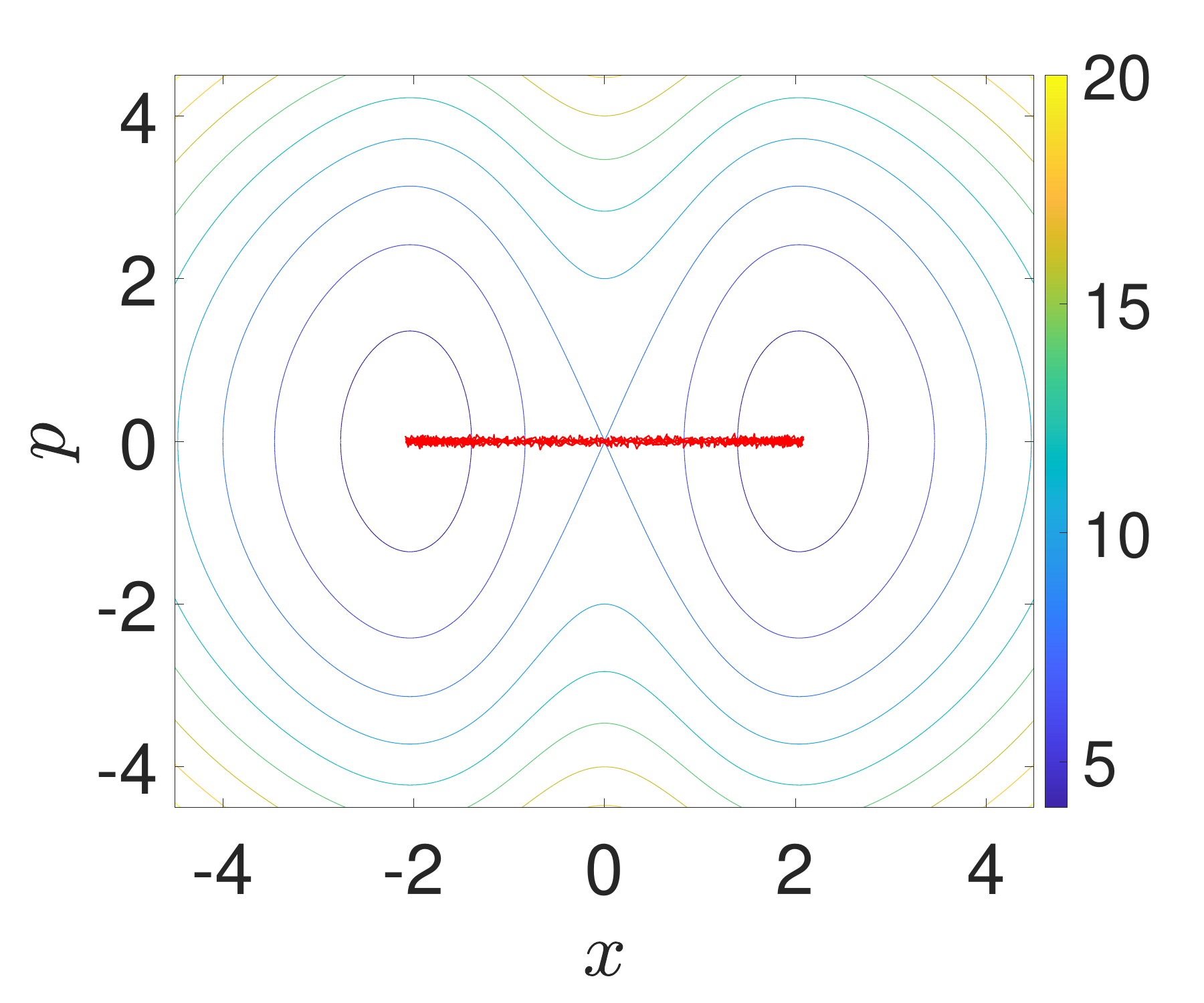}
 	 \caption{}\label{fig:G_Deep_HGaussContour_tunnel}
\end{subfigure}&
\begin{subfigure}[c]{.23\textwidth} 
	  \includegraphics[width=\textwidth]{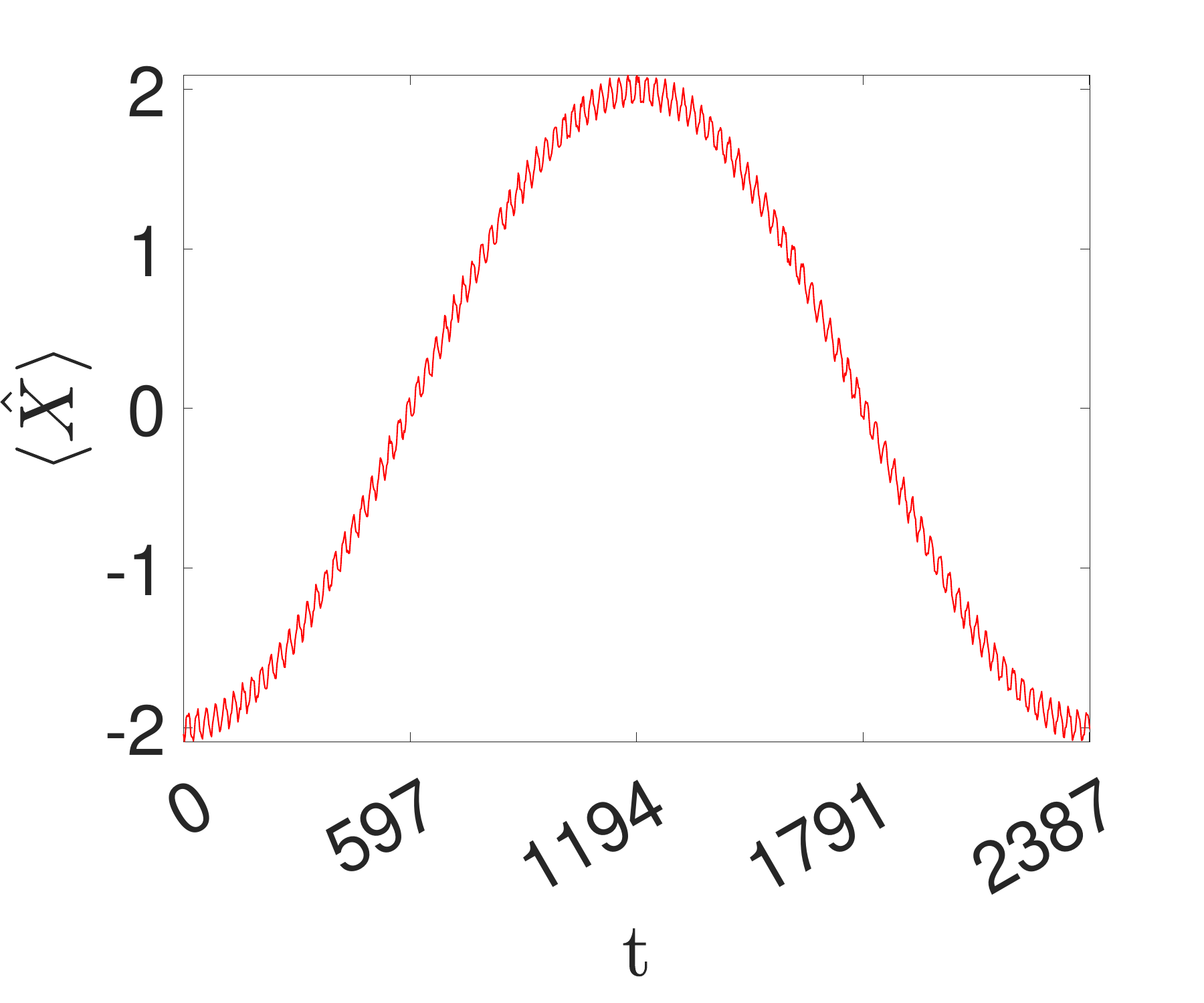}
 	 \caption{}\label{fig:G_Deep_XDoubleHam}
\end{subfigure}&
\begin{subfigure}[c]{.23\textwidth} 
	  \includegraphics[width=\textwidth]{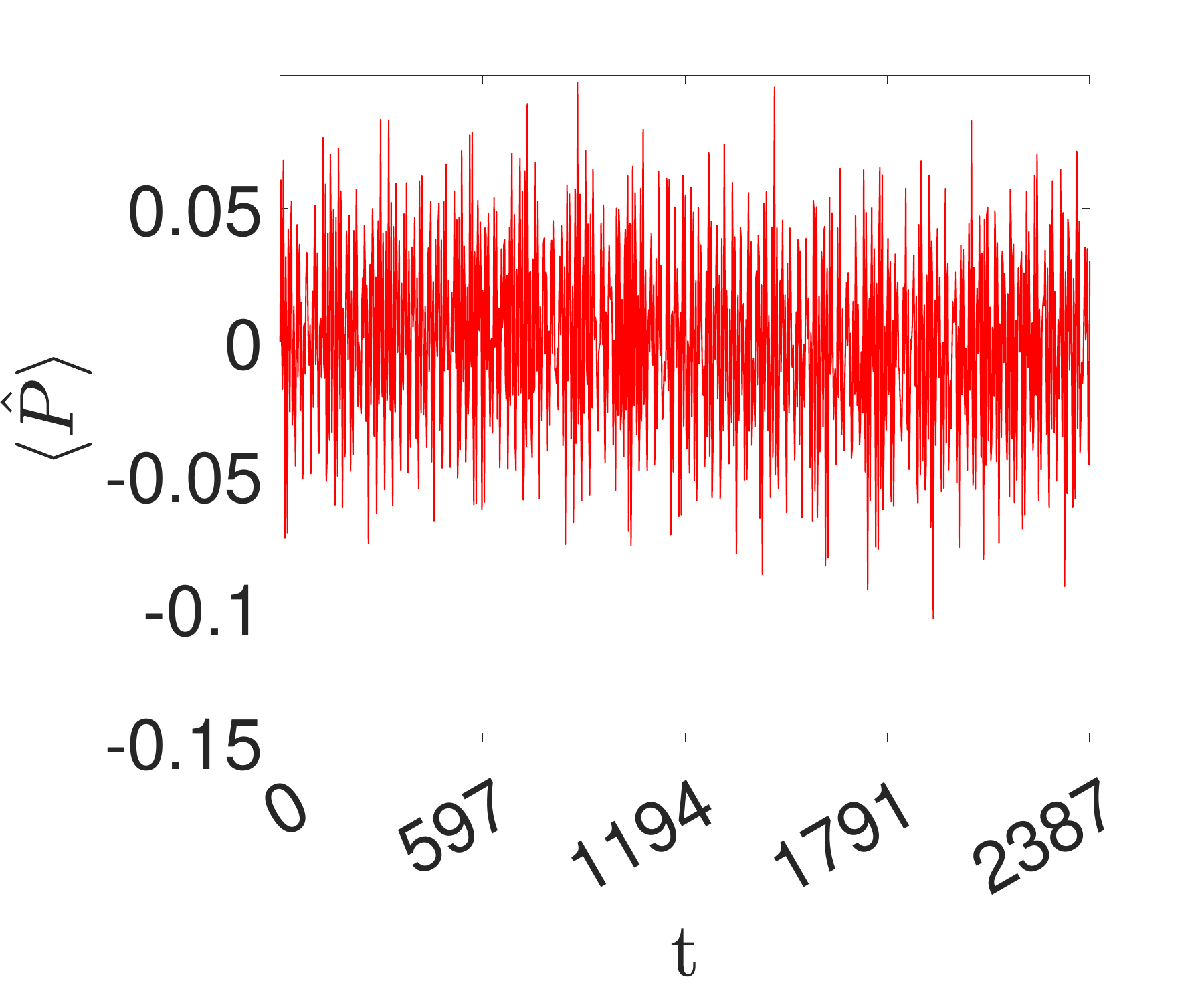}
 	 \caption{}\label{fig:G_Deep_PDoubleHam}
\end{subfigure}&
\begin{subfigure}[c]{.23\textwidth} 
	  \includegraphics[width=\textwidth]{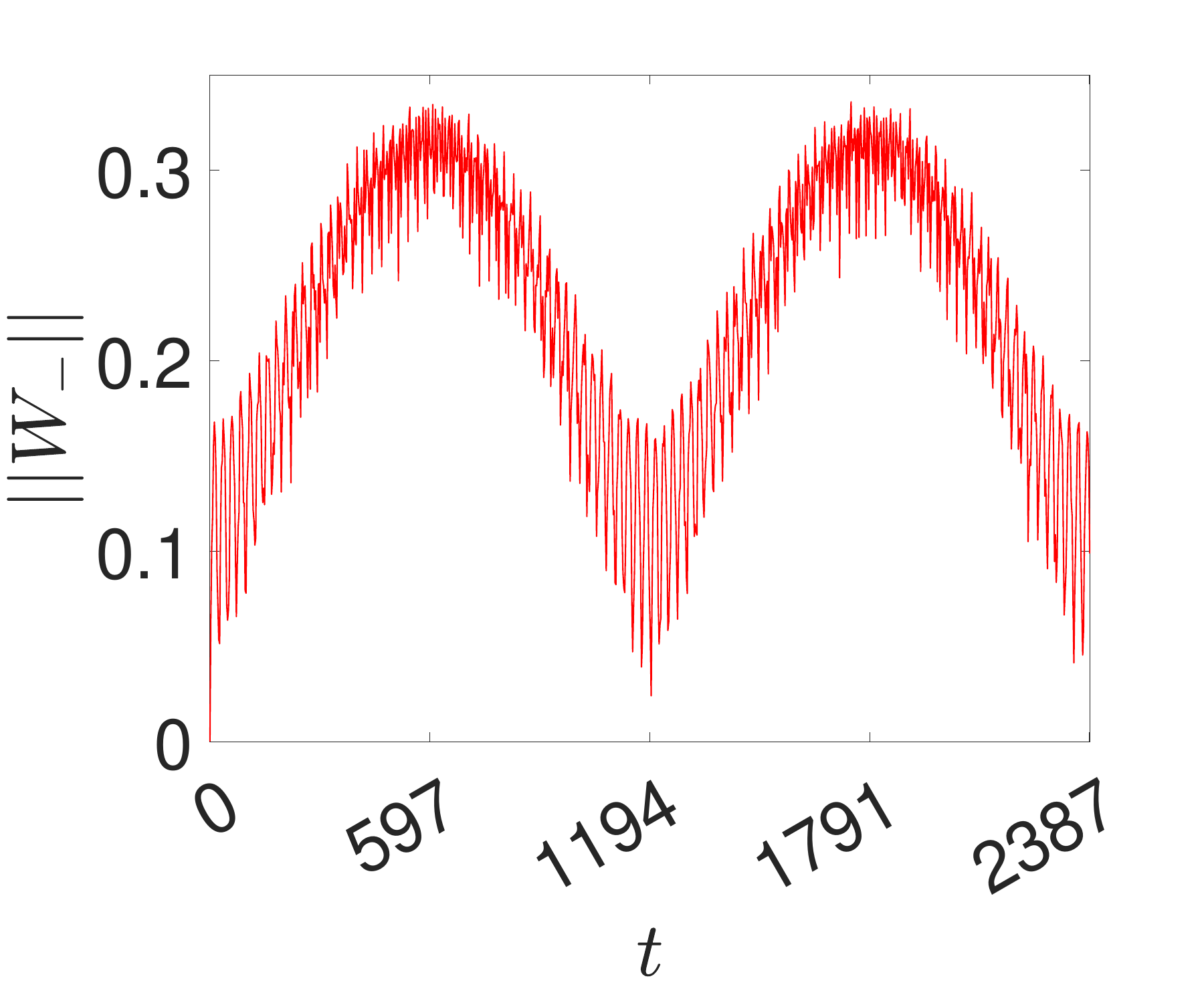}
 	 \caption{}\label{fig:G_Deep_WigNegDoubleHam}
\end{subfigure}\\
\begin{subfigure}[c]{.22\textwidth} \label{fig:G_Deep_WigStartDoubleHam}
	  \includegraphics[width=\textwidth]{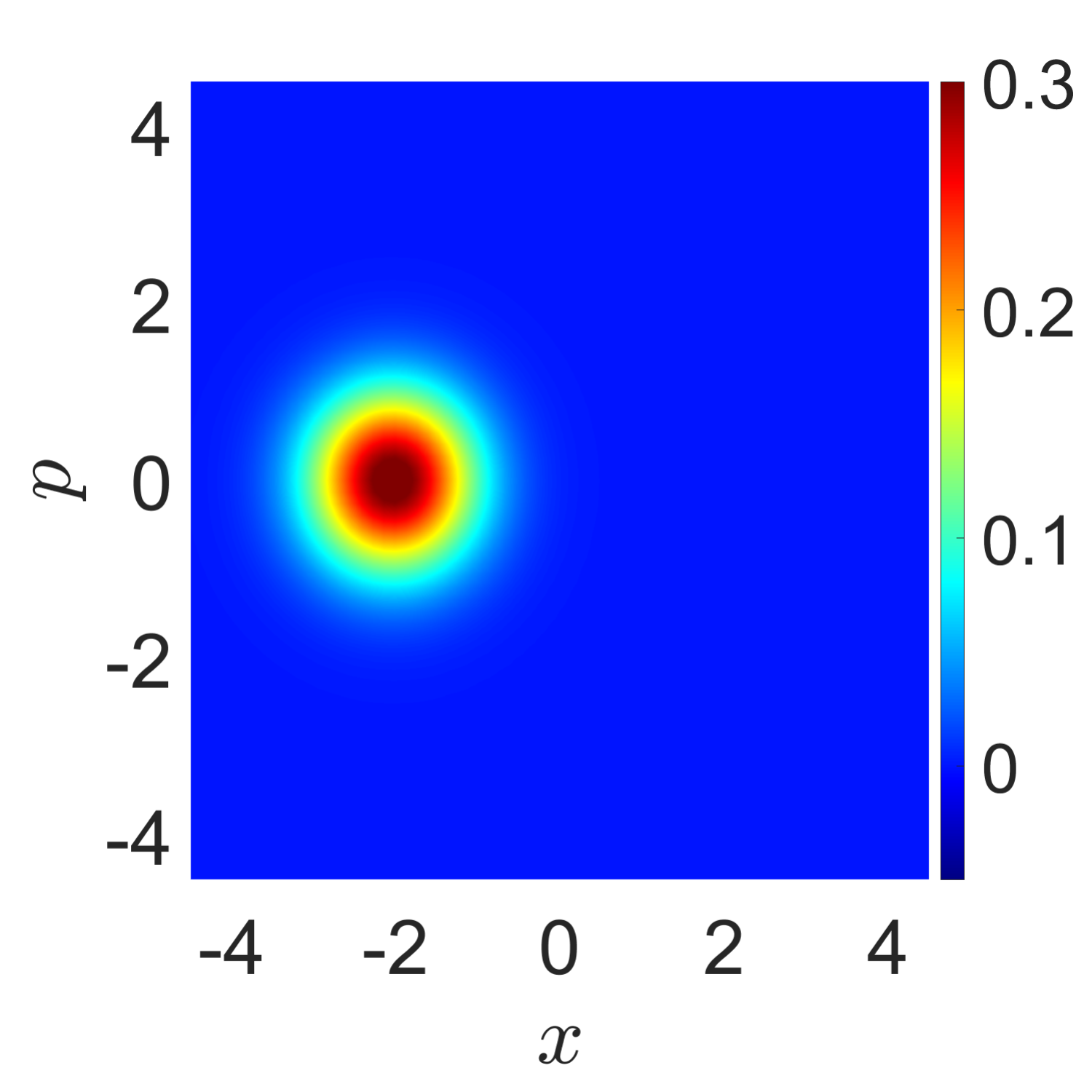}
 	 \caption{: t=0}
\end{subfigure}&
\begin{subfigure}[c]{.22\textwidth} \label{fig:G_Deep_WigMidDoubleHam}
	  \includegraphics[width=\textwidth]{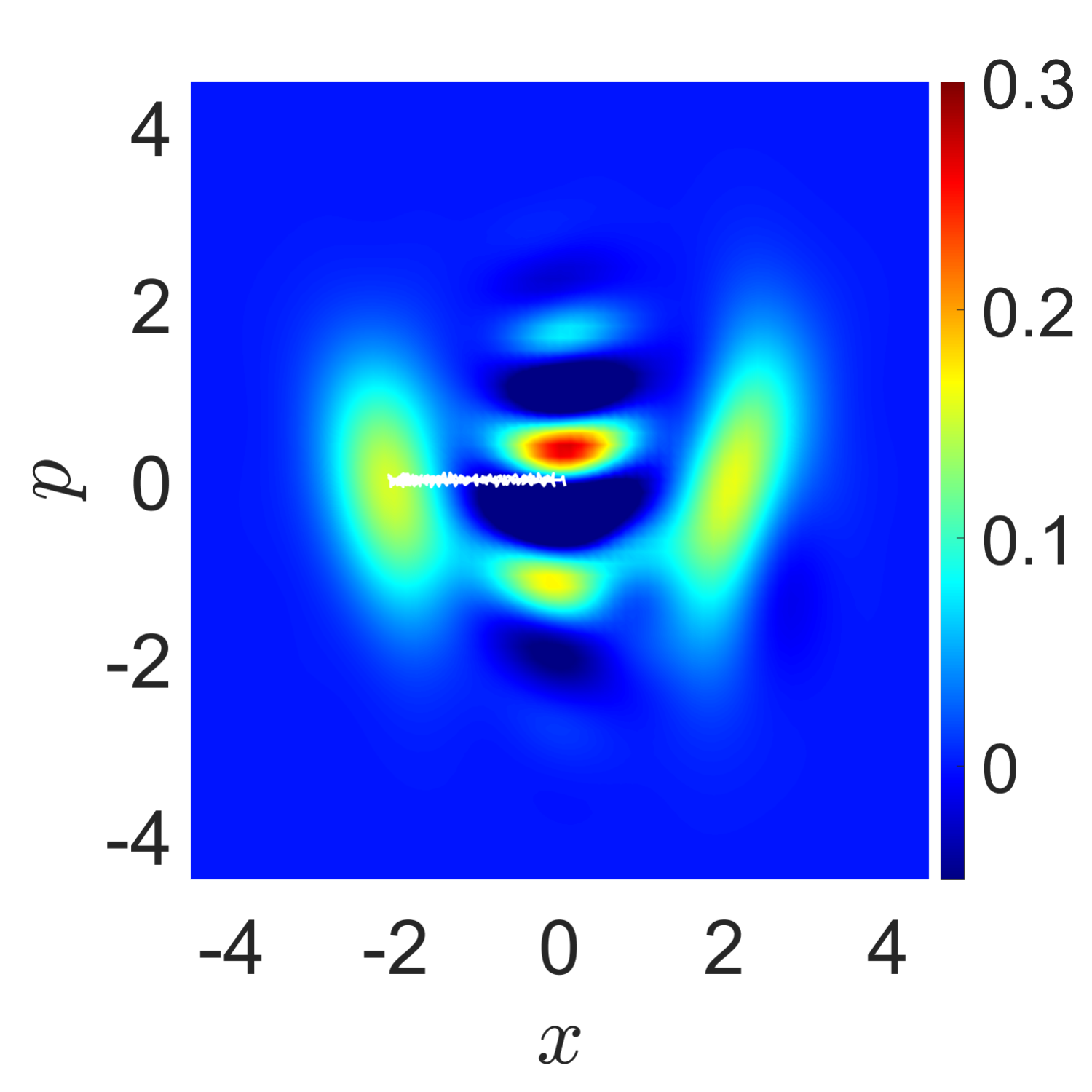}
 	 \caption{: t=597}
\end{subfigure}&
\begin{subfigure}[c]{.22\textwidth} \label{fig:G_Deep_WigEndDoubleHam}
	  \includegraphics[width=\textwidth]{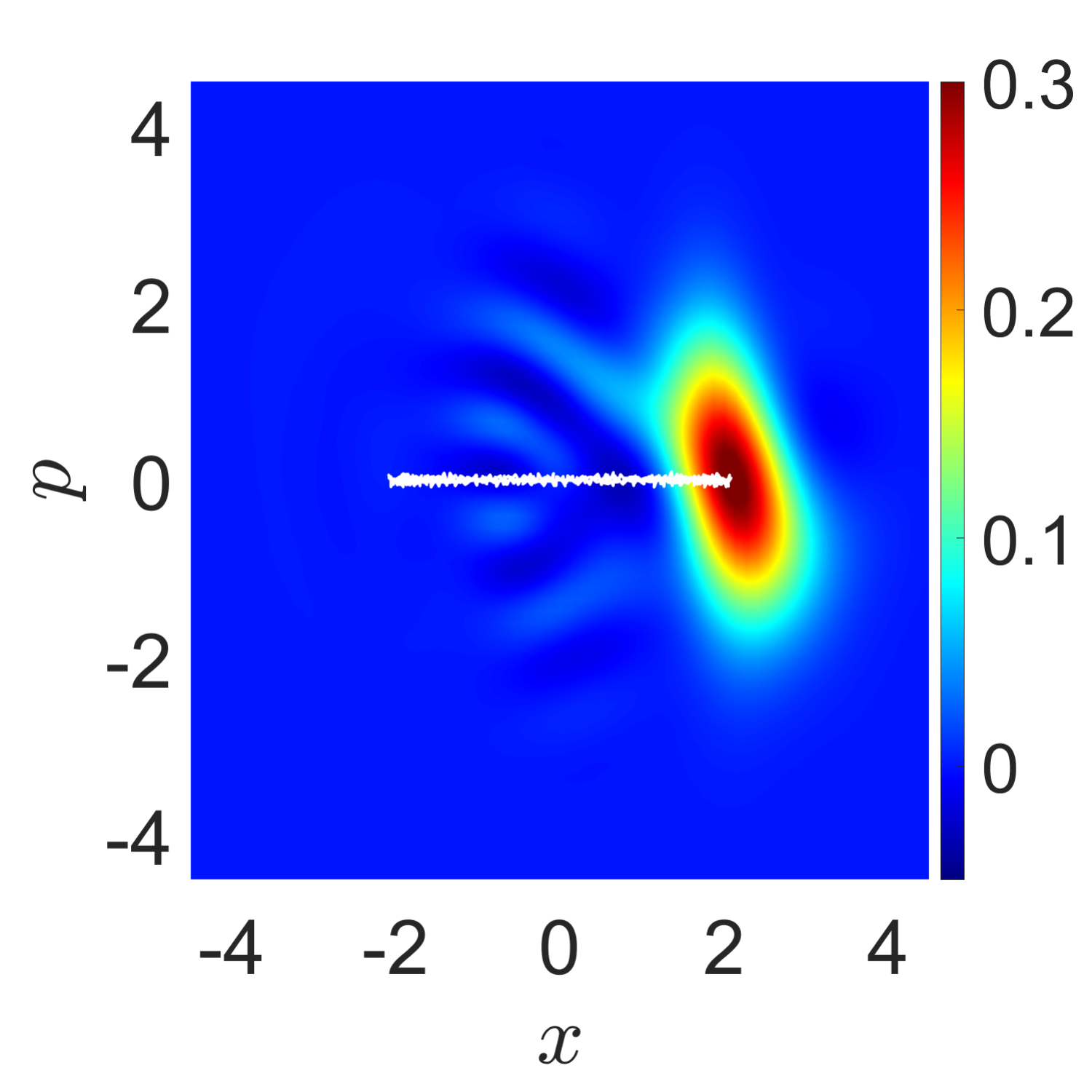}
 	 \caption{: t=1194}
\end{subfigure}&
\begin{subfigure}[c]{.22\textwidth} \label{fig:G_Deep_WigEnd2DoubleHam}
	  \includegraphics[width=\textwidth]{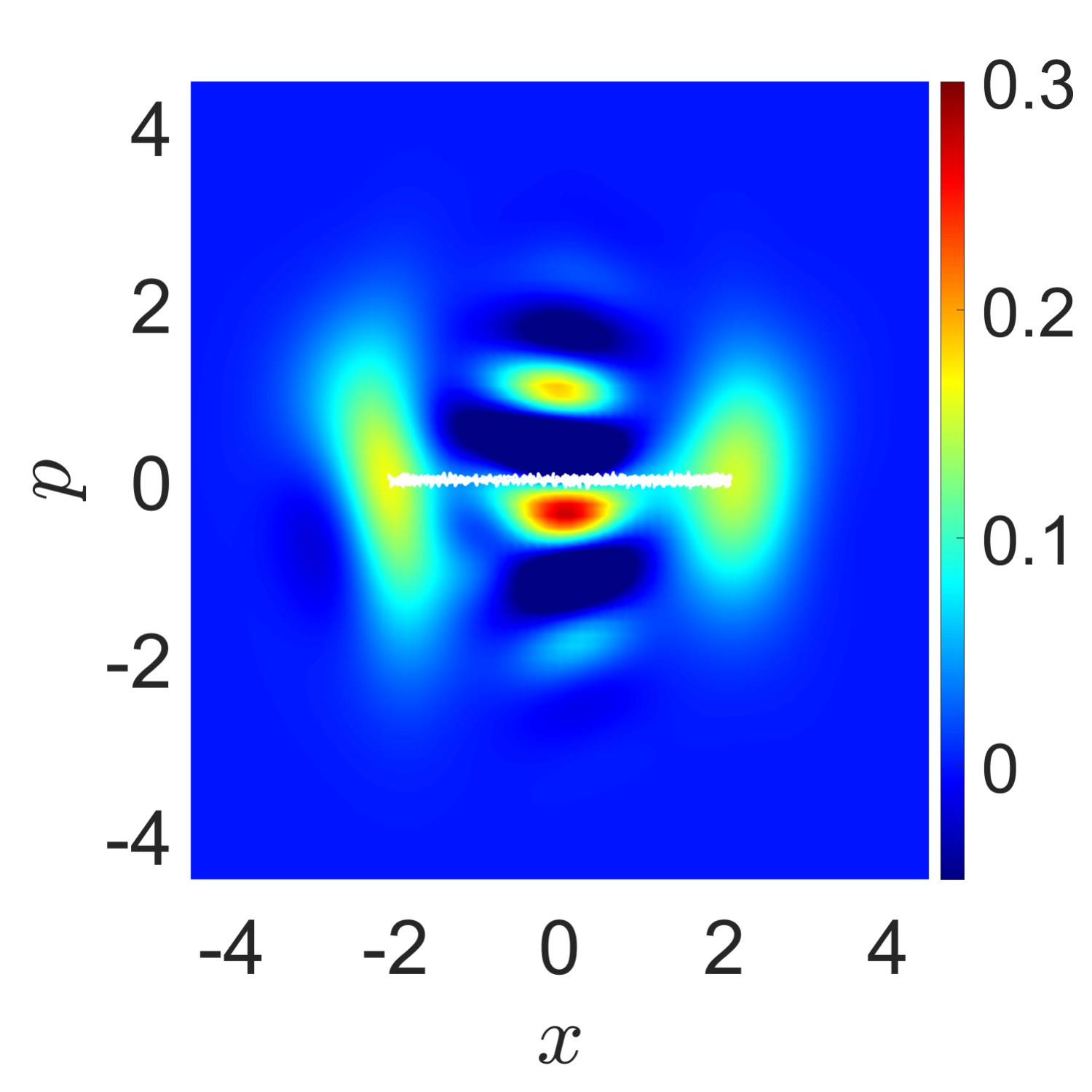}
 	 \caption{: t=1791}
\end{subfigure}
\end{tabular}
\caption{As figure \cref{fig:TunnelPlots2}, for $A=8$ and $\sigma=1$. The tunnel time here is given by $t_{tunnel}\approx 1200$.  \label{fig:TunnelPlots5} }
\end{figure}

Figure \ref{fig:TunnelPlots5} depicts the tunnelling for an initial Gaussian wave packet centred in the left well for a deeper double-well potential. We observe the same overall behaviour as for the shallow well, however, on vastly longer time scales, as expected. The momentum acquired during the transition is so small that the overall oscillation in the momentum expectation value is hardly visible in the surrounding noise stemming from the contribution from higher excited states. We note that the maximum of the negativity of the Wigner function here is even higher than in the case of the shallow well. 

In what follows we shall investigate the modification of the quantum tunnelling dynamics under the influence of a finite temperature as described by the stochastic Schr\"odinger equation arising for the model \cref{eq:heatbath}. 

\FloatBarrier

\subsection{Quantum transitions at finite temperatures} \label{sec:Transition}

\begin{figure}[t]
\centering 
\begin{tabular}{c c c c}
\begin{subfigure}[c]{.23\textwidth} 
	  \includegraphics[width=\textwidth]{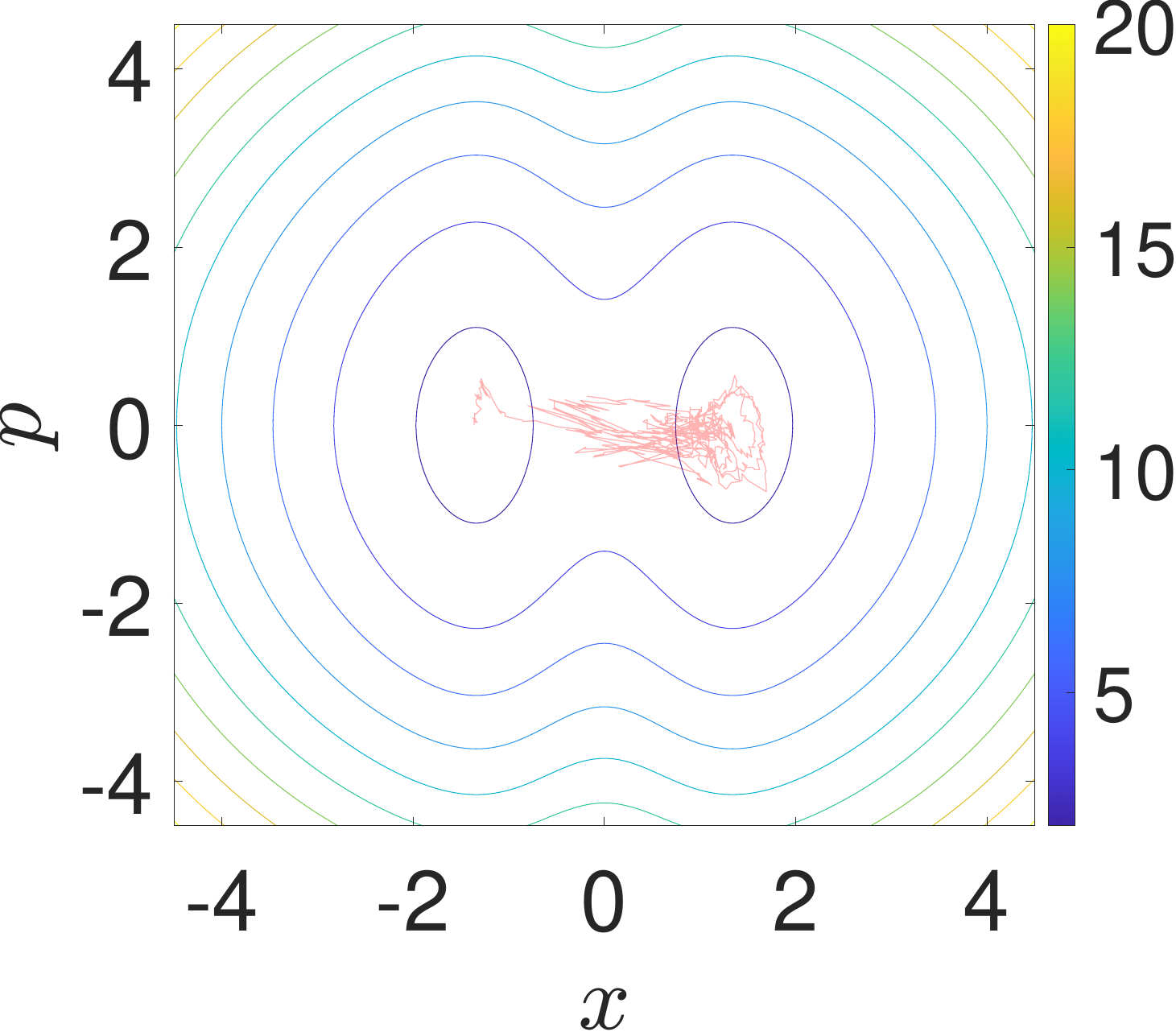}
 	 \caption{}\label{fig:QSDContourLow.pdf}
\end{subfigure}&
\begin{subfigure}[c]{.23\textwidth} 
	  \includegraphics[width=\textwidth]{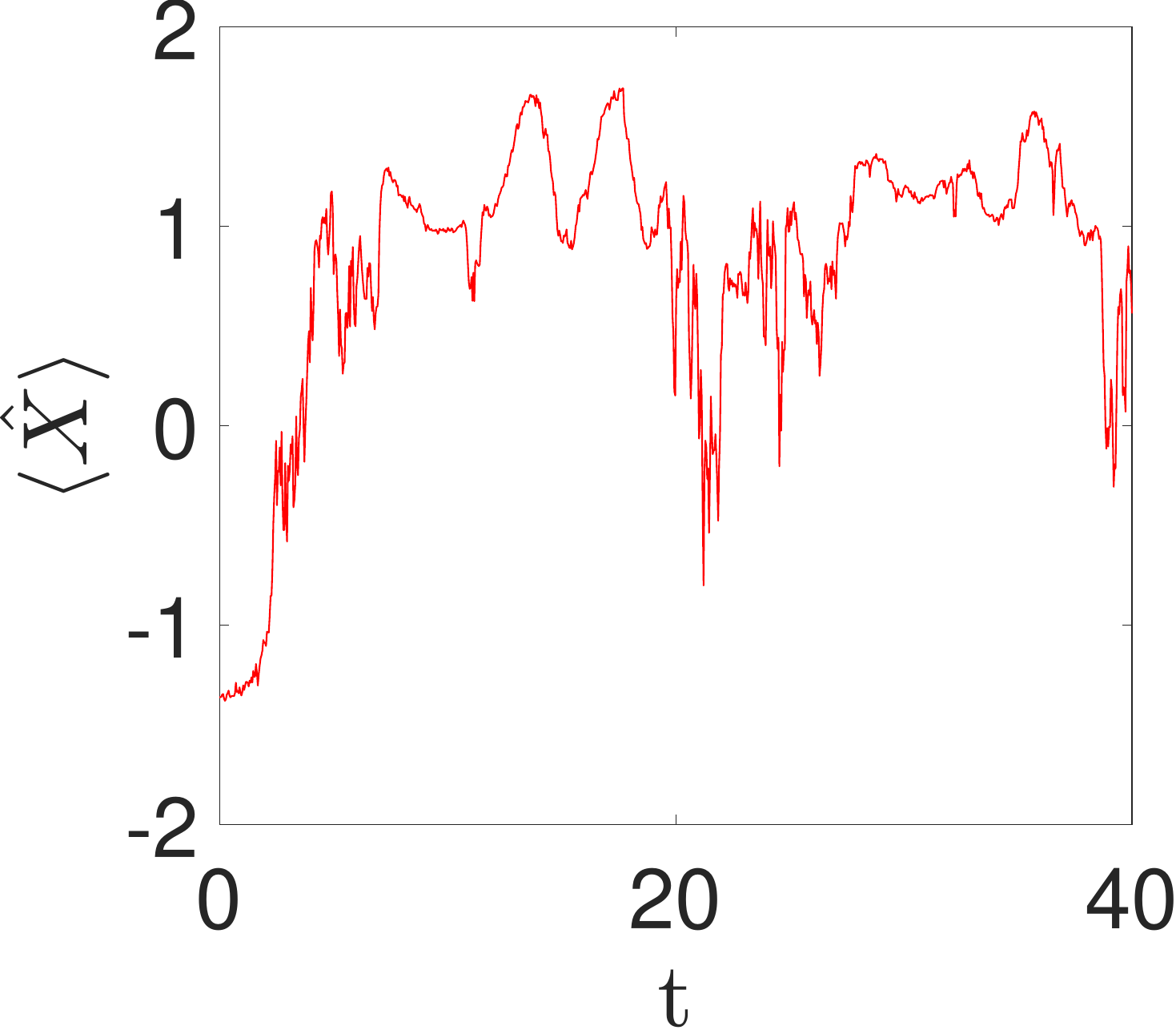}
 	 \caption{}\label{fig:XDoubleQSDLow}
\end{subfigure}&
\begin{subfigure}[c]{.23\textwidth} 
	  \includegraphics[width=\textwidth]{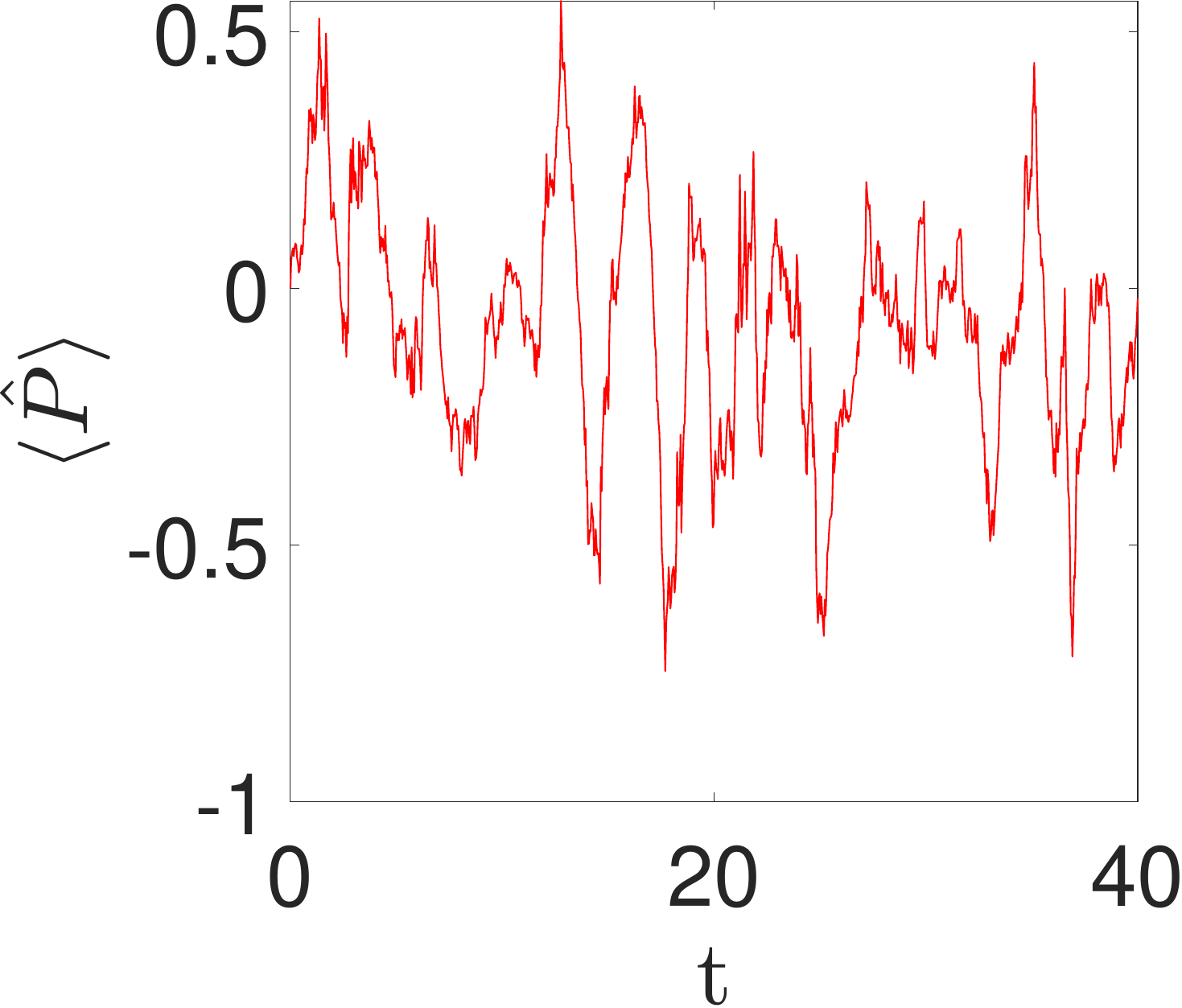}
 	 \caption{}\label{fig:PDoubleQSDLow}
\end{subfigure}&
\begin{subfigure}[c]{.23\textwidth} 
	  \includegraphics[width=\textwidth]{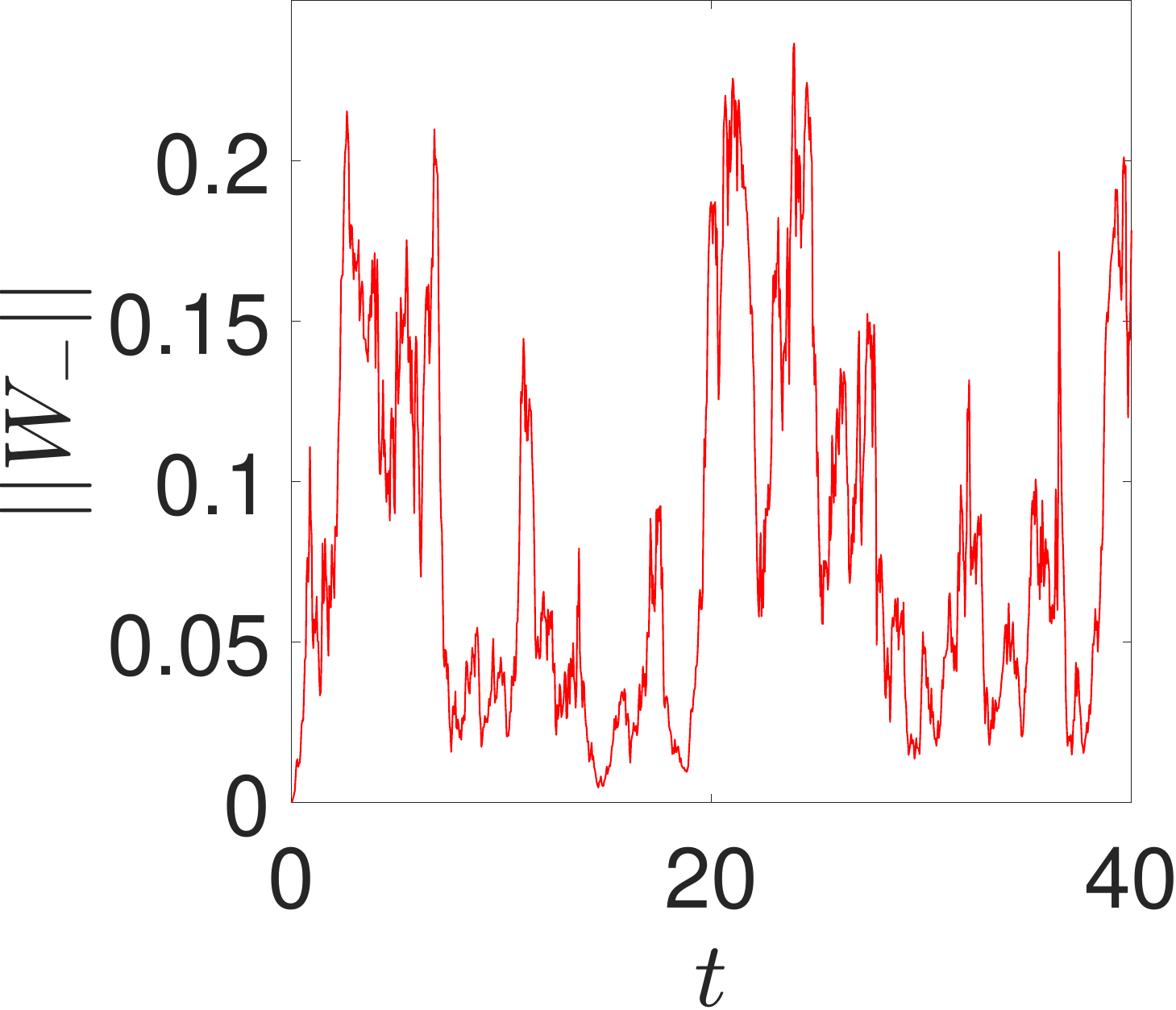}
 	 \caption{}\label{fig:WigNegDoubleQSDLow}
\end{subfigure}\\
\begin{subfigure}[c]{.23\textwidth} 
	  \includegraphics[width=\textwidth]{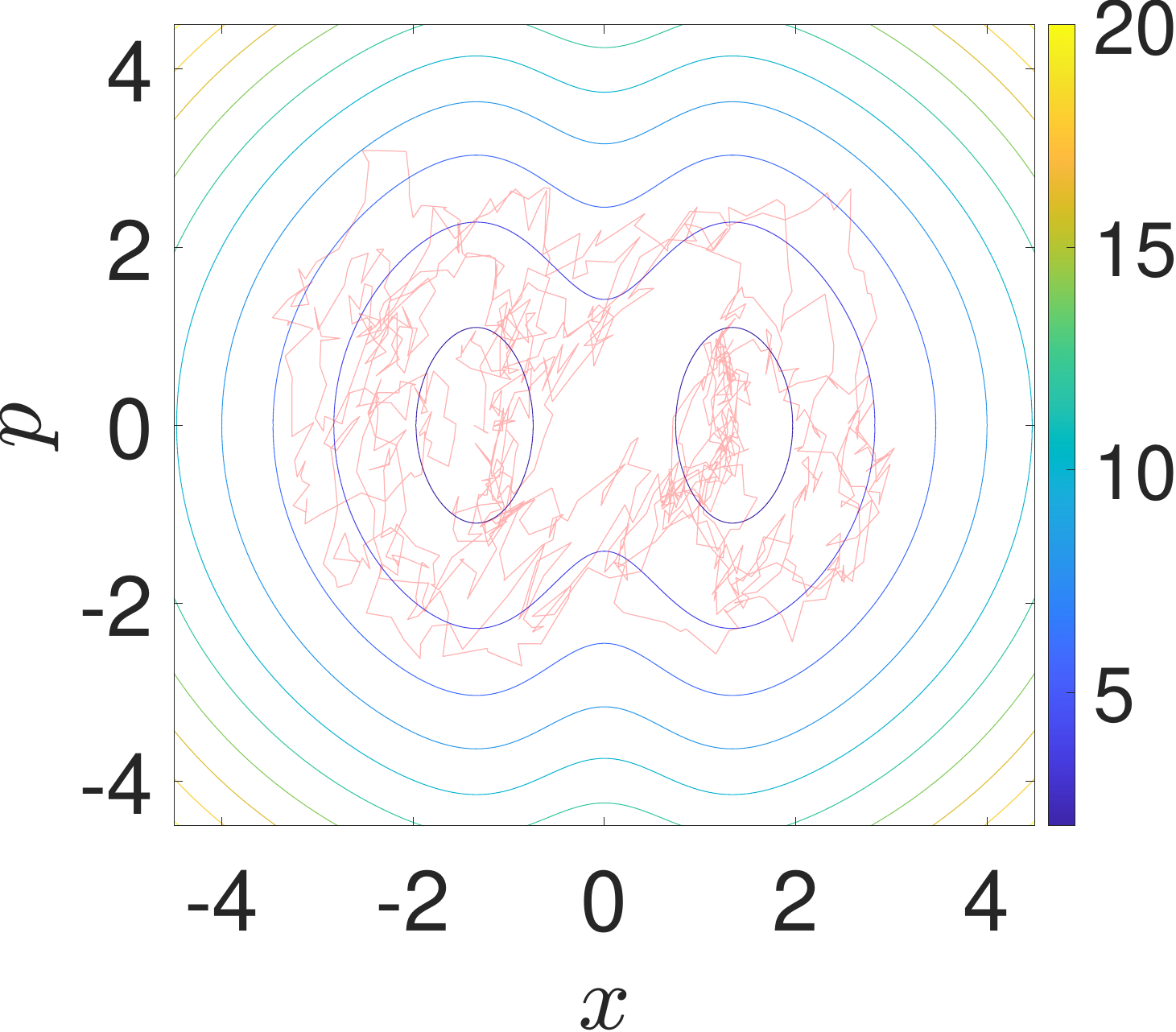}
 	 \caption{}\label{fig:QSDContourHigh.pdf}
\end{subfigure}&
\begin{subfigure}[c]{.23\textwidth} 
	  \includegraphics[width=\textwidth]{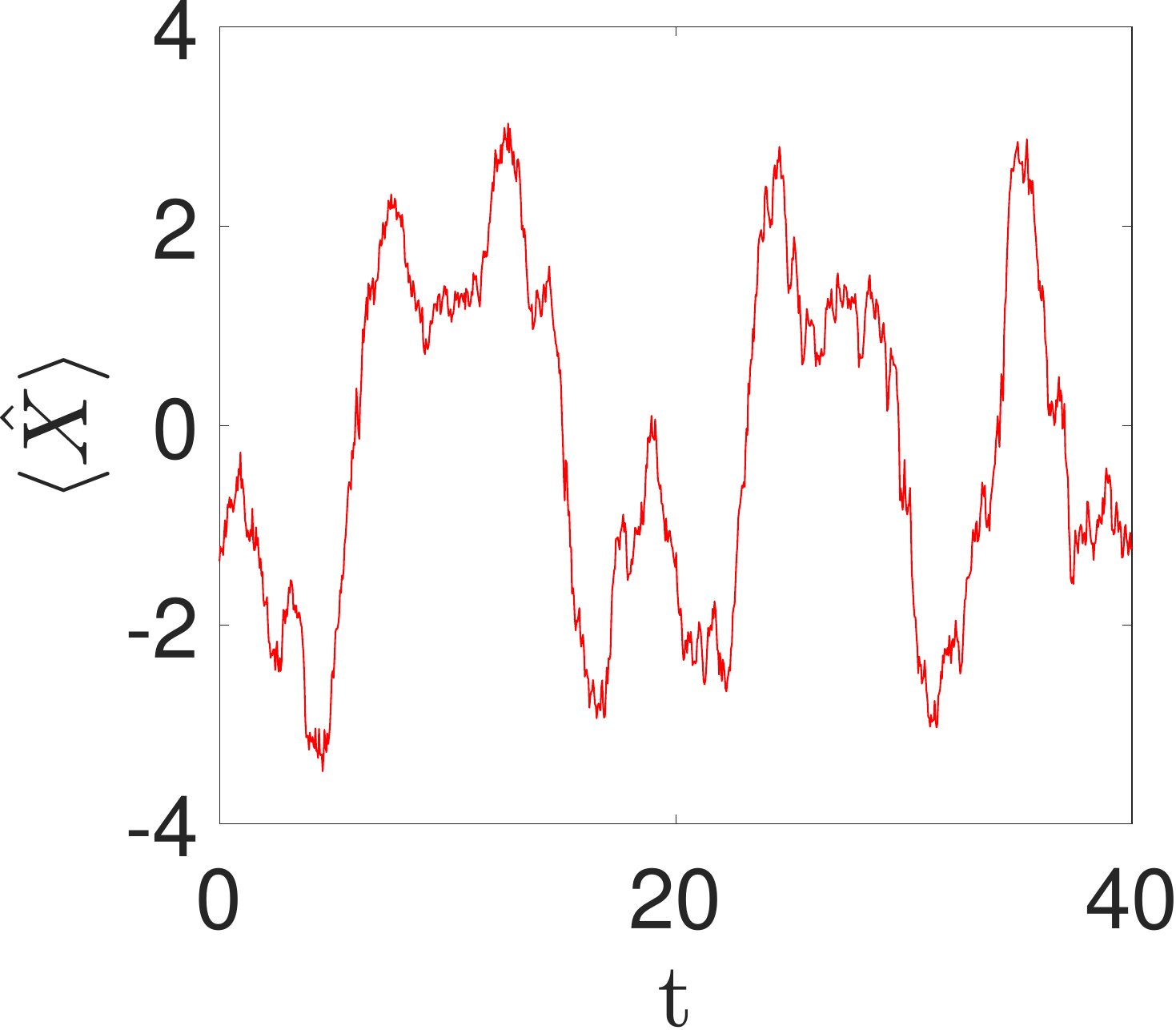}
 	 \caption{}\label{fig:XDoubleQSDHigh}
\end{subfigure}&
\begin{subfigure}[c]{.23\textwidth} 
	  \includegraphics[width=\textwidth]{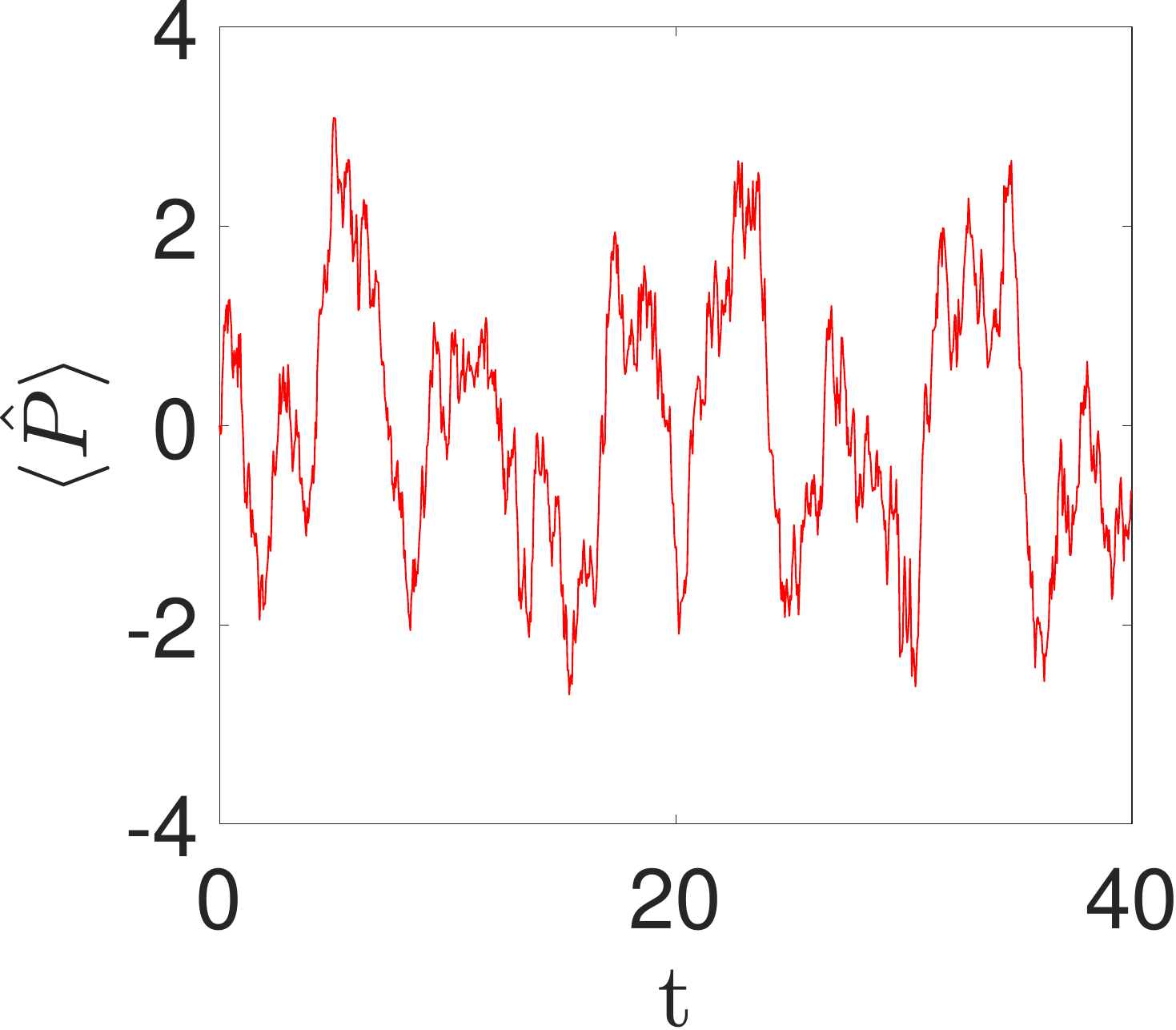}
 	 \caption{}\label{fig:PDoubleQSDHigh}
\end{subfigure}&
\begin{subfigure}[c]{.23\textwidth} 
	  \includegraphics[width=\textwidth]{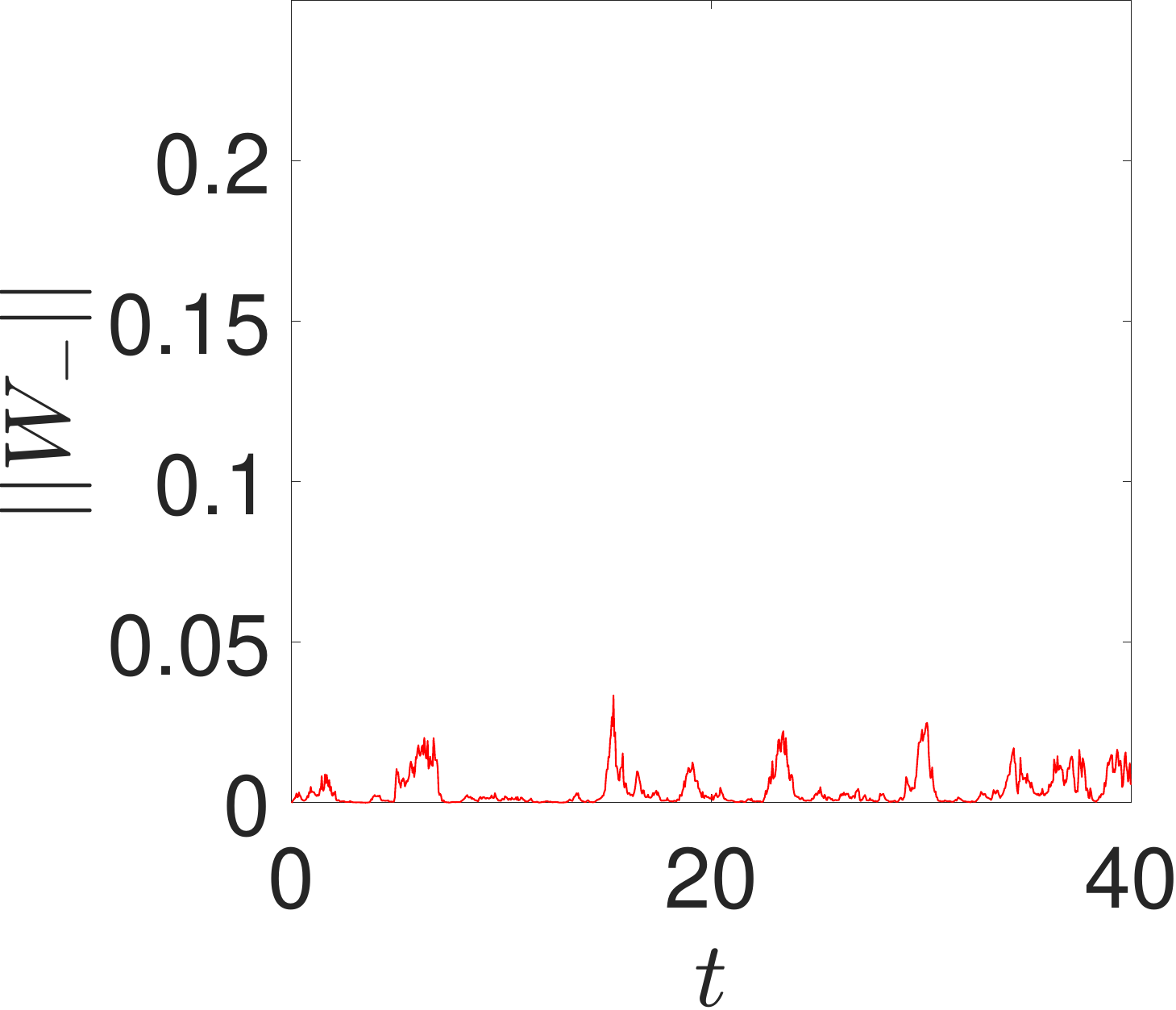}
 	 \caption{}\label{fig:WigNegDoubleQSDHigh}
\end{subfigure}
\end{tabular}
\caption{SSE trajectories of an initial coherent state in the minimum of the left well \cref{eq:gausswell} with $A=3$,$\gamma=0.25$,  and $\sigma=\tfrac{1}{\sqrt2}$ for different temperatures. From left to right, we show the phase space expectation values over a contour plot of the Hamiltonian, the time dependence of the position expectation $\ev*{\hat X}$, the momentum expectation $\ev*{\hat P}$, and the negativity of the Wigner function. The top row corresponds to $T=0.2$ and the bottom row to $T=3$.\label{fig:TunnelPlots2QSD}}
\end{figure}

In \cref{fig:TunnelPlots2QSD}, we show examples of typical quantum trajectories arising from the stochastic Schr\"odinger equation (\ref{eq:SSE}) with Hamiltonian and Lindbladian (\ref{eq:heatbath})  in a shallow double well at temperatures $T=0.2$ (top row), and $T=3$ (bottom row). The left plot depicts the trajectories in phase space, the second panel from the left depicts the dynamics of the expectation value of the position, the next panel depicts the dynamics of the momentum expectation value, and the right panel depicts the negativity of the Wigner function. Comparing the dynamics to the classical Langevin dynamics in figure \ref{fig:classical_trajectories} and the quantum dynamics at zero temperature in figure \ref{fig:TunnelPlots2}, we observe a combination of tunnelling and classical transitions, resulting in an increase in transitions compared to both the classical and zero temperature quantum dynamics on their own.  

relatively low temperature $T=0.2$ corresponding classical dynamics would only very rarely transition (rate of $\approx 0.01$), we observe one full and two further attempted transitions in the quantum trajectory, associated with high negativity of the Wigner function. We do not observe coherent tunnelling, however, which would happen on slightly longer time scales. It appears as if the classical and quantum transition mechanisms work together here to lead to more frequent transitions than corresponding to either on their own.

\begin{figure}[t] 
\centering
\begin{tabular}{c c c c}
\begin{subfigure}[c]{.22\textwidth} \label{fig:Olap0QSD}
	  \includegraphics[width=\textwidth]{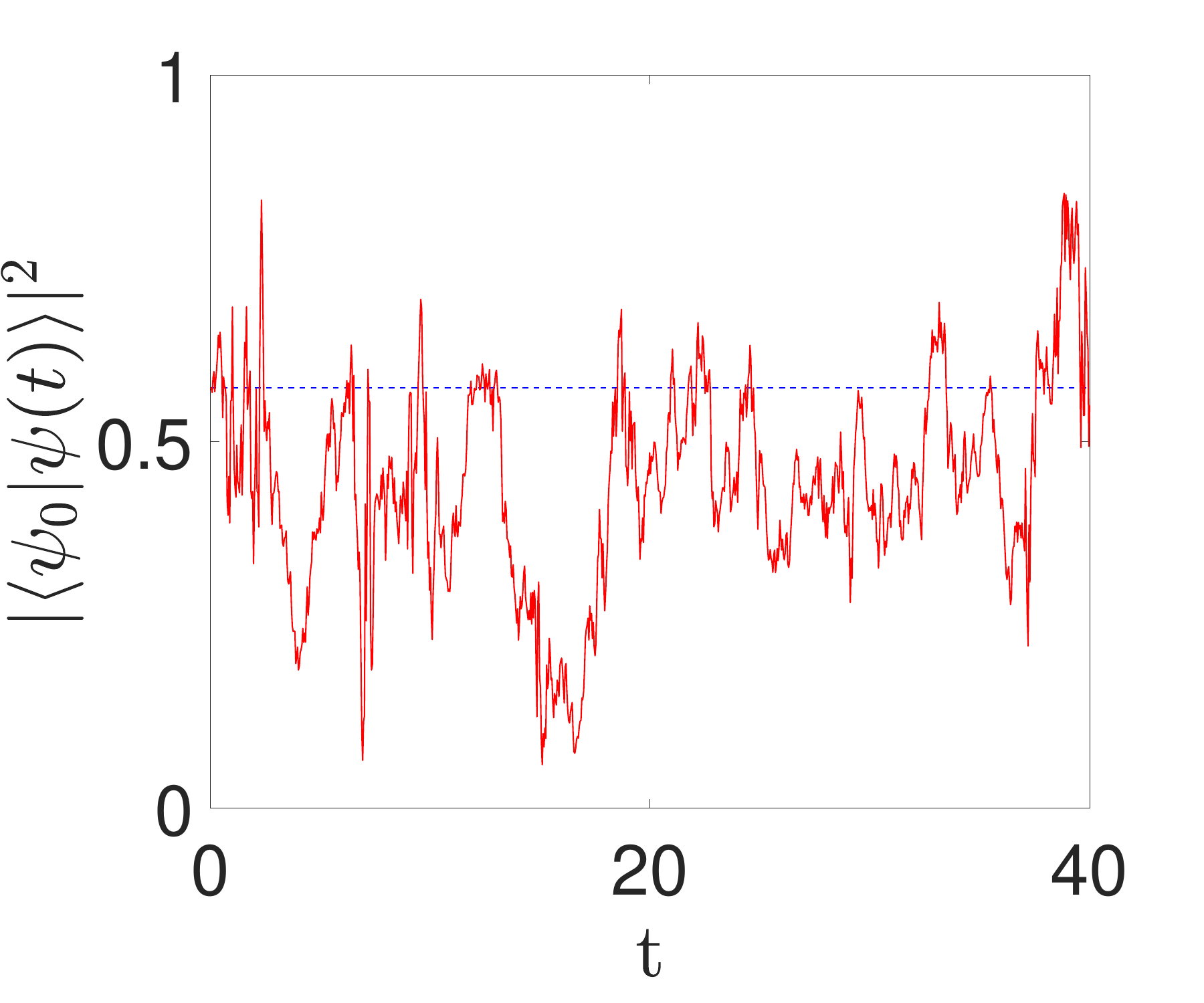}
 	 \caption{}
\end{subfigure}&
\begin{subfigure}[c]{.22\textwidth} \label{fig:Olap1QSD}
	  \includegraphics[width=\textwidth]{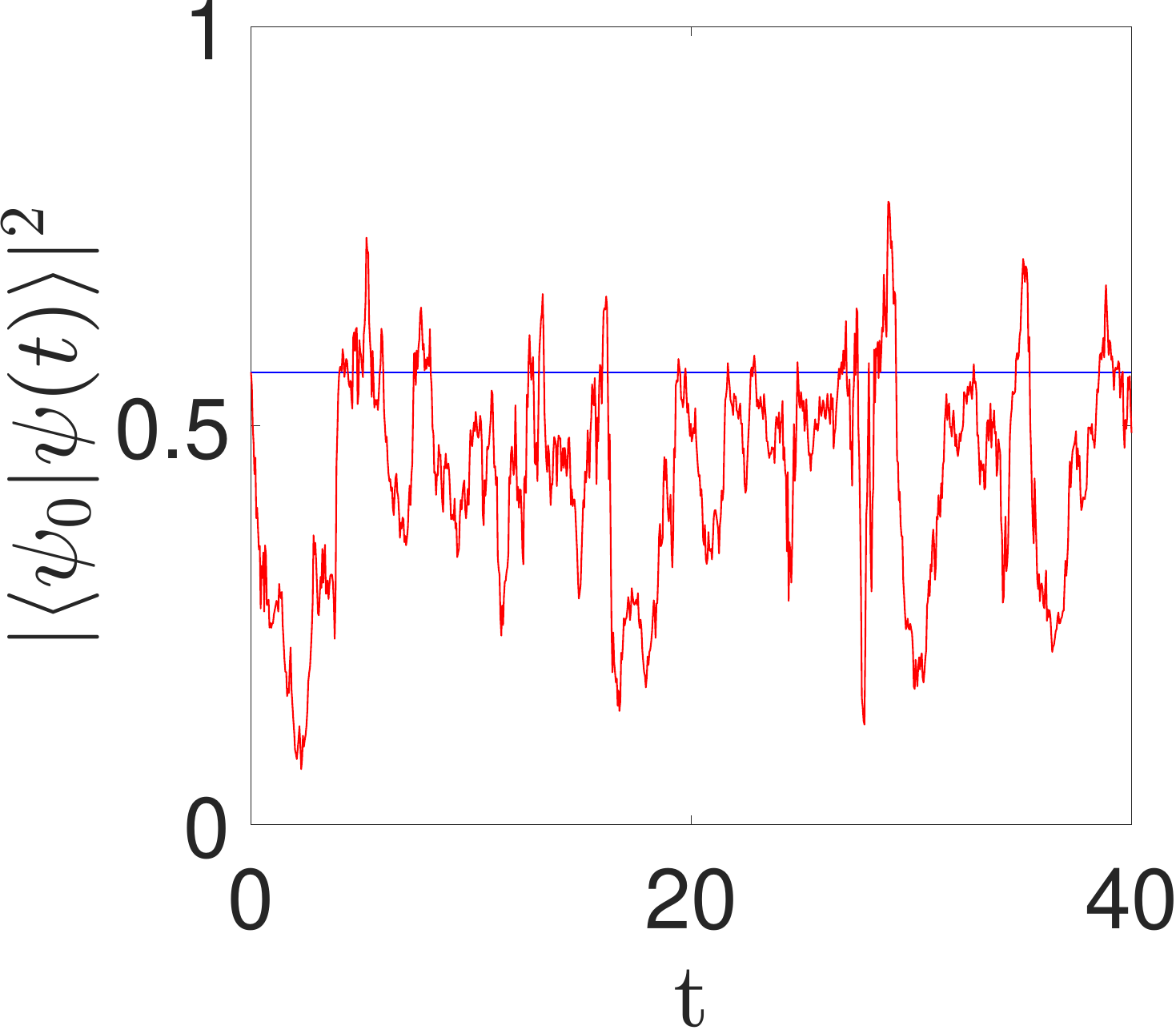}
 	 \caption{}
\end{subfigure}&
\begin{subfigure}[c]{.22\textwidth} \label{fig:Olap2QSD}
	  \includegraphics[width=\textwidth]{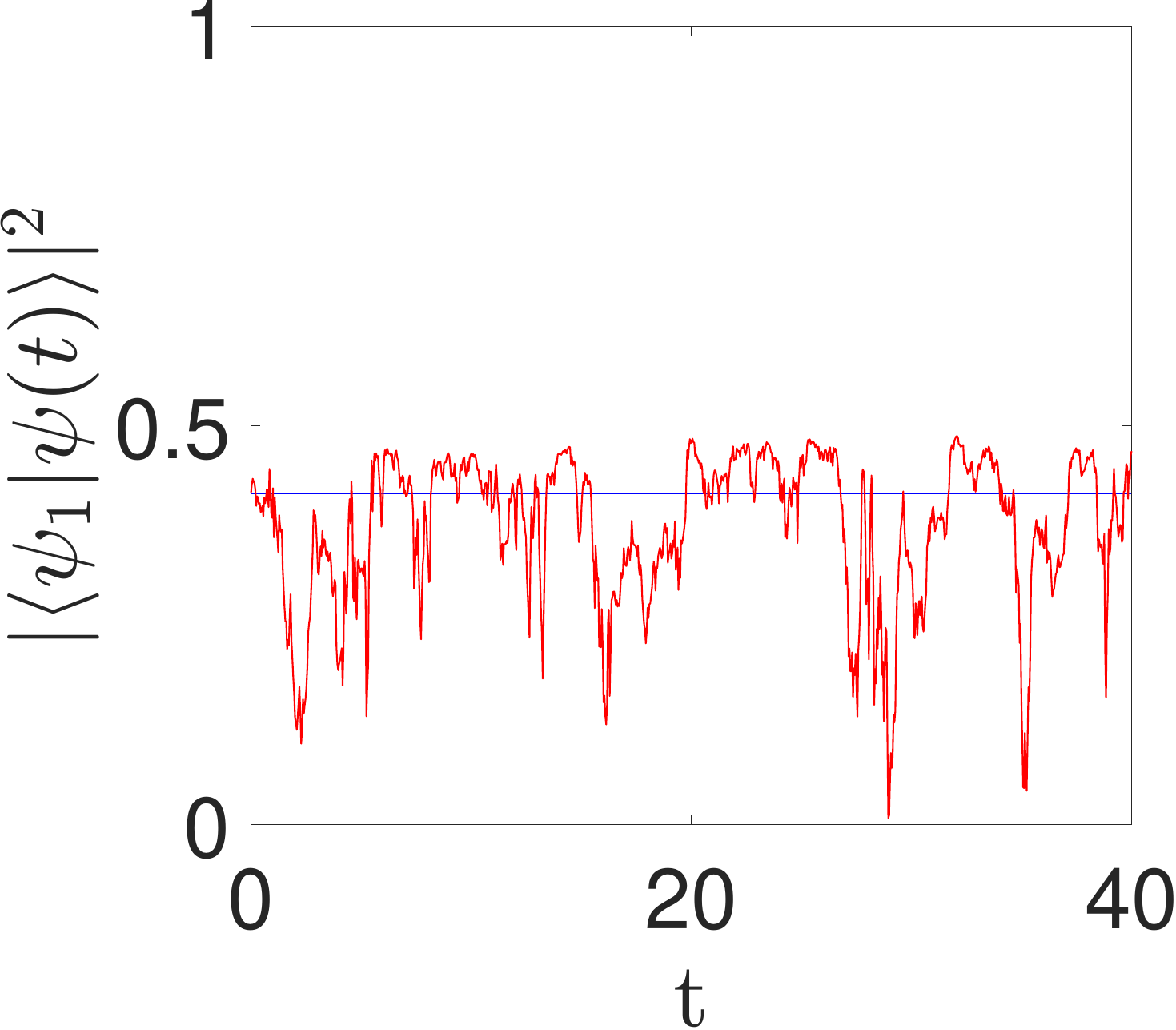}
 	 \caption{}
\end{subfigure}&
\begin{subfigure}[c]{.22\textwidth} \label{fig:Olap3QSD}
	  \includegraphics[width=\textwidth]{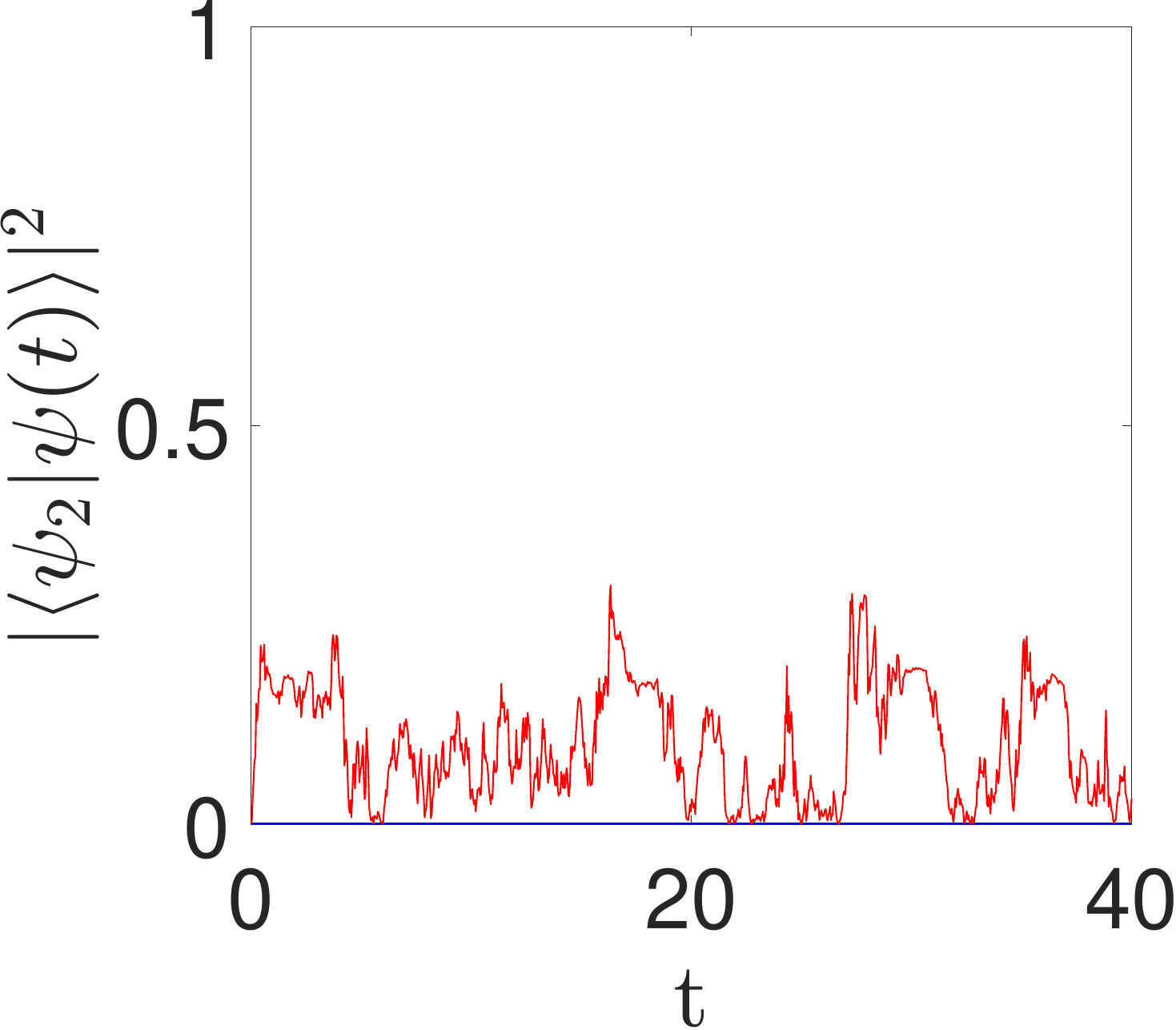}
 	 \caption{}
\end{subfigure}\\
\begin{subfigure}[c]{.22\textwidth} 
	  \includegraphics[width=\textwidth]{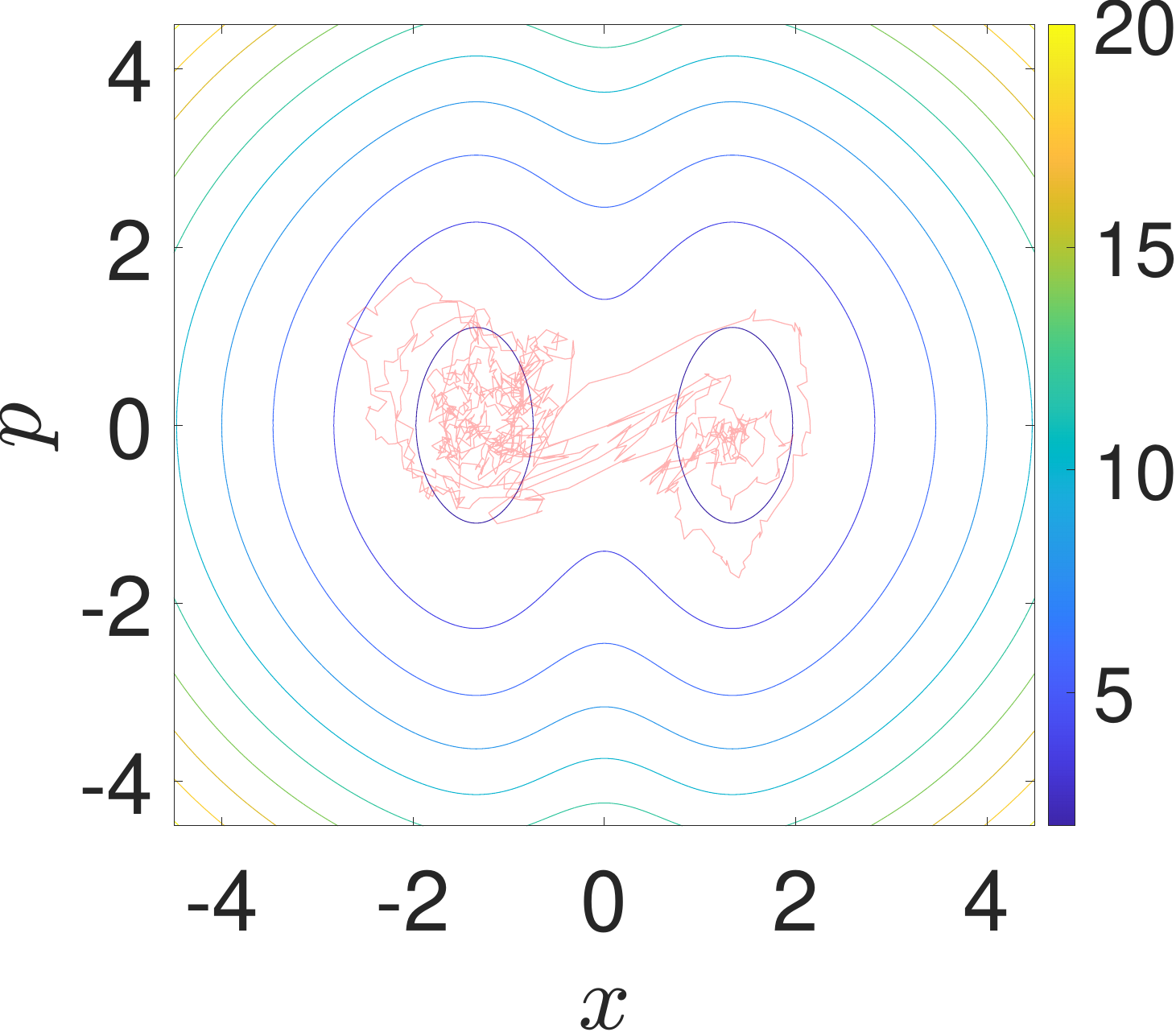}
 	 \caption{}\label{fig:QSDContour.pdf}
\end{subfigure}&
\begin{subfigure}[c]{.22\textwidth} 
	  \includegraphics[width=\textwidth]{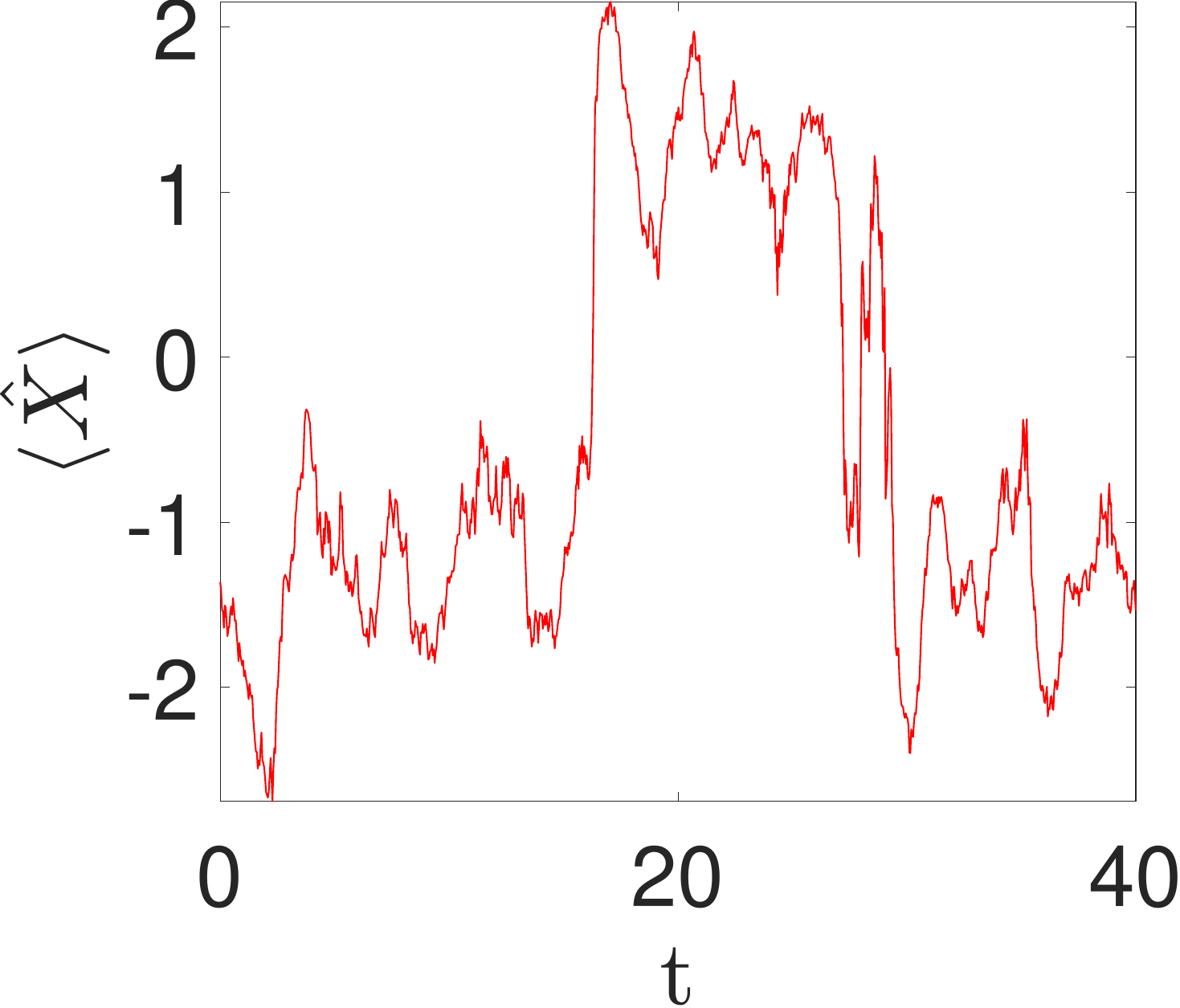}
 	 \caption{}\label{fig:XDoubleQSD}
\end{subfigure}&
\begin{subfigure}[c]{.22\textwidth} 
	  \includegraphics[width=\textwidth]{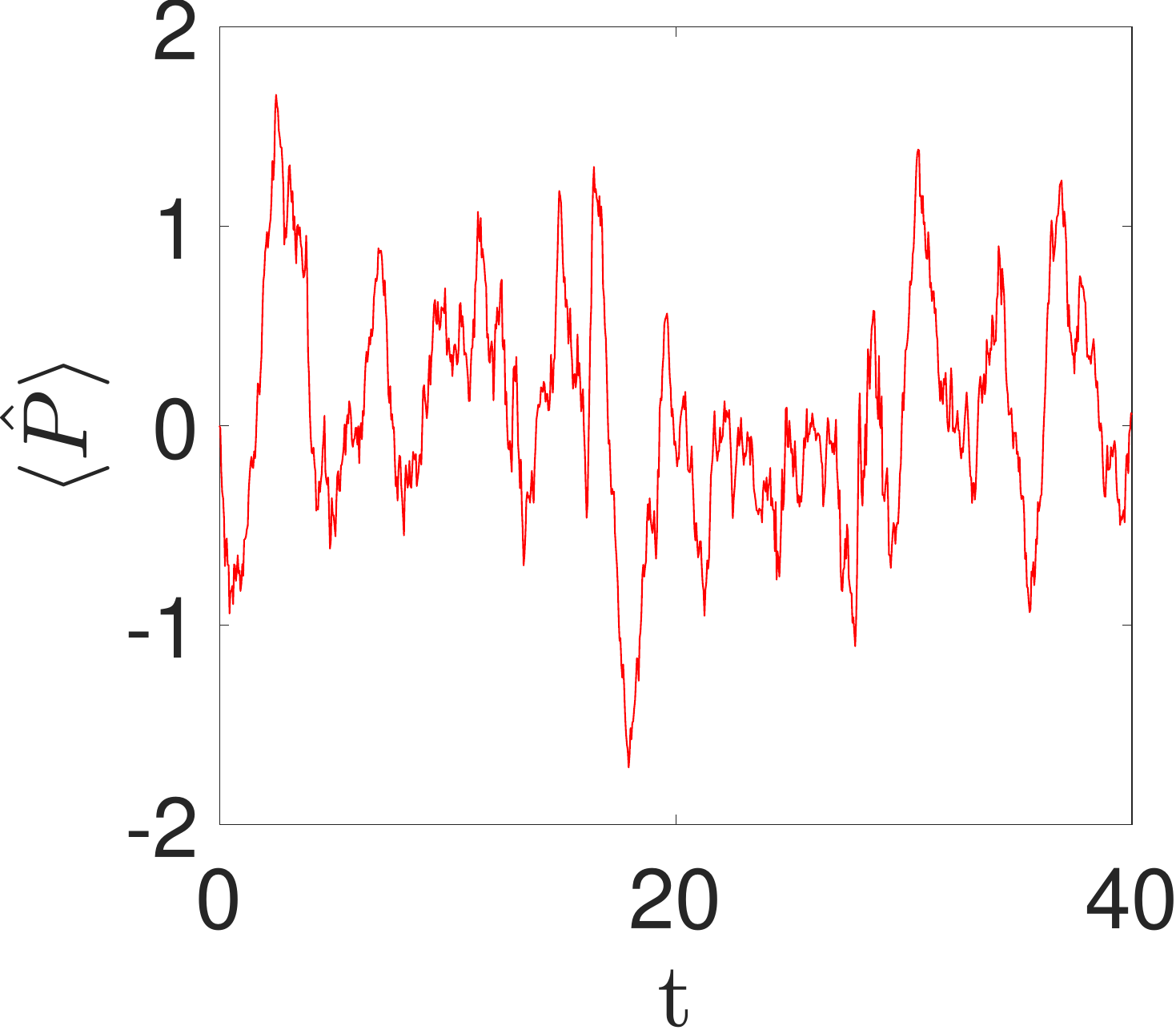}
 	 \caption{}\label{fig:PDoubleQSD}
\end{subfigure}&
\begin{subfigure}[c]{.22\textwidth} 
	  \includegraphics[width=\textwidth]{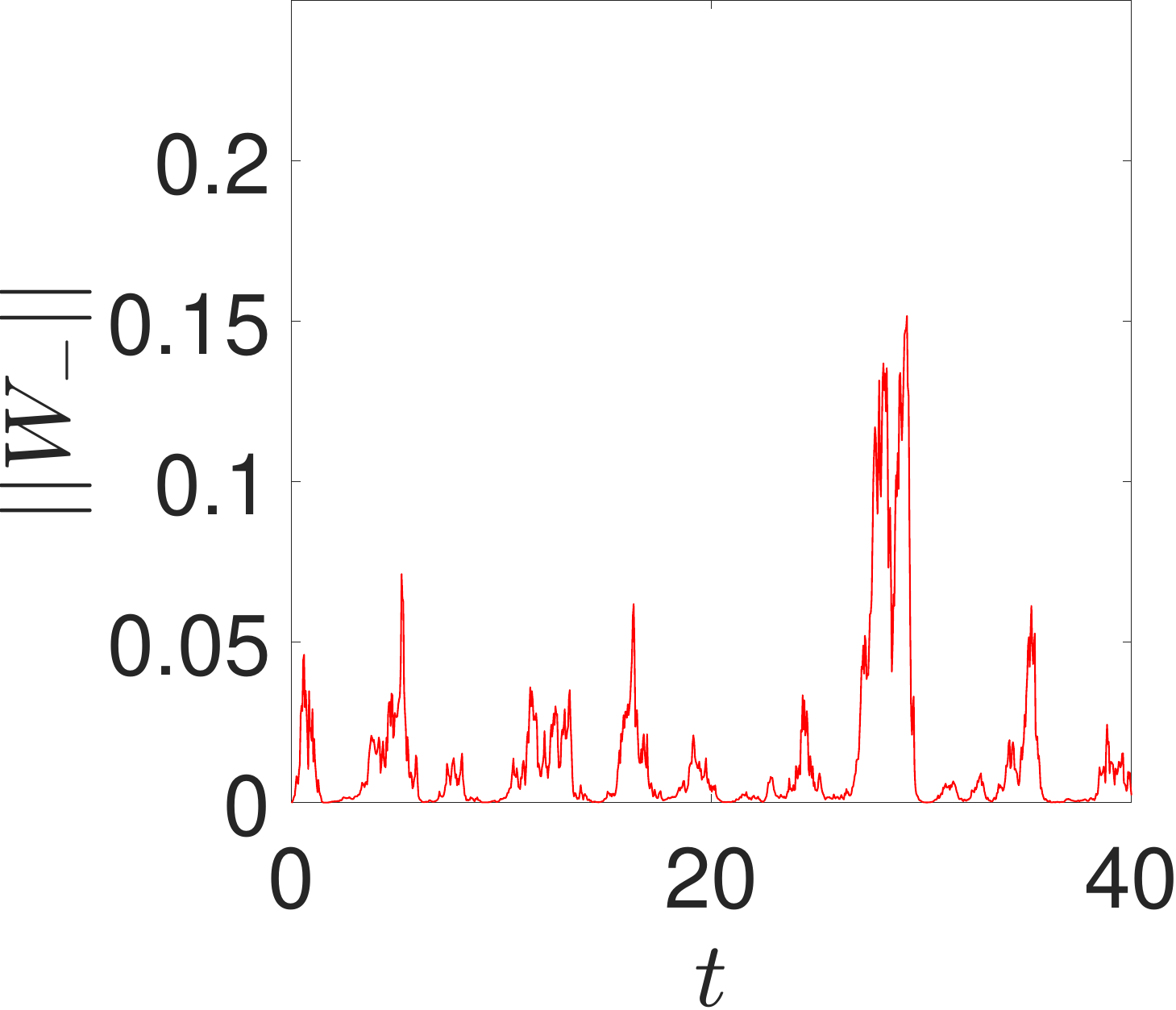}
 	 \caption{}\label{fig:WigNegDoubleQSD}
\end{subfigure}\\
\begin{subfigure}[c]{.22\textwidth} \label{fig:WigStartDoubleQSD}
	  \includegraphics[width=\textwidth]{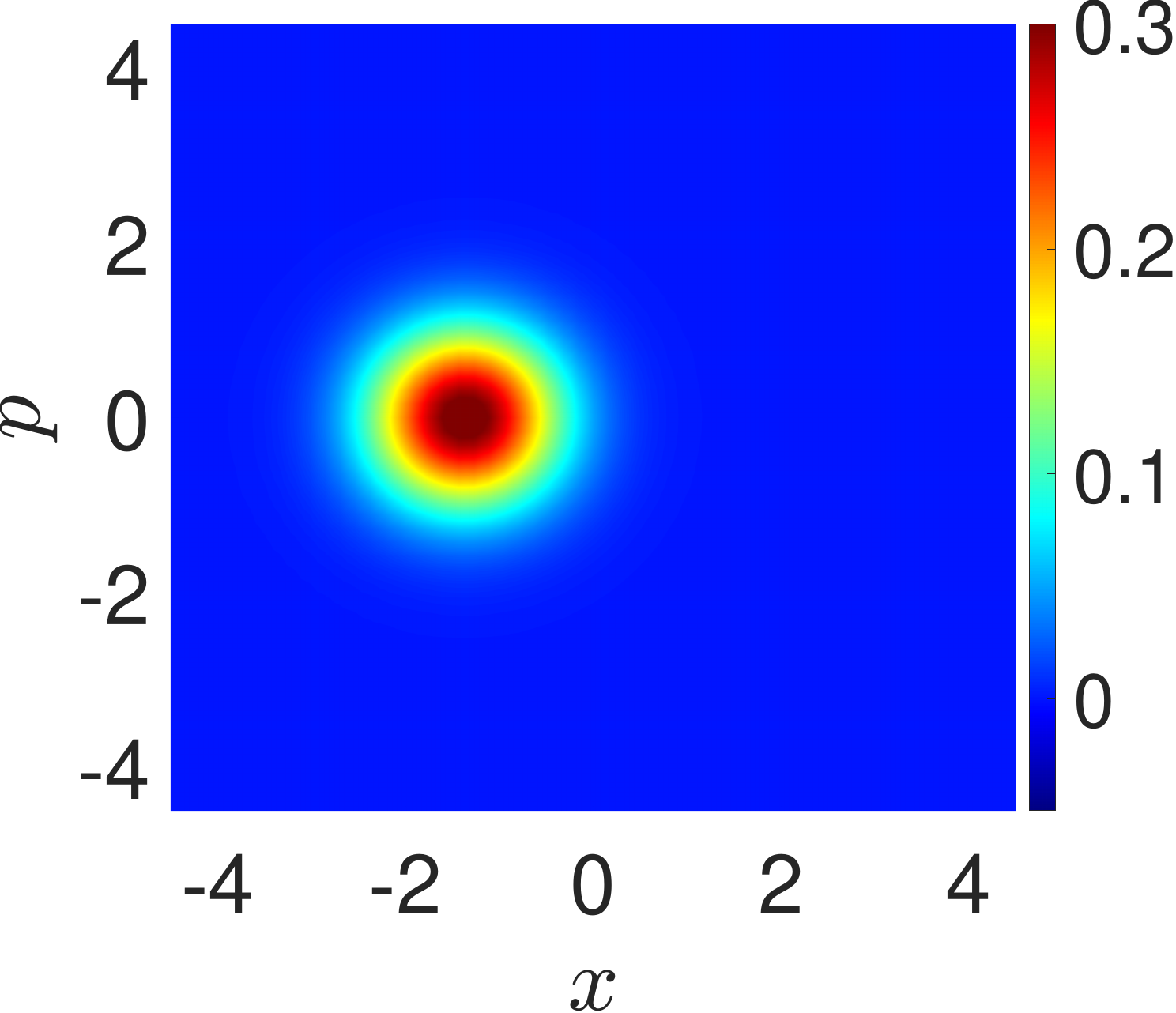}
 	 \caption{: t=0}
\end{subfigure}&
\begin{subfigure}[c]{.22\textwidth} \label{fig:WigMidDoubleQSD}
	  \includegraphics[width=\textwidth]{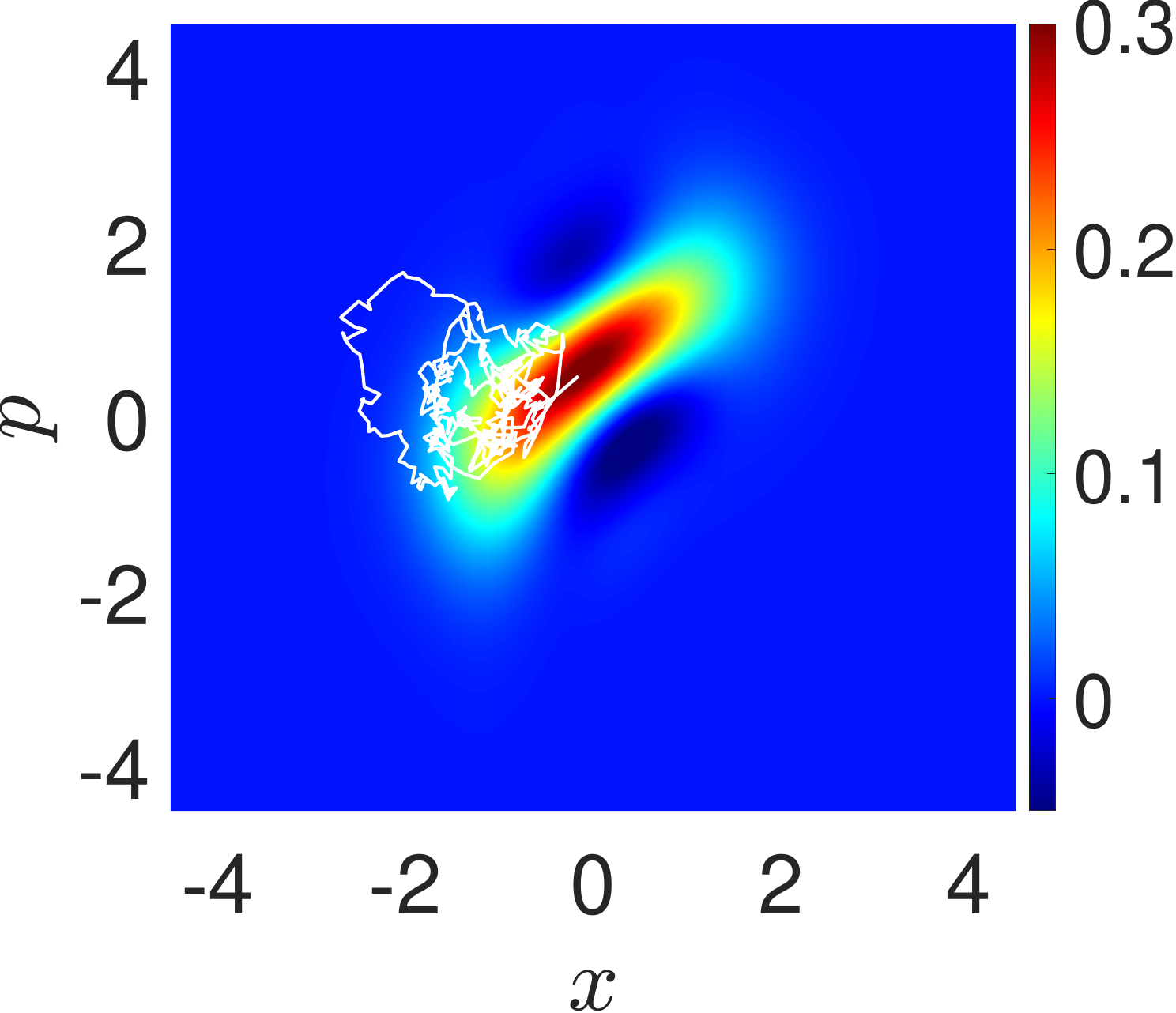}
 	 \caption{: t=16.28}
\end{subfigure}&
\begin{subfigure}[c]{.22\textwidth} \label{fig:WigEndDoubleQSD}
	  \includegraphics[width=\textwidth]{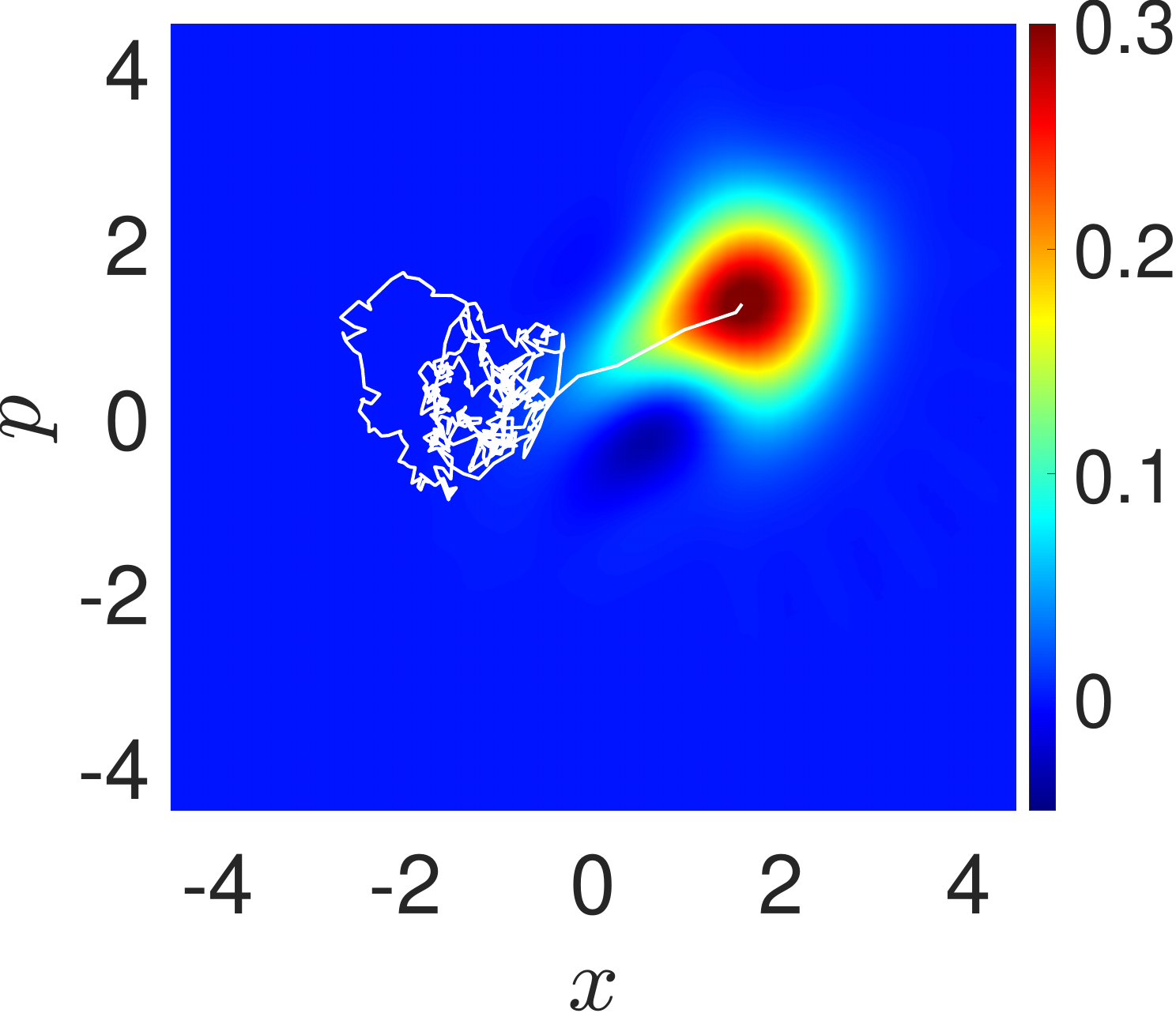}
 	 \caption{: t=16.48}
\end{subfigure}&
\begin{subfigure}[c]{.22\textwidth} \label{fig:WigEnd2DoubleQSD}
	  \includegraphics[width=\textwidth]{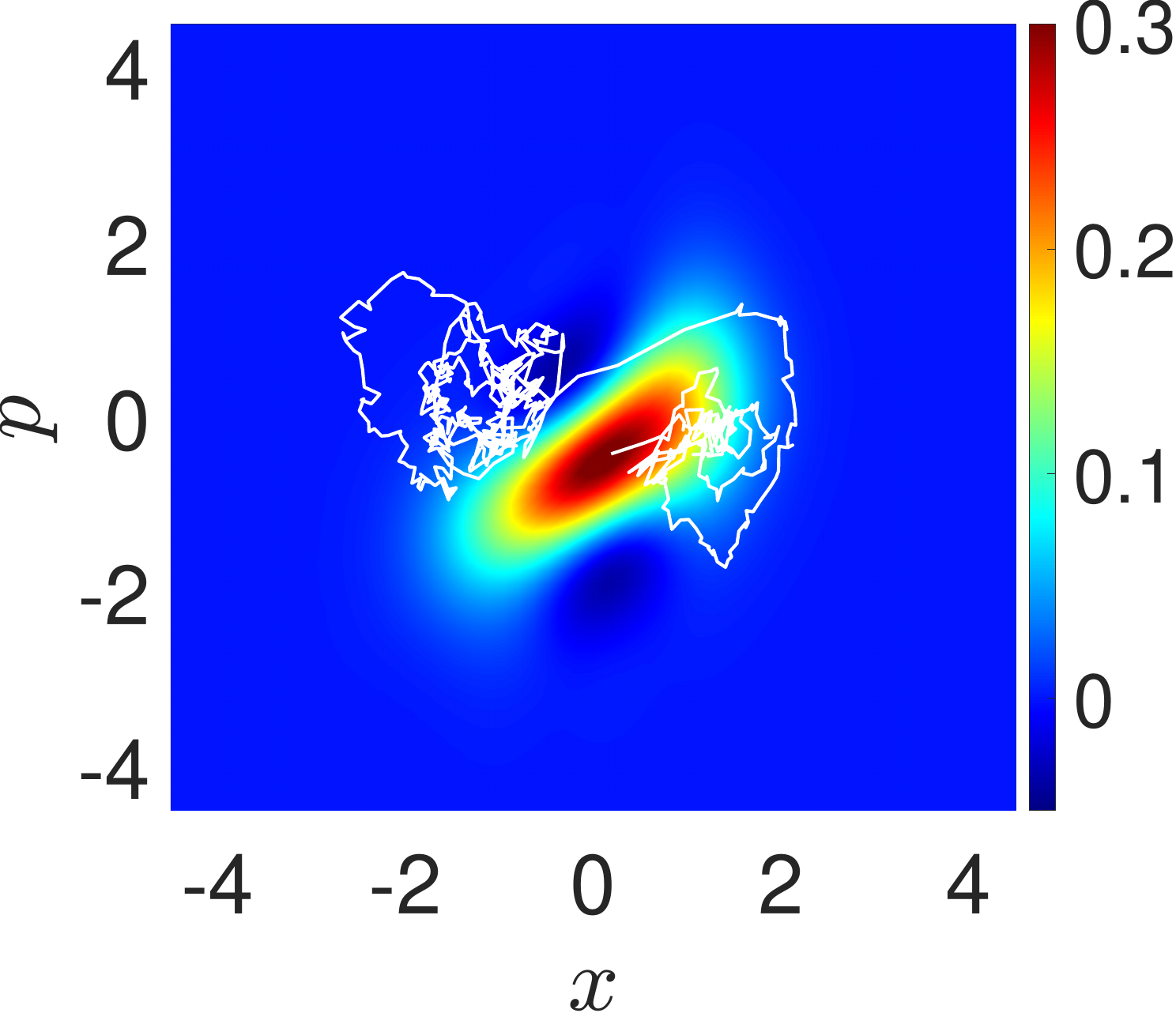}
 	 \caption{: t=27.16}
\end{subfigure}
\end{tabular}
\caption{The top row corresponds to the inner product of the SSE state with the first four well eigenstates plotted against time for $T=1$. The blue line is the initial value $|\ev*{\psi_i|\psi(0)}|^2$, allowing easy comparison with the coherent Hamiltonian tunnelling. Bottom row  Wigner functions of these SSE dynamics at selected times, from left to right, we show the initial state, midway through the first transition, the state localised in the right well, and midway through the second transition. \label{fig:TunnelPlots3QSD}} 
\end{figure}

Even for the relatively low temperature of $T=0.2$, the SSE transitions no longer mimic the coherent Hamiltonian transitions observed in \cref{fig:XDoubleHam}, given their aperiodic behaviour and shortened transition timescales. The transitions observed in SSE encapsulate a hybrid of classical transition mechanisms and quantum tunnelling. We can see the right column of \cref{fig:TunnelPlots2QSD} transitions still correspond to spikes in Wigner function negativity; however, as the energy eigenstate overlap plots in the top row of \cref{fig:TunnelPlots3QSD} show, we also have the added classical transition mechanism of the heat bath partially exciting eigenstates above the barrier height (see \cref{fig:GaussWellEnergyLevels}).

To better understand these transitions, in the bottom row of \cref{fig:TunnelPlots3QSD}, we show the Wigner function during the first two transitions of the trajectory in the middle row of \cref{fig:TunnelPlots2QSD}. The Wigner function exhibits a dumbbell shape during the transitions, which is similar to the shape of the Hamiltonian Weyl symbol shown in \cref{fig:HamPhase}, yet different from the concave disk shape observed during pure-Hamiltonian tunnelling in \cref{fig:TunnelPlots2}. As the SSE can exchange energy with the bath, this dumbbell shape minimizes the expectation value of the Hamiltonian while the state is at the top of the barrier.

Interestingly comparing \cref{fig:WigNegDoubleQSDLow} with the coherent tunnelling in \cref{fig:WigNegDoubleHam} we see that the peak negativity of the Wigner function is higher in the low-temperature system than the Hamiltonian system disconnected from the heat bath; this suggests that for this model quantum interference effects may be enhanced by coupling to the environment, this property of the SSE echoes characteristics of the anti-Zeno effect \cite{AntiZeno1, AntiZeno2}.
\begin{figure}[h]
\centering 
\begin{tabular}{c c c}
\begin{subfigure}[c]{.25\textwidth} 
	  \includegraphics[width=\textwidth]{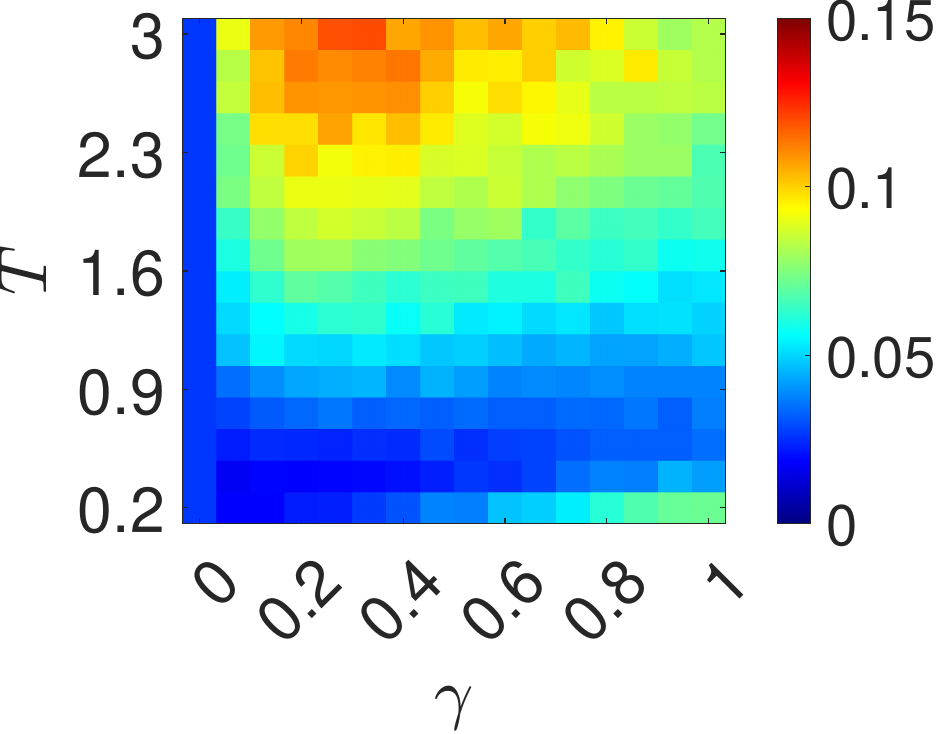}
 	 \caption{}\label{fig:G_Shall_RateFreeCombined}
\end{subfigure}&
\begin{subfigure}[c]{.25\textwidth} 
	  \includegraphics[width=\textwidth]{G_Shall_RateLangevinCombined.pdf}
 	 \caption{}
\end{subfigure}&
\begin{subfigure}[c]{.25\textwidth} 
	  \includegraphics[width=\textwidth]{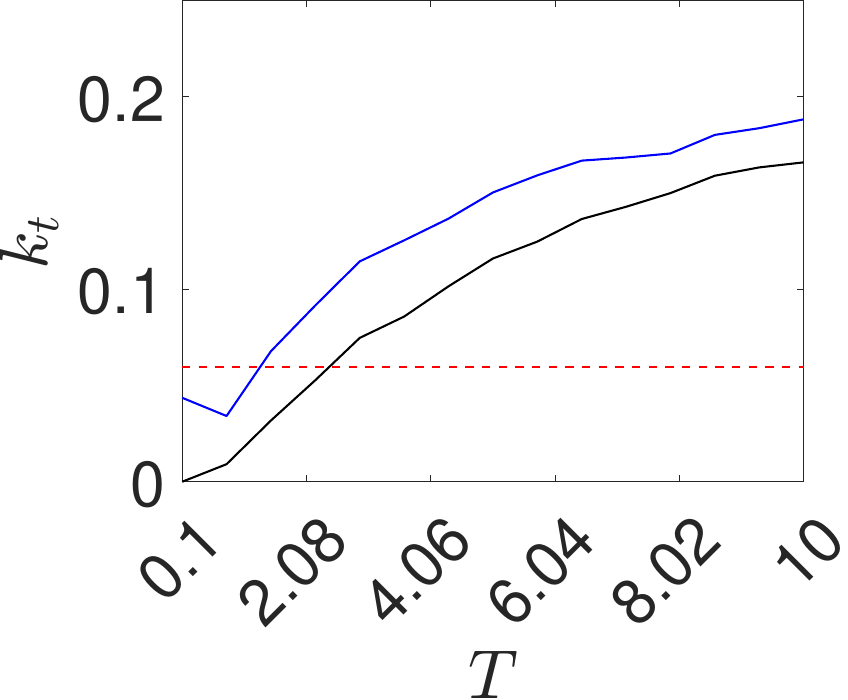}
 	 \caption{}\label{fig:G_Shall_RatesVsTempBigFig}
\end{subfigure}\\
\begin{subfigure}[c]{.25\textwidth} 
	  \includegraphics[width=\textwidth]{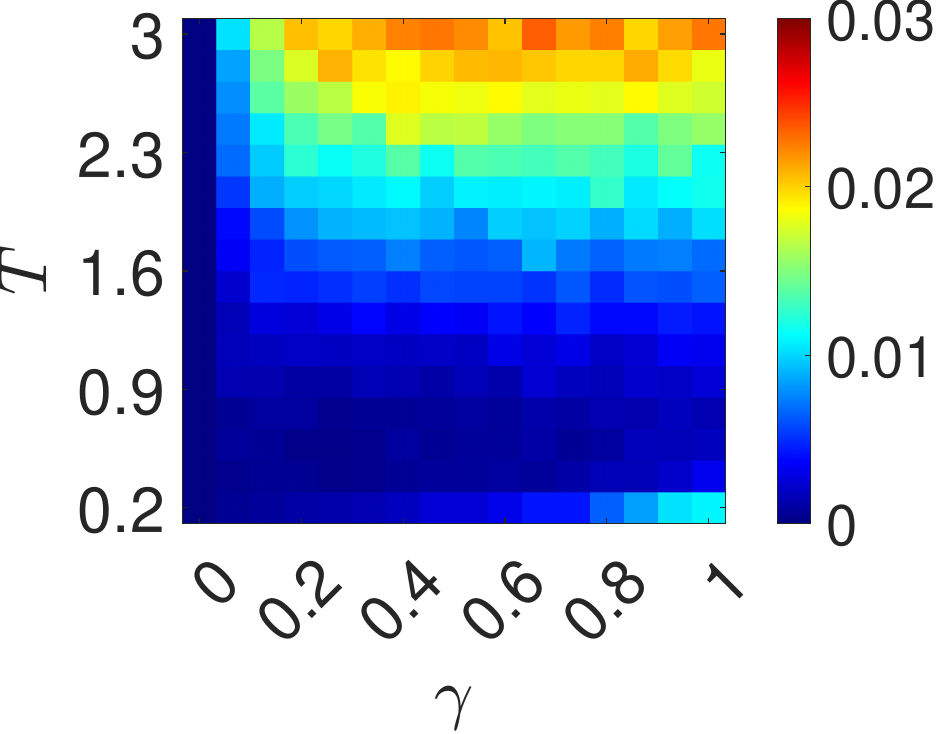}
 	 \caption{}\label{fig:G_Deep_RateFreeCombined}
\end{subfigure}&
\begin{subfigure}[c]{.25\textwidth} 
	  \includegraphics[width=\textwidth]{G_Deep_RateLangevinCombined.pdf}
 	 \caption{}
\end{subfigure}&
\begin{subfigure}[c]{.25\textwidth} 
	  \includegraphics[width=\textwidth]{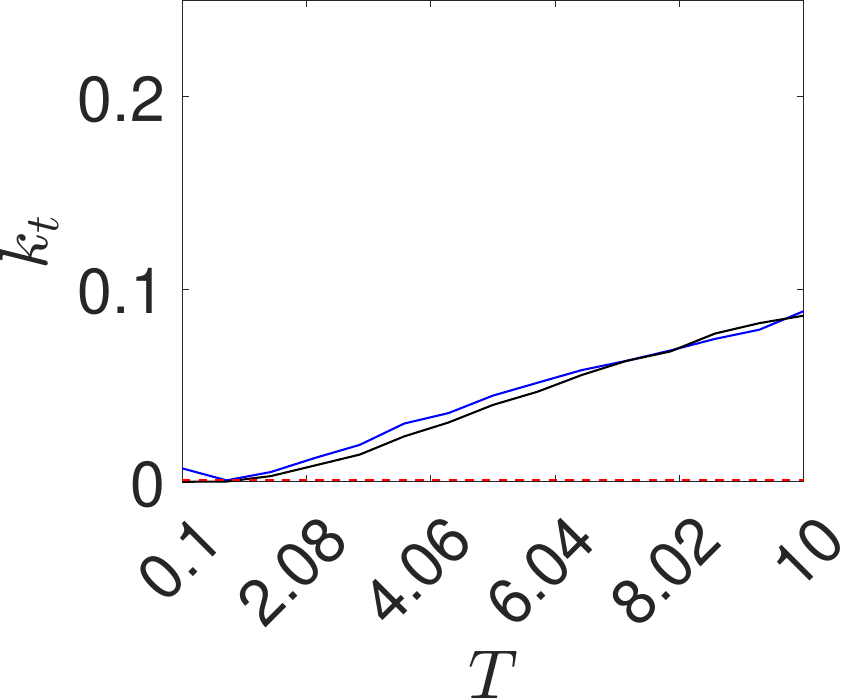}
 	 \caption{}\label{fig:G_Deep_RatesVsTempBigFig}
\end{subfigure}
\end{tabular}
\caption{Transition rates for the quantum (Petruccione) and classical (Langevin) dynamics in several different potentials. We have the left column consisting of heatmaps of the quantum transition rate with respect to coupling strength $\gamma$ and temperature with the middle column consisting of these same plots for the Langevin rate. The right column is the transition rate of both the classical (black) and quantum (blue) transition rates with respect to temperature at fixed $\gamma=0.25$. The dashed red line is the top row corresponds to a shallow double well ($A=3$, and $\sigma=\tfrac{1}{\sqrt2})$ and the bottom row a deep double well ($A=8$, and $\sigma=1$). \label{fig:transition-rate-big-fig}}
\end{figure}

In \cref{fig:transition-rate-big-fig}, we show heatmaps of the transition rate with respect to the temperature $T$ and the bath coupling $\gamma$ for the Caldeira Leggett SSE model in the left column; in the middle column, we do the same for the classical Langevin model. In the right column, we compare the classical and quantum transition rates at fixed coupling strength $\gamma=0.25$. The rows \cref{fig:transition-rate-big-fig} correspond to a shallow double well (top) and a deeper double well (bottom) respectively. The limit $\gamma=0$ on the left border for the quantum models corresponds to the pure tunnelling dynamics; since the classical Langevin dynamics are not able to tunnel, the transition rate at this border is zero in the classical case. Moving away from the $\gamma=0$ border of the heatmaps, we notice the same overall shape on all the heatmaps consistent with Kramers rate theory \cite{pavliotis2014stochastic} with the highest rates found at high temperature and moderate coupling. As a general feature, we observe that the quantum model has higher transition rates at low temperatures than the Langevin dynamics, which can be explained by tunnelling effects. We also note a significant increase at very low temperatures for the quantum model. This very low-temperature increase is likely an unphysical consequence of the singular behaviour of the model \eqref{eq:heatbath} as $T\to0$. We will return to this limit and a possible modification to this model with more desirable low-temperature properties in \cref{sec:modified}.

In the right column of \cref{fig:transition-rate-big-fig}, we can see the quantum and classical rates over a wide temperature range, the dashed red line is the zero temperature tunnelling rate. We observe that the quantum rate is greater than the classical rate by an amount approximately proportional to the tunnelling rate.

\FloatBarrier
\subsection{Master equation dynamics} \label{sec:Lindblad} 
\begin{figure}
\centering	  
\begin{tabular}{c c c}
     \begin{subfigure}[c]{.3\textwidth}
	  \includegraphics[width=\textwidth]{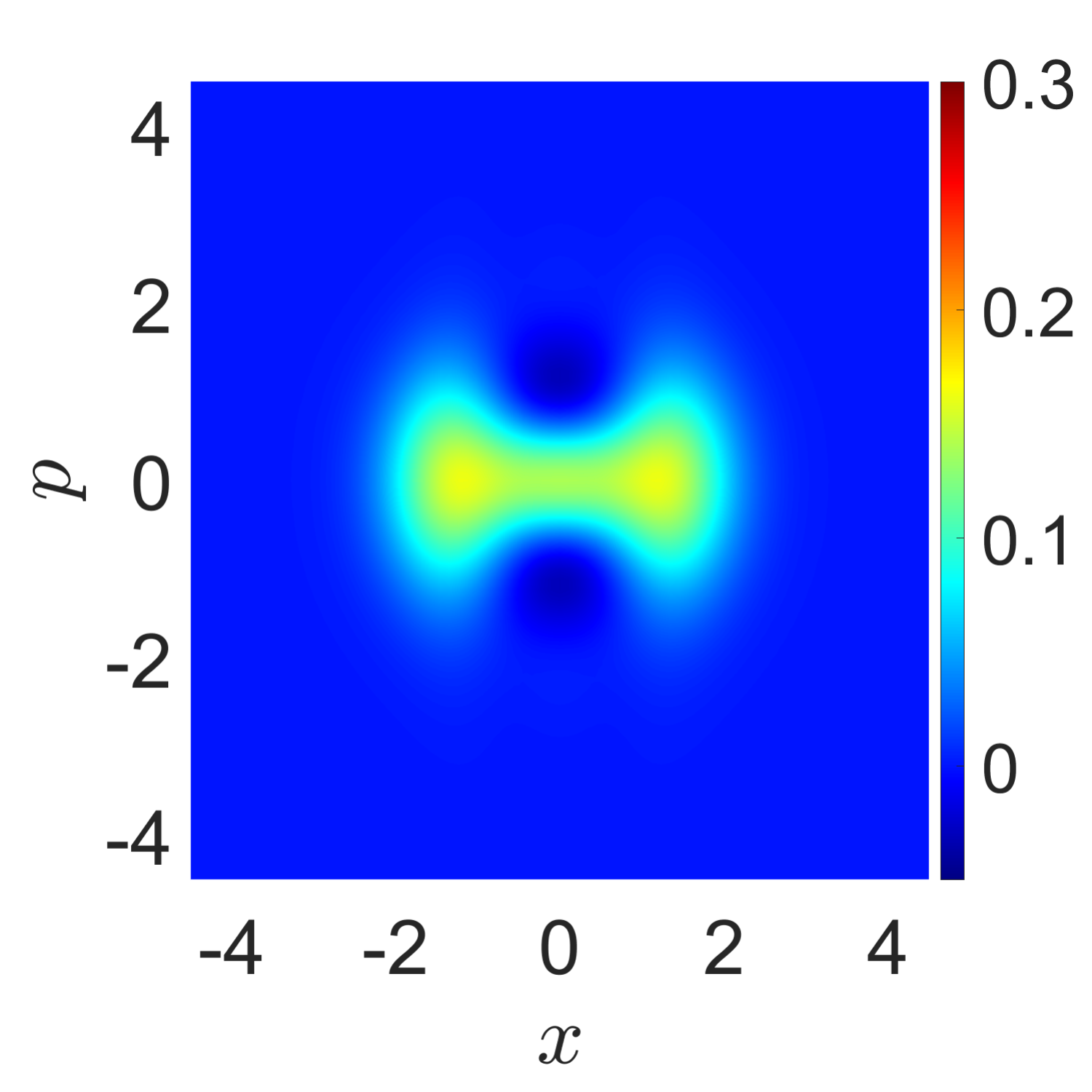}\\
	  \includegraphics[width=\textwidth]{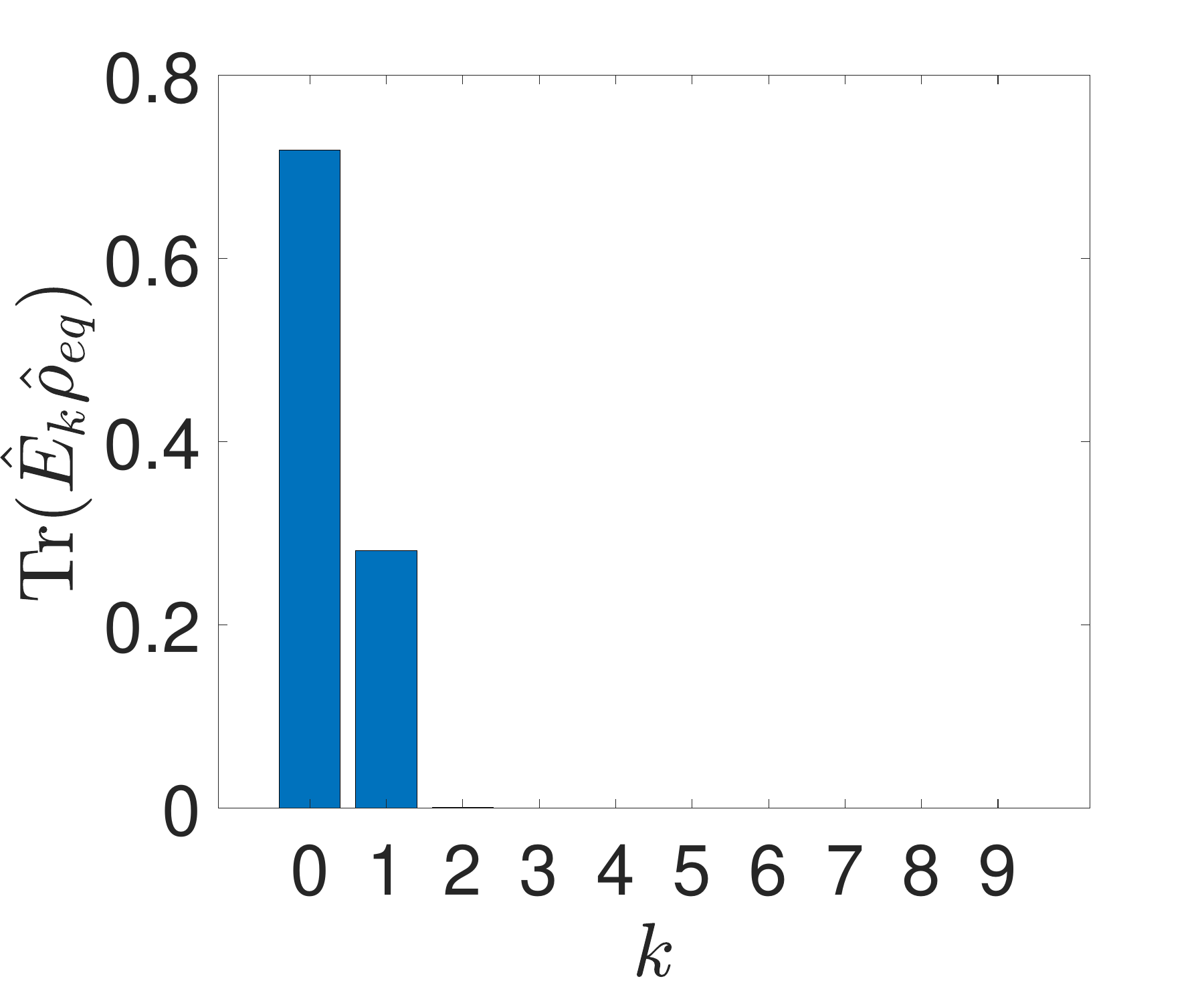}
 	 \caption{$T=0.2$}
    \end{subfigure}&
         \begin{subfigure}[c]{.3\textwidth}
	  \includegraphics[width=\textwidth]{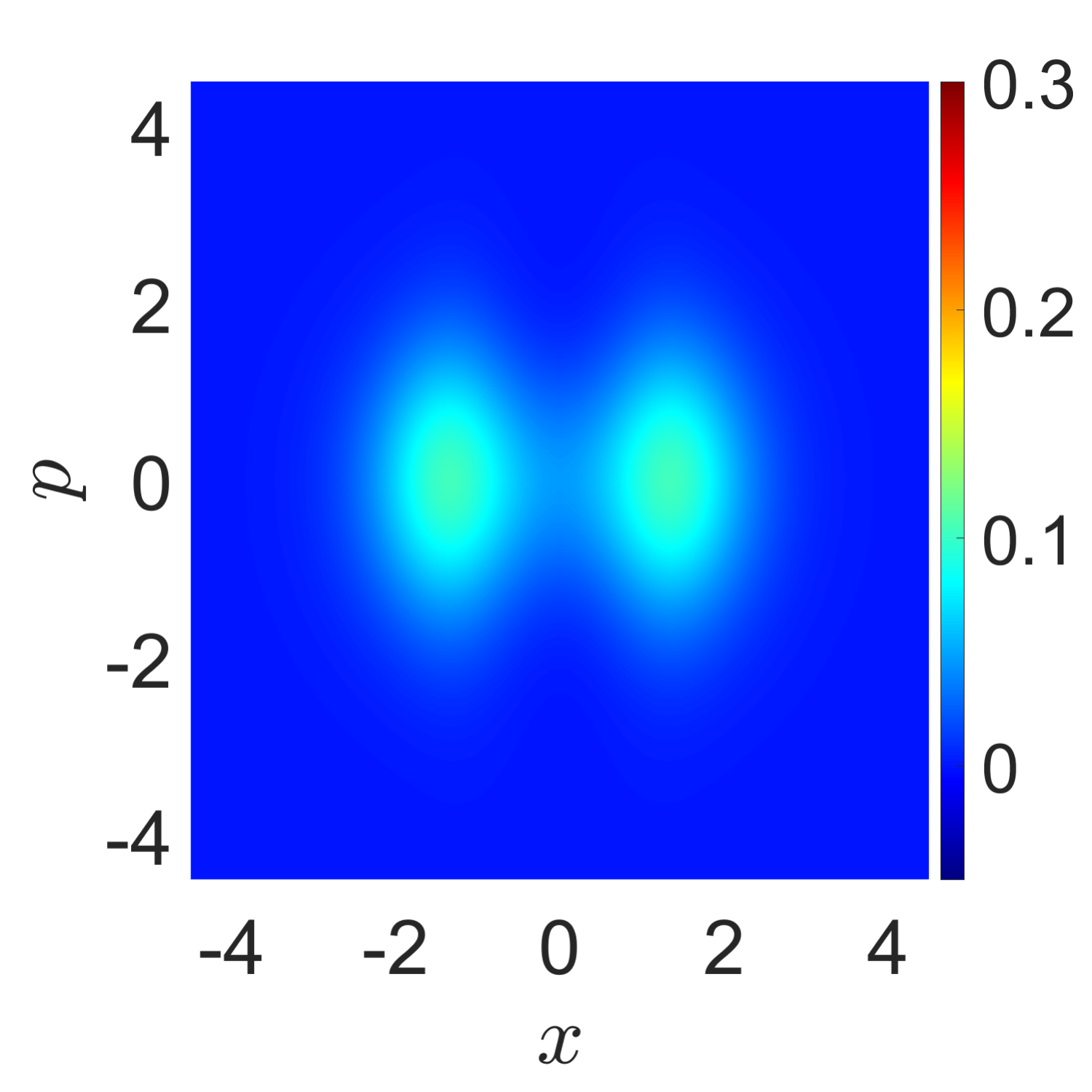}\\
	  \includegraphics[width=\textwidth]{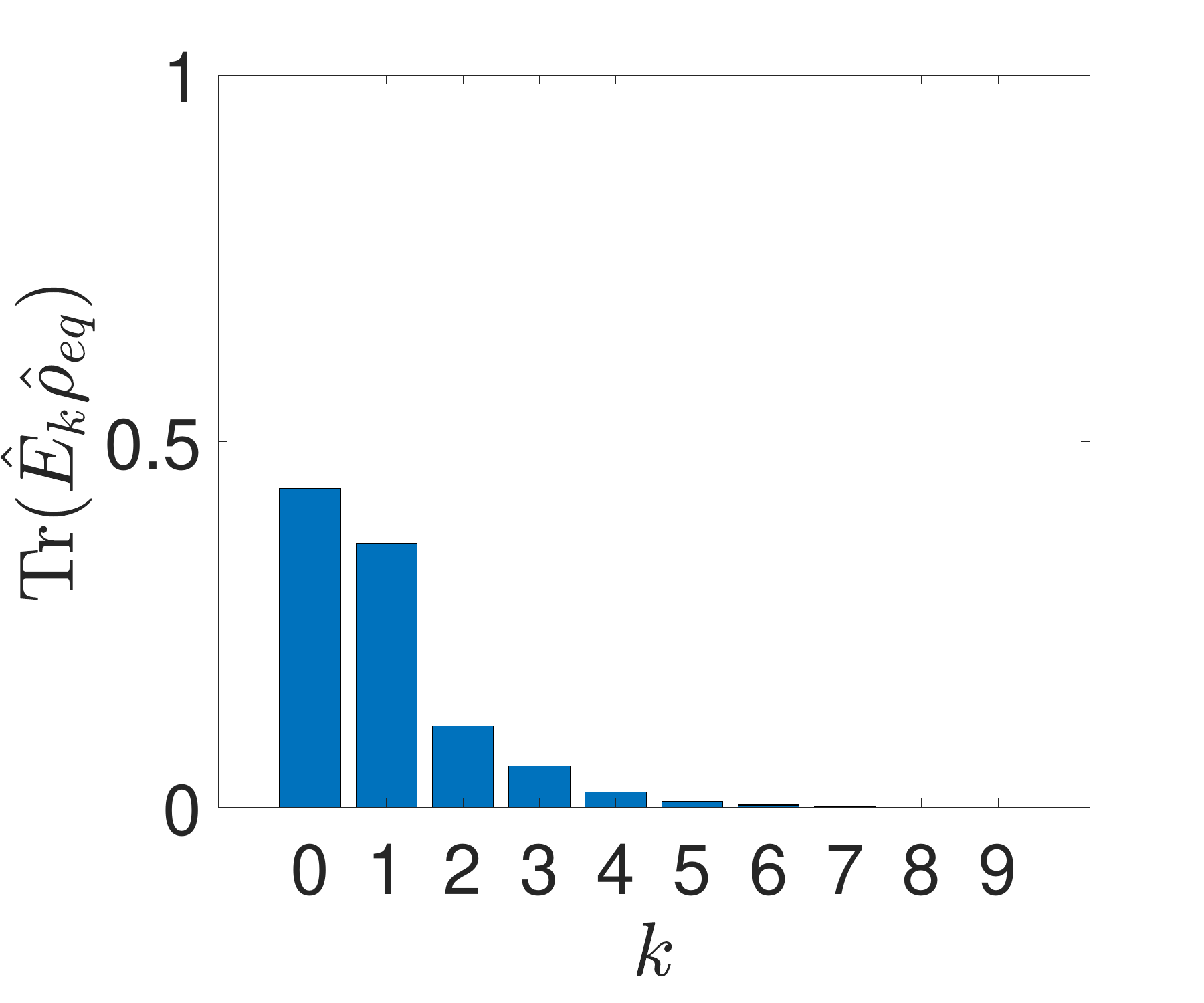}
 	 \caption{$T=1$}
    \end{subfigure}&
         \begin{subfigure}[c]{.3\textwidth}
	  \includegraphics[width=\textwidth]{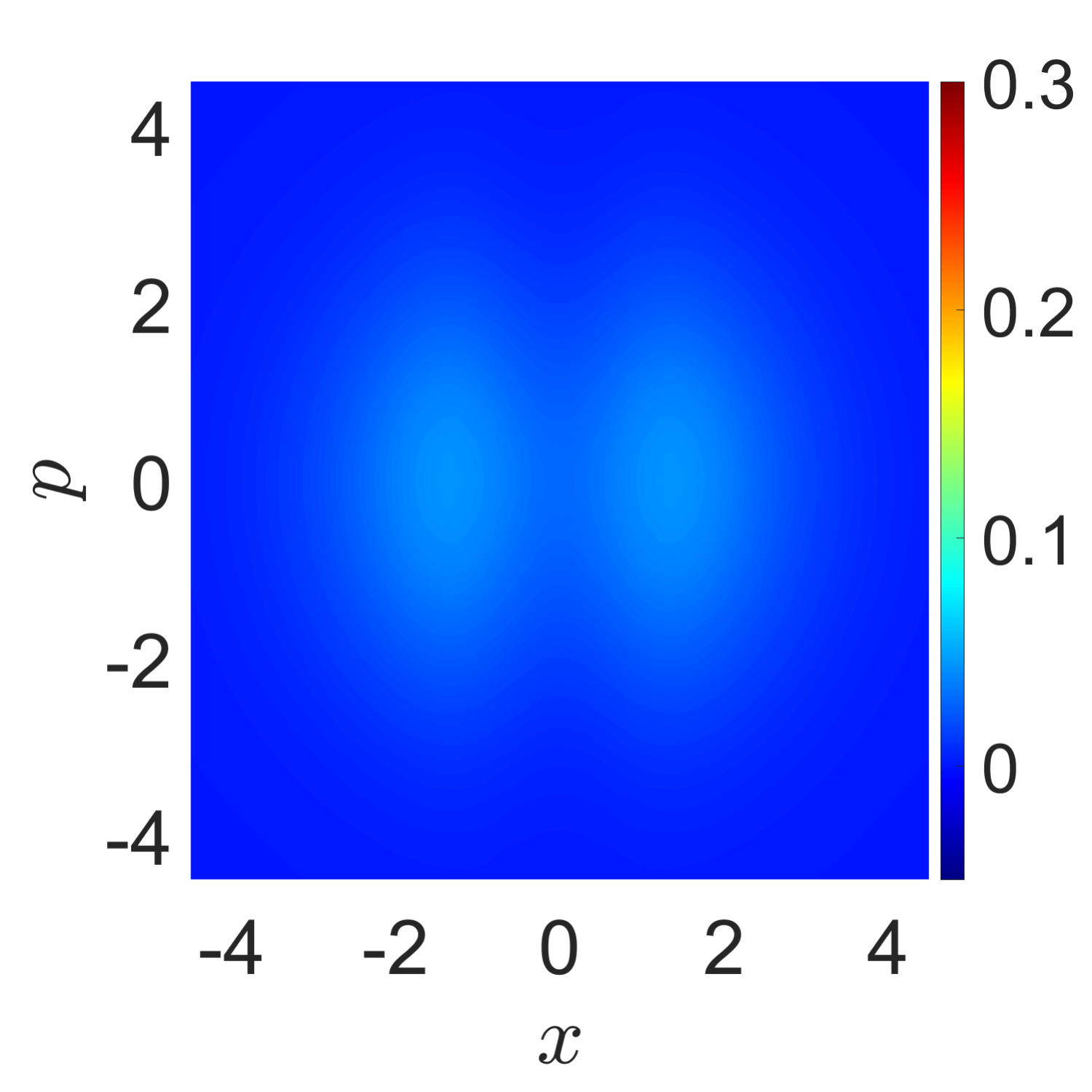}\\
	  \includegraphics[width=\textwidth]{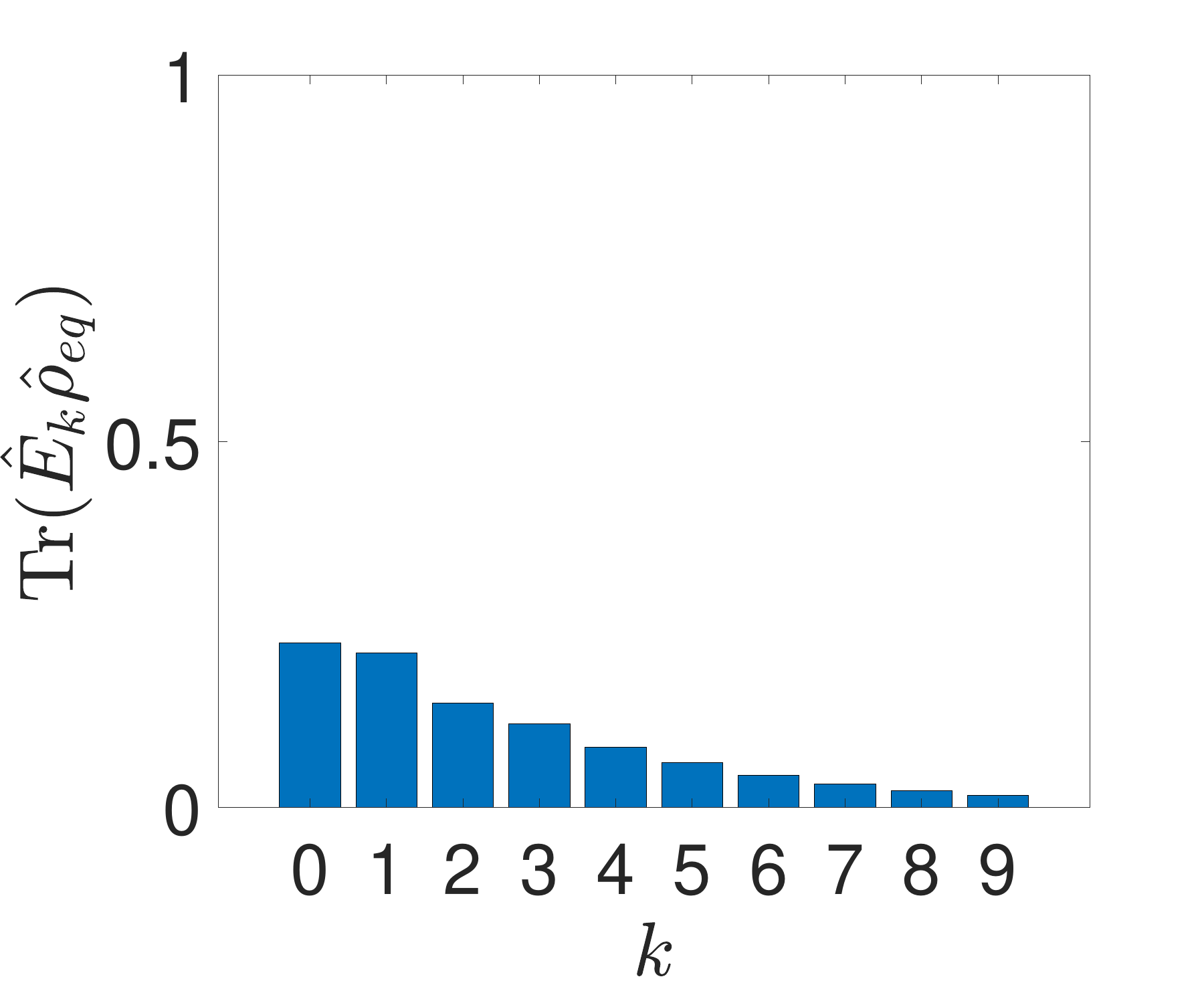}
 	 \caption{$T=3$}
    \end{subfigure}
	 \end{tabular}
\caption{Top row: Wigner functions of the thermal state $\hat \rho_{T}$ defined in \eqref{eq:WigTherm} at different temperatures. Bottom row: relative contributions of the first ten Hamiltonian eigenstates the thermal state \eqref{eq:WigTherm}.} \label{fig:ThermalWig}
\end{figure}
\begin{figure}
\centering 
\begin{tabular}{c c c}
\begin{subfigure}[c]{.3\textwidth} 
	  \includegraphics[width=\textwidth]{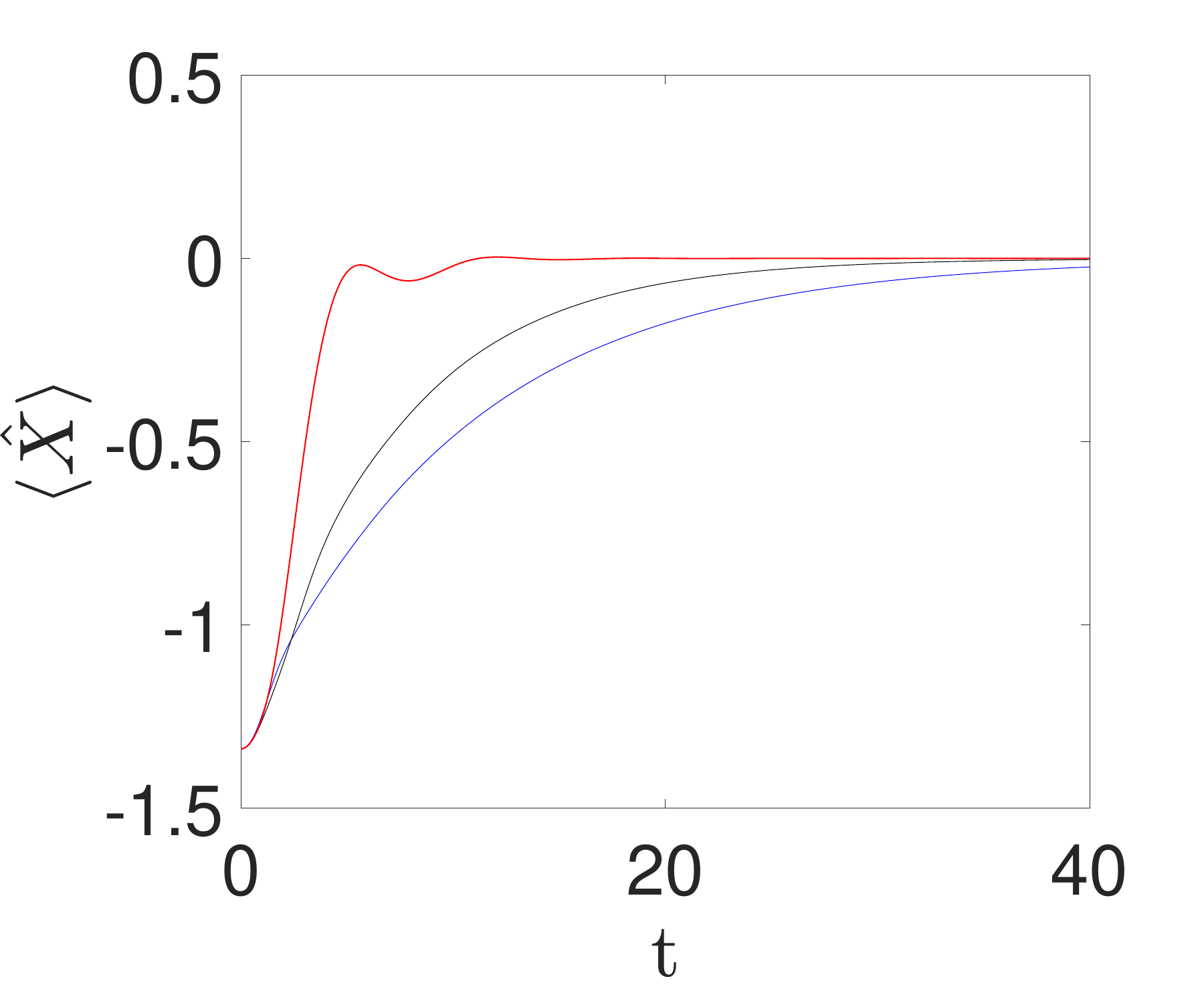}
 	 \caption{}\label{fig:XDoubleLind}
\end{subfigure}&
\begin{subfigure}[c]{.3\textwidth} 
	  \includegraphics[width=\textwidth]{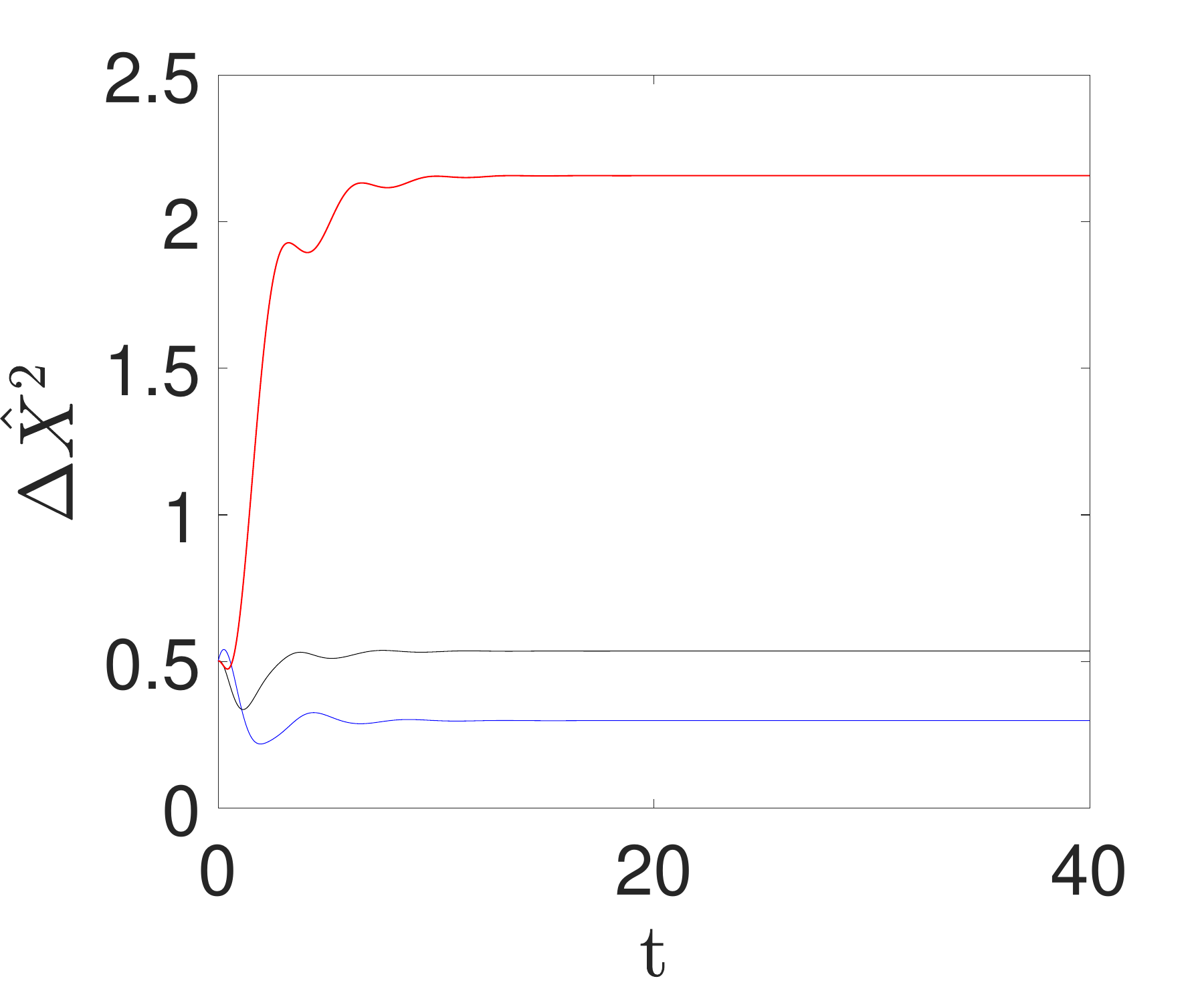}
 	 \caption{}\label{fig:varXLind}
\end{subfigure}&
\begin{subfigure}[c]{.3\textwidth} 
	  \includegraphics[width=\textwidth]{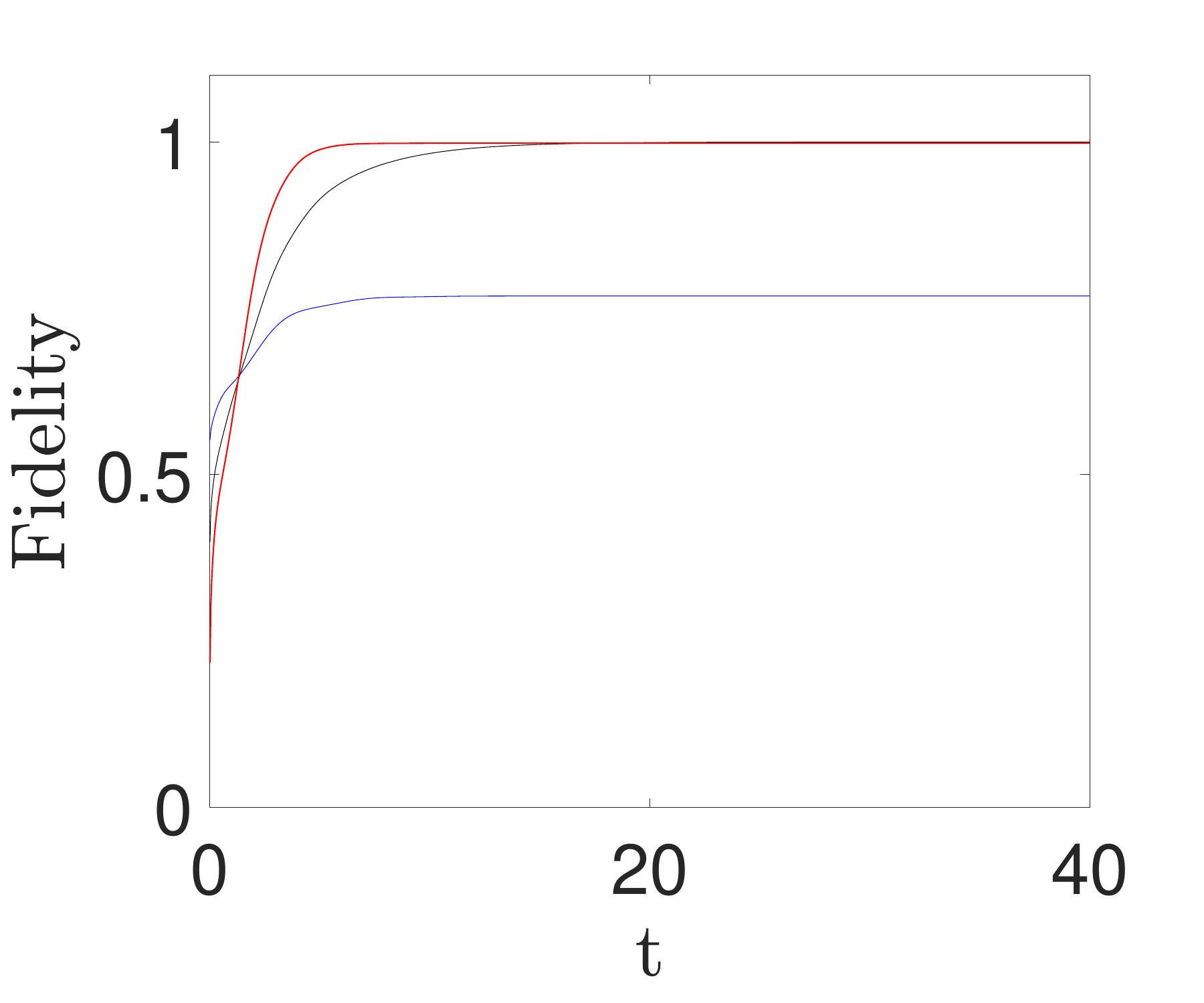}
 	 \caption{}\label{fig:FidelityLind}
\end{subfigure}
\end{tabular}
\caption{Figures for the Lindblad evolution of an initially coherent centred at the minima of the left well at different temperatures. The blue line corresponds to $T=0.2$, the black line $T=1$, and the red line $T=3$. The potential considered is the shallow double well with parameters $A=3$,$\gamma=0.25$ and $\sigma=1/\sqrt2$. We show the time dependence of the position expectation $\ev*{\hat X}$ in (a), the position variance $\Delta \hat X^2$ in (b), and the fidelity of the density operator with respect to the thermal state \eqref{eq:WigTherm} in (c).\label{fig:LindPlots}}
\end{figure}
An essential property of the classical Fokker Planck equation \cref{eq:Fokker} is that it admits a unique stationary distribution given by the Gibbs state \cite{pavliotis2014stochastic}
\begin{equation} \label{eq:Classical_Therm}
    \rho_T(x,p)=\frac{\exp(-H(x,p)/k_B T)}{\int_{-\infty}^{\infty}\int_{-\infty}^{\infty}  \exp(-H(x,p)/k_B T) dx dp}.
\end{equation}
Analogously to be consistent with statistical mechanics we expect a quantum model of Brownian motion to asymptotically attain and preserve the quantum Gibbs state, given by 
\begin{equation} \label{eq:WigTherm}
    \hat \rho_{T}=\frac{\sum_{n} e^{-\frac{E_n}{k_B T}}\hat E_{n}}{\sum_{n}e^{-\frac{E_n}{k_B T}}},
\end{equation}
where $\hat E_n$ is the projection operator corresponding to the n-th Hamiltonian eigenstate. We briefly recap the properties of the quantum Gibbs state in a double well in \cref{fig:ThermalWig} before moving to its role in quantum Brownian motion. In the top row of \cref{fig:ThermalWig} we plot the Wigner function corresponding to $\hat \rho_{T}$ thermal state at three temperatures, $T=0.2$, $T=1$, and $T=3$. We observe this thermal state is increasingly delocalized at higher temperatures with the Wigner function approaching the classical thermal state probability density at high temperatures. In the bottom row of \cref{fig:LindPlots}, we show a histogram of the contribution of the Hamiltonian energy eigenstates to the thermal equilibrium state. At higher temperatures, we have increasing contributions of the higher excited states to $\hat \rho_T$ as defined in \cref{eq:WigTherm}.

Moving to the Lindblad dynamics of \eqref{eq:heatbath}. In \cref{fig:XDoubleLind}, we plot the position expectation $\ev*{\hat X}$ with respect to time; we found that this travels to the centre of the symmetric potential at a rate that increases with temperature. \Cref{fig:varXLind} we plot the position variance $\Delta \hat X^2$; after an initial transient period, we observe the variance to reach an asymptotic fixed value, which again increases with temperature. \Cref{fig:FidelityLind} shows the fidelity (a measure of density operator similarity \cite{Fidelity}) with respect to the Gibbs state over time. We observe the Lindblad dynamics drive the system into the thermal equilibrium state at a rate increasing with temperature, however at low temperatures the ground state energy is comparable to $k_B T$ and the Lindblad dynamics approach an asymptotic fixed point distinct from the Gibbs state as shown in the right panel of \cref{fig:FidelityLind}. This low-temperature behaviour is an artefact of pushing \eqref{eq:heatbath} beyond its intended scope and violating the assumption \cref{eq:TempCondition}. In the next section, we propose a modification of \cref{eq:heatbath} that results in more sensible low-temperature dynamics using the harmonic oscillator thermal state preserving dynamics derived in \ref{sec:QHO} and the harmonic approximation of a double well.

\section{Harmonic approximation of Brownian motion} \label{sec:modified}
\begin{figure}
\centering 
\begin{tabular}{c c c}
\begin{subfigure}[c]{.3\textwidth} 
	  \includegraphics[width=\textwidth]{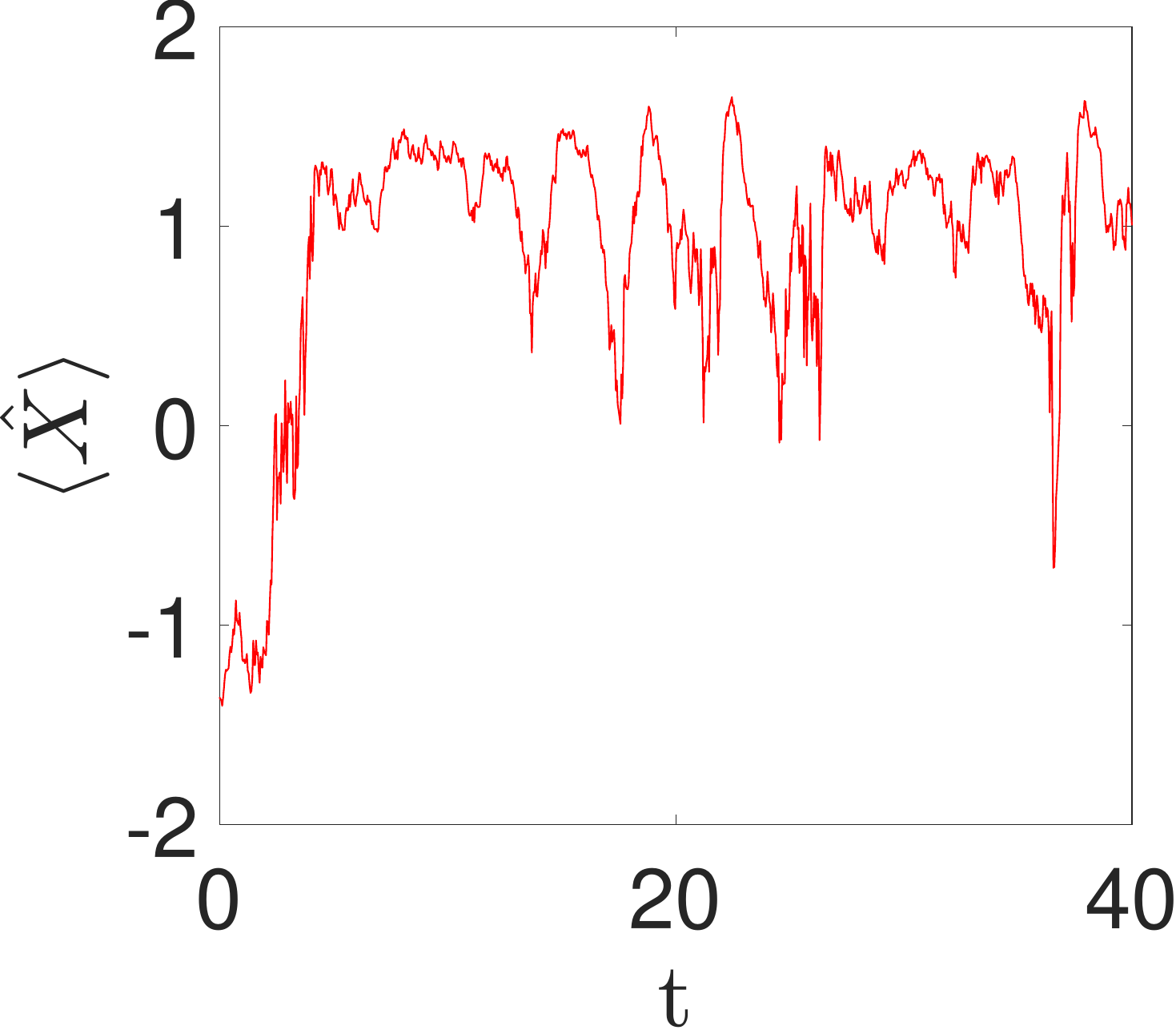}
 	 \caption{}\label{fig:XDoubleQHOLow}
\end{subfigure}&
\begin{subfigure}[c]{.3\textwidth} 
	  \includegraphics[width=\textwidth]{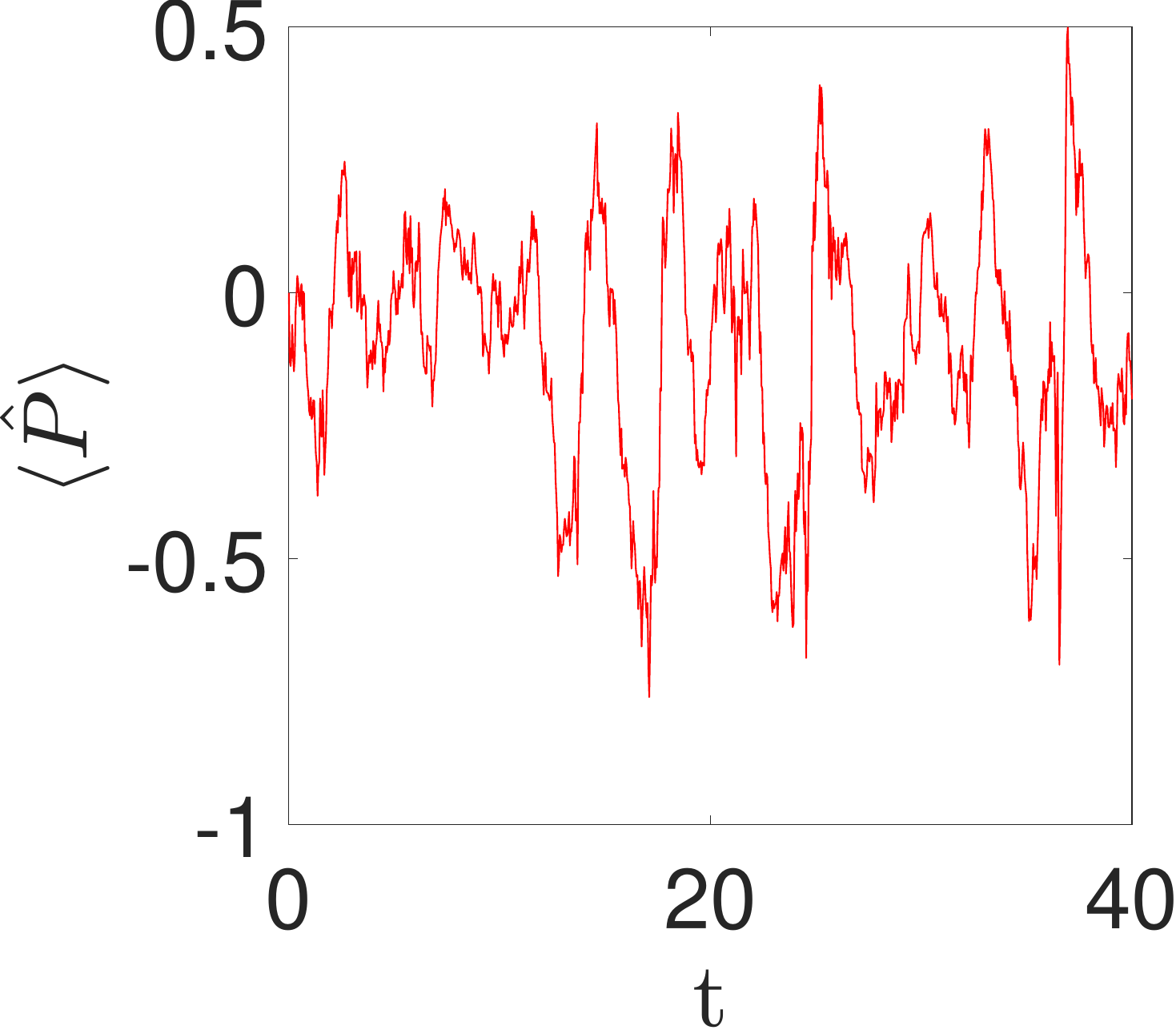}
 	 \caption{}\label{fig:PDoubleQHOLow}
\end{subfigure}&
\begin{subfigure}[c]{.3\textwidth} 
	  \includegraphics[width=\textwidth]{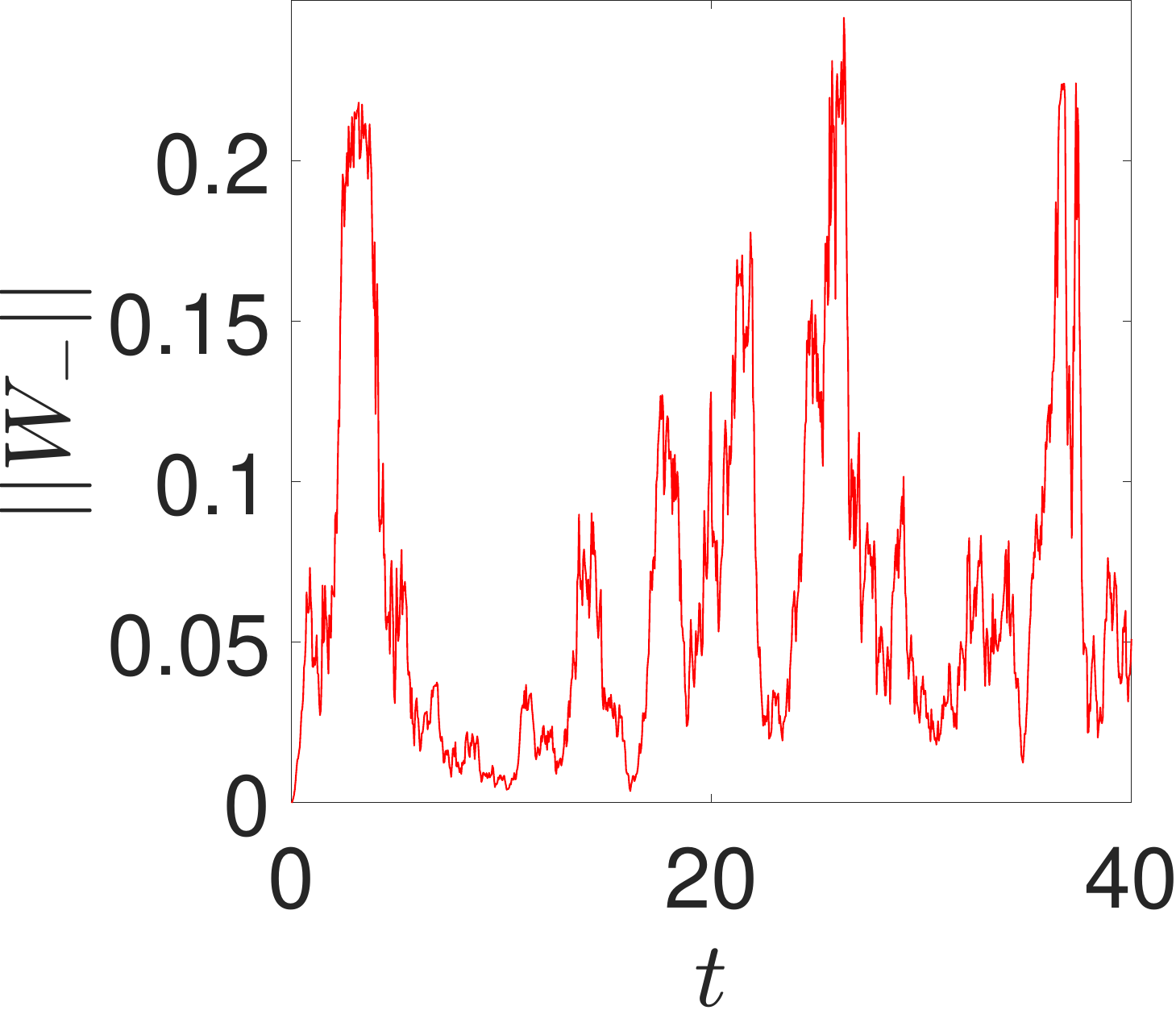}
 	 \caption{}\label{fig:WigNegDoubleQHOLow}
\end{subfigure}\\
\begin{subfigure}[c]{.3\textwidth} 
	  \includegraphics[width=\textwidth]{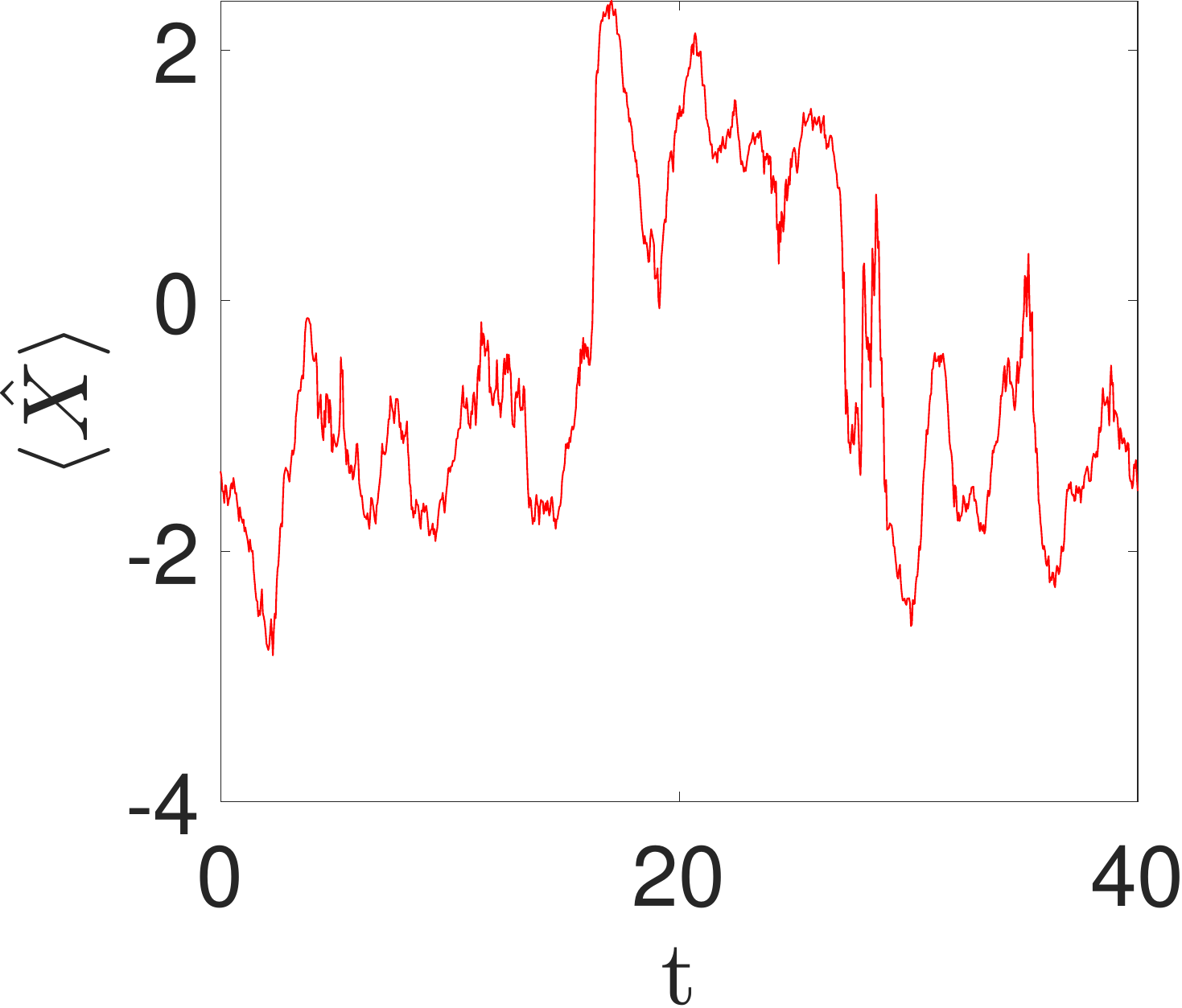}
 	 \caption{}\label{fig:XDoubleQHO}
\end{subfigure}&
\begin{subfigure}[c]{.3\textwidth} 
	  \includegraphics[width=\textwidth]{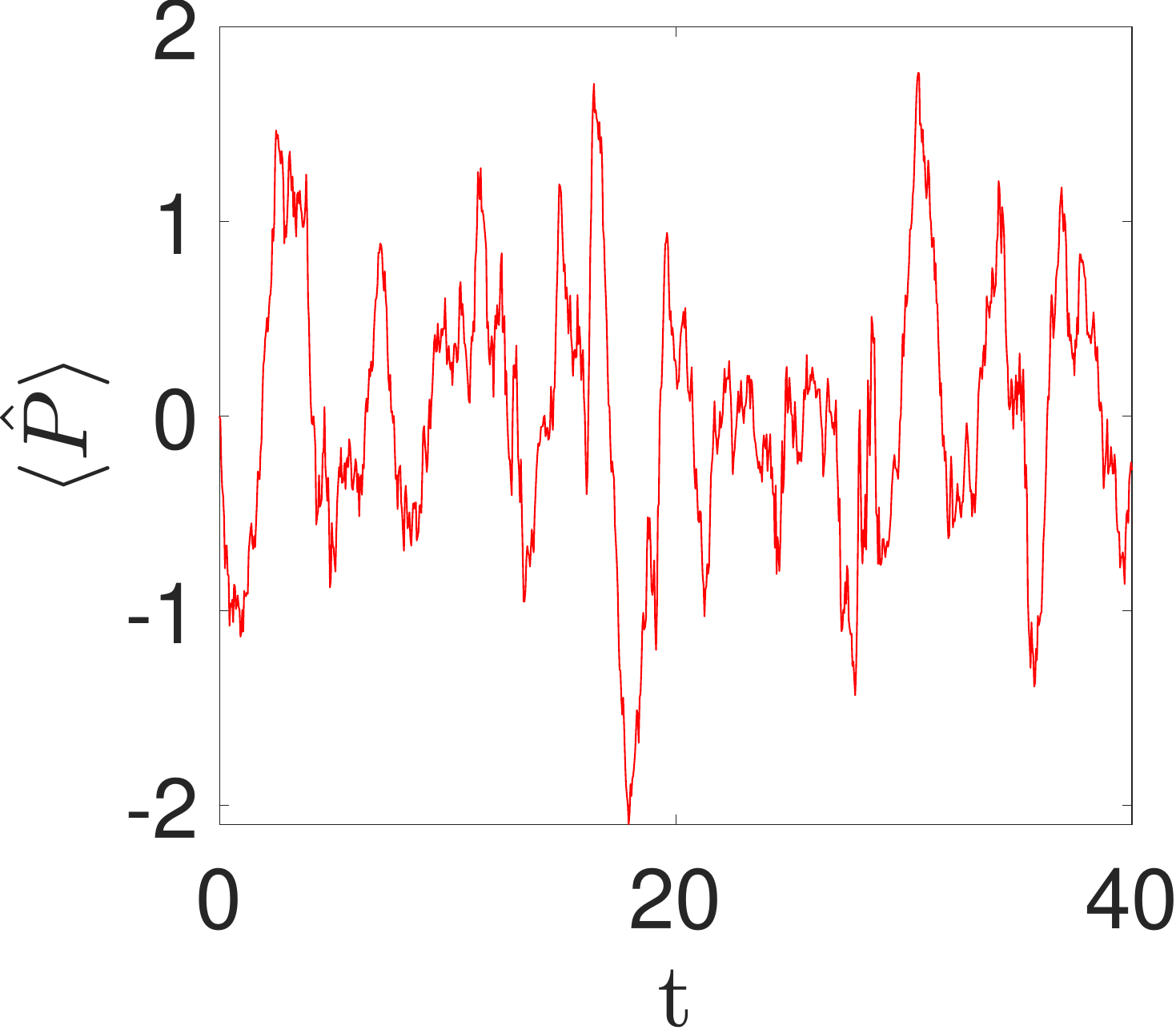}
 	 \caption{}\label{fig:PDoubleQHO}
\end{subfigure}&
\begin{subfigure}[c]{.3\textwidth} 
	  \includegraphics[width=\textwidth]{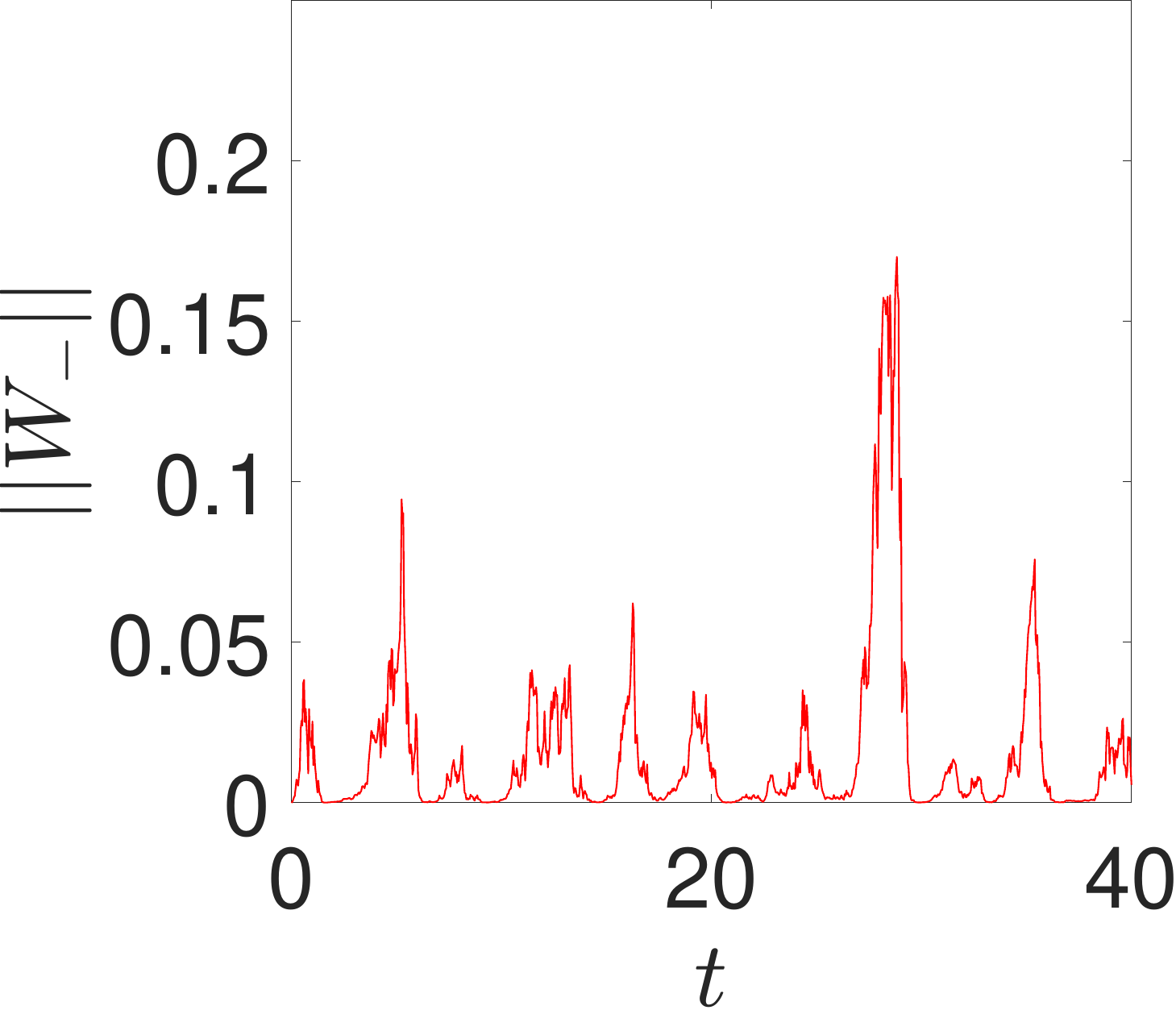}
 	 \caption{}\label{fig:WigNegDoubleQHO}
\end{subfigure}\\
\begin{subfigure}[c]{.3\textwidth} 
	  \includegraphics[width=\textwidth]{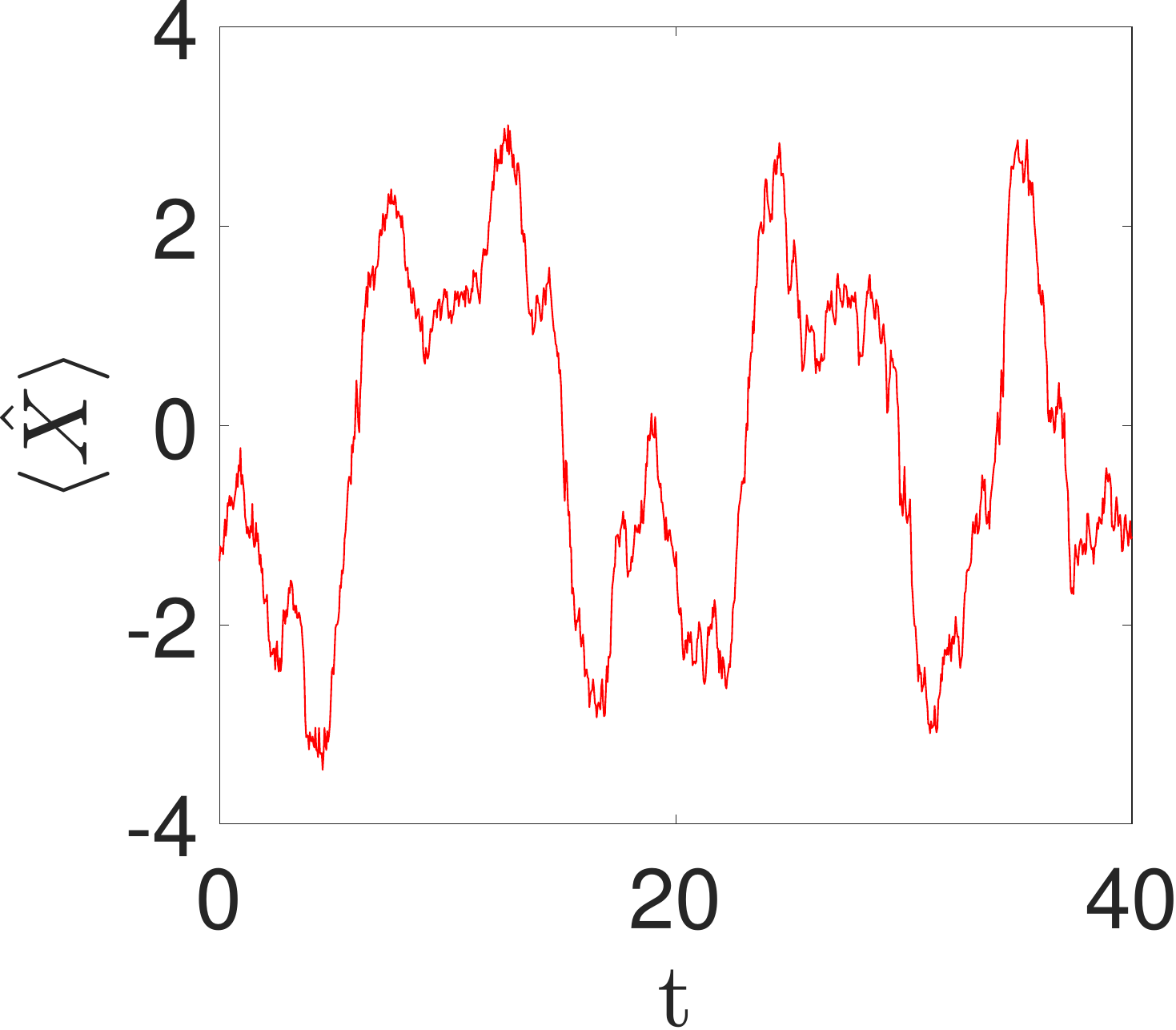}
 	 \caption{}\label{fig:XDoubleQHOHigh}
\end{subfigure}&
\begin{subfigure}[c]{.3\textwidth} 
	  \includegraphics[width=\textwidth]{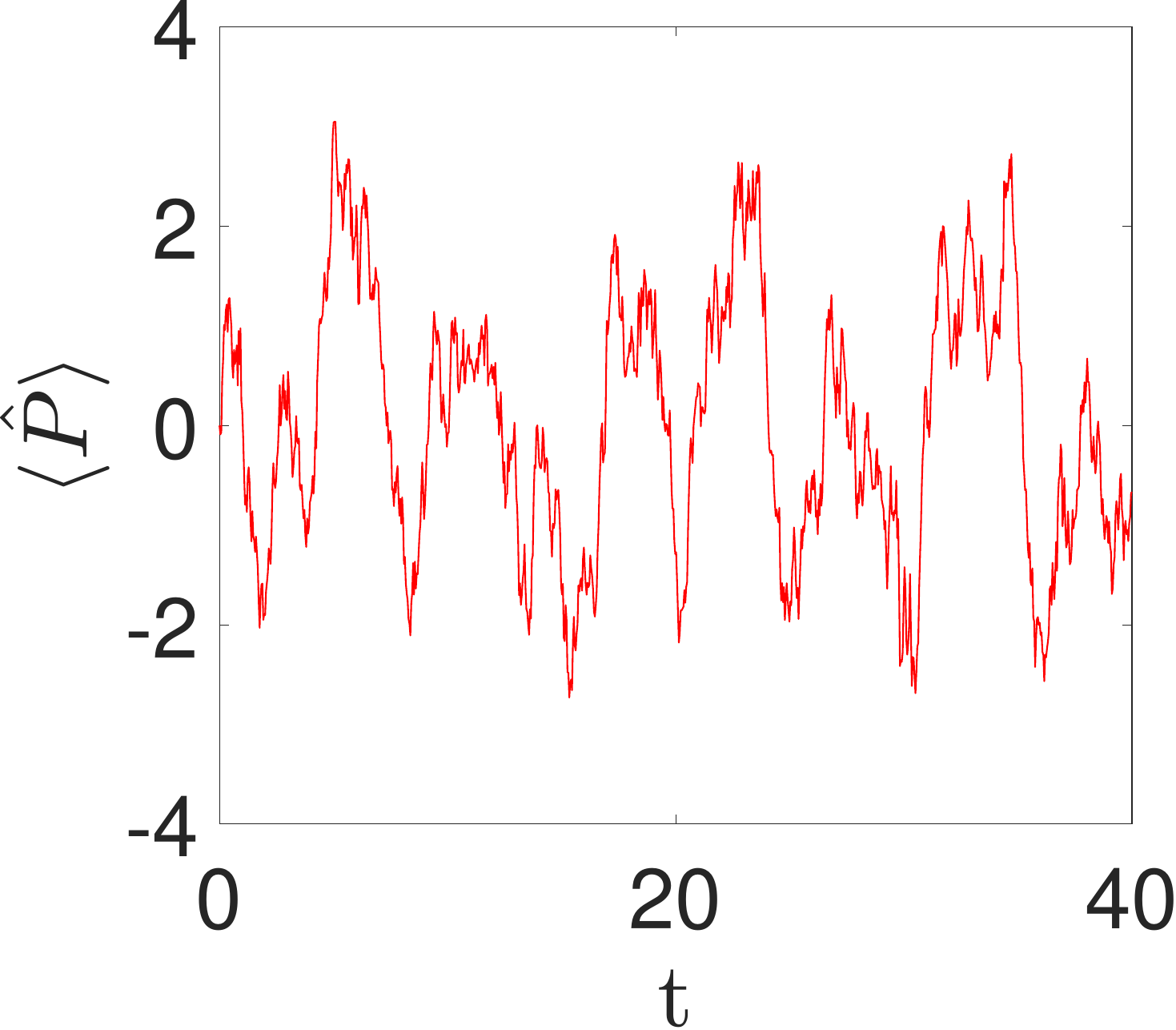}
 	 \caption{}\label{fig:PDoubleQHOHigh}
\end{subfigure}&
\begin{subfigure}[c]{.3\textwidth} 
	  \includegraphics[width=\textwidth]{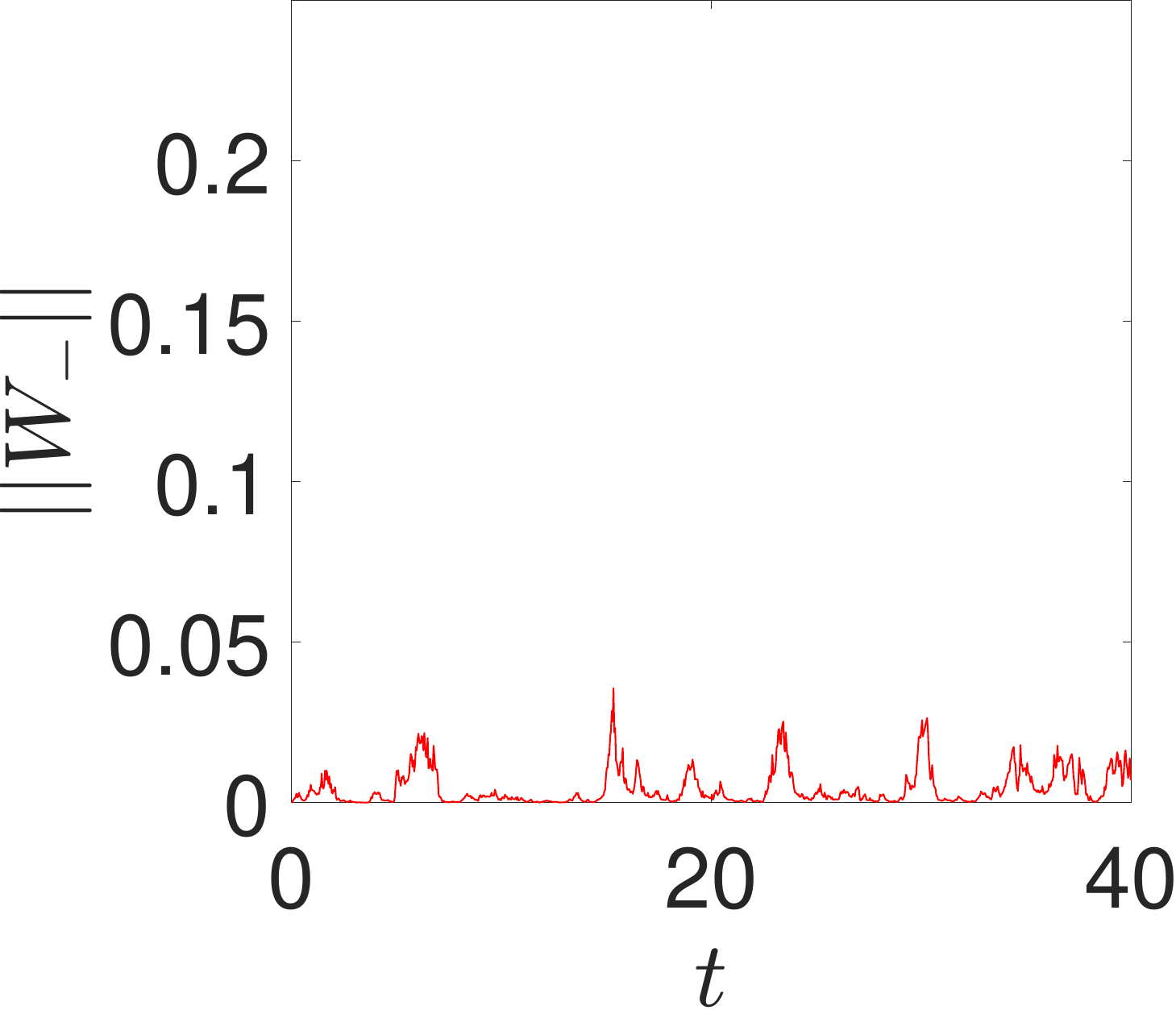}
 	 \caption{}\label{fig:WigNegDoubleQHOHigh}
\end{subfigure}
\end{tabular}
\caption{Three SSE trajectories under the harmonic approximation model with an initial coherent state in the minimum of the left well \cref{eq:gausswell} with $A=3$,$\gamma=0.25$,  and $\sigma=\tfrac{1}{\sqrt2}$ for different temperatures. We show the time dependence of the position expectation $\ev*{\hat X}$ (left), the momentum expectation $\ev*{\hat P}$ (middle), and the negativity of the Wigner function (right). The top row corresponds to $T=0.2$, the middle row to $T=1$ and the bottom row to $T=3$.
 \label{fig:TunnelPlotsQHO}}
\end{figure}
In this section, we delve into an alternative formulation of quantum Brownian motion which aims to better preserve the thermal state at low temperatures. We derive this model in \ref{sec:QHO} as a modification to the Petruccione model which exactly preserves the thermal state of a harmonic oscillator system. For an arbitrary Hamiltonian $\hat H_0$ we write the modified Hamiltonian 
\begin{equation} \label{eq:HarmonicH}
    \hat H=\hat H_0 + \frac{2  \gamma k_B T}{\hbar \omega_e} \sech\left(\frac{\hbar \omega_e}{2 k_B T}\right) \tanh\left(\frac{\hbar \omega_e}{4 k_B T}\right) (\hat X \hat P+\hat P \hat X).
\end{equation}
and Lindblad operator
\begin{equation} \label{eq:HarmonicL}
    \hat L=\sqrt{\frac{4 \gamma m k_B T }{\hbar}} \hat{X} + i \sqrt{\frac{4 \gamma k_B T }{\hbar m \omega_e^2}} \tanh\left(\frac{h \omega}{4 k_B T}\right) \hat{P}.
\end{equation}
The parameter $\omega_e$ is the effective frequency, for the double well we take this as the value of the second derivative of the potential at the well minima giving
\begin{equation}
    \omega_e=2 \log(A/\sigma^2).
\end{equation}
To avoid confusing the two Brownian motion models given by \eqref{eq:heatbath} and \eqref{eq:HarmonicH}\eqref{eq:HarmonicL}, we refer to the model given in \eqref{eq:heatbath} as the minimally invasive model as it is said to be a minimally invasive modification to the Caldeira Leggett master equation \eqref{eq:Caldeira-Master} in \cite{petruccione}. We refer to the model derived in \ref{sec:QHO} and given by \eqref{eq:HarmonicH} and \eqref{eq:HarmonicL} as the harmonic approximation model. A notable benefit of the harmonic approximation is that the singularity at $T=0$ is removed. We can see this in \cref{fig:QHOTerms} comparing the minimally invasive model and the harmonic approximation model.

\begin{figure}[t] 
\centering
\begin{tabular}{c c}
\begin{subfigure}[c]{.4\textwidth} 
	  \includegraphics[width=\textwidth]{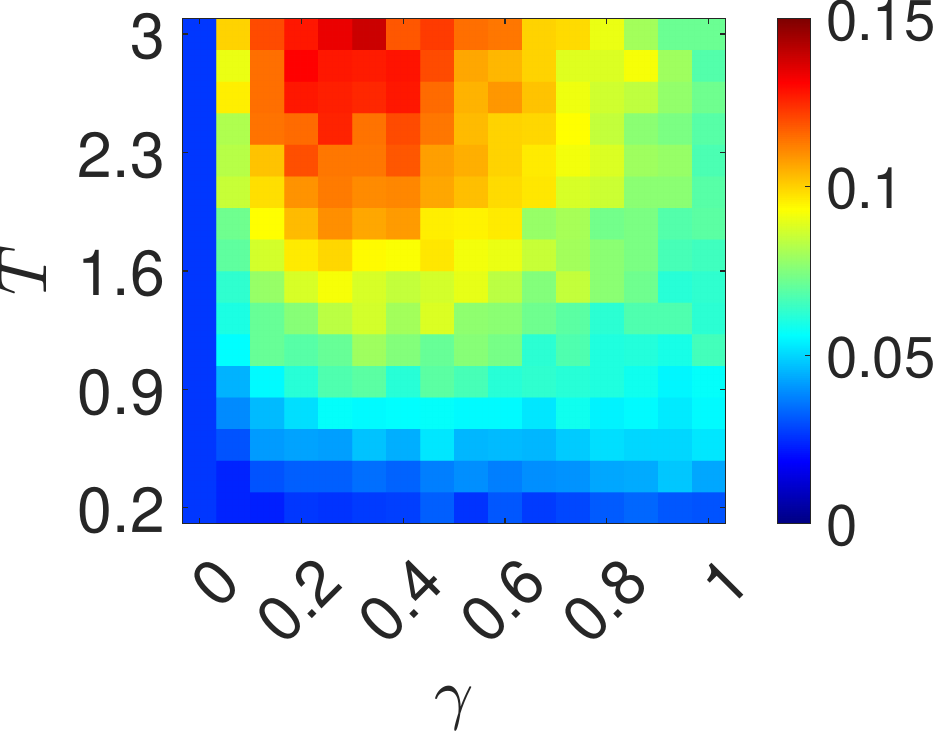}
 	 \caption{\label{fig:G_Shall_RatesQHOCombined}}
\end{subfigure}&
\begin{subfigure}[c]{.4\textwidth} 
	  \includegraphics[width=\textwidth]{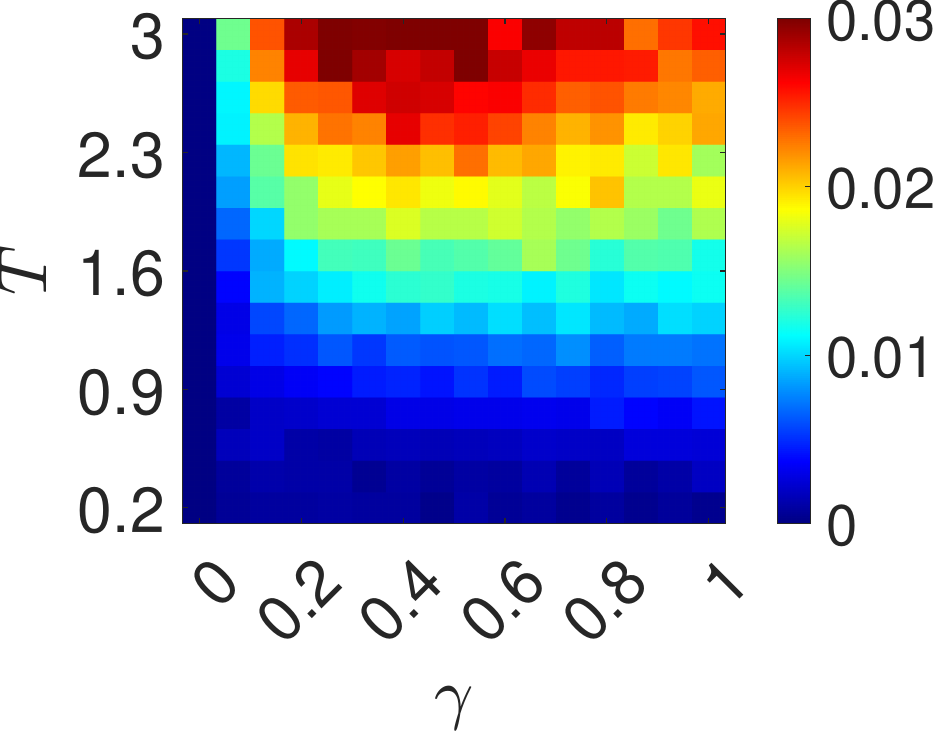}
 	 \caption{\label{fig:G_Deep_RatesQHOCombined}}
\end{subfigure}
\end{tabular}
\caption{Top row: Heatmaps of the quantum transition rate with respect to coupling strength $\gamma$ and temperature for the harmonic approximation model. The columns correspond to different choices of potential on the left a shallow double well ($A=3$, and $\sigma=\tfrac{1}{\sqrt2})$ and on the right, a deep double well ($A=8$, and $\sigma=1$) \label{fig:QHORatesCombined}} 
\end{figure}

To directly compare the harmonic approximation trajectories with those of the minimally invasive model as seen in \cref{fig:TunnelPlots2QSD}, we present plots of typical trajectories at low, medium, and high temperatures in \cref{fig:TunnelPlotsQHO}. We notice a qualitatively similar behaviour to the minimally invasive model, with the transitions of the position expectation being correlated with spikes in the Wigner negativity, especially noticeable at low temperatures. In \cref{fig:QHORatesCombined}, we illustrate the transition rates in relation to temperature and coupling strength for the harmonic approximation model.

By directly comparing \cref{fig:QHORatesCombined} with the minimally invasive and Langevin rates in \cref{fig:transition-rate-big-fig}, the same overall pattern predicted by Kramers rate theory \cite{kramers1940brownian} is evident, with low-temperature rates exceeding those of the Langevin dynamics due to tunnelling effects. However, at very low temperatures, compared to the minimally invasive model, we do not observe an increase in transition rate. This absence is likely attributed to the harmonic approximation Lindblad operator, which does not contain a zero-temperature singularity.
\begin{figure}
\centering 
\begin{tabular}{c c c}
\begin{subfigure}[c]{.3\textwidth} 
	  \includegraphics[width=\textwidth]{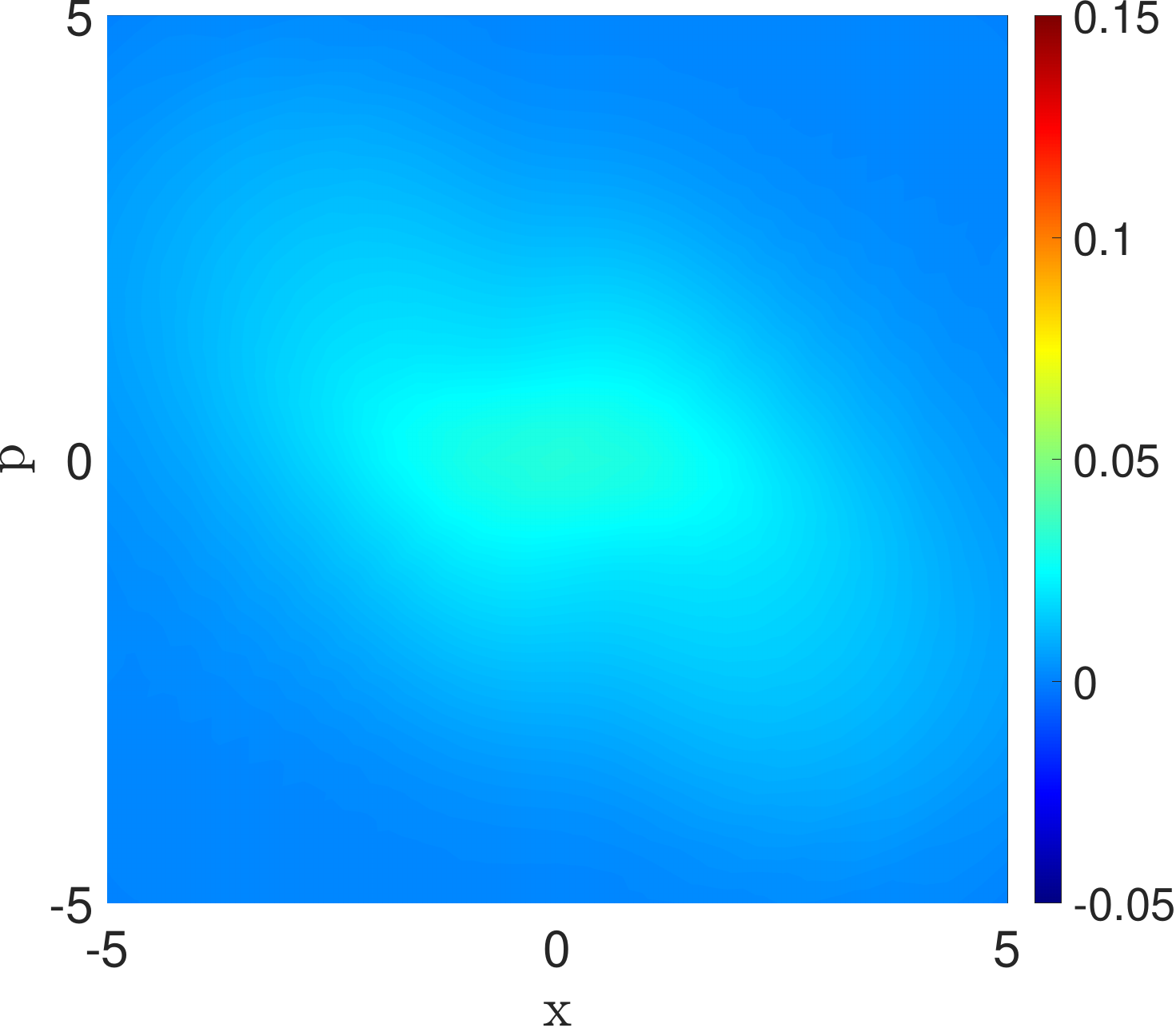}
 	 \caption{}\label{fig:GammaAsymStateWignerG}
\end{subfigure}&
\begin{subfigure}[c]{.3\textwidth} 
	  \includegraphics[width=\textwidth]{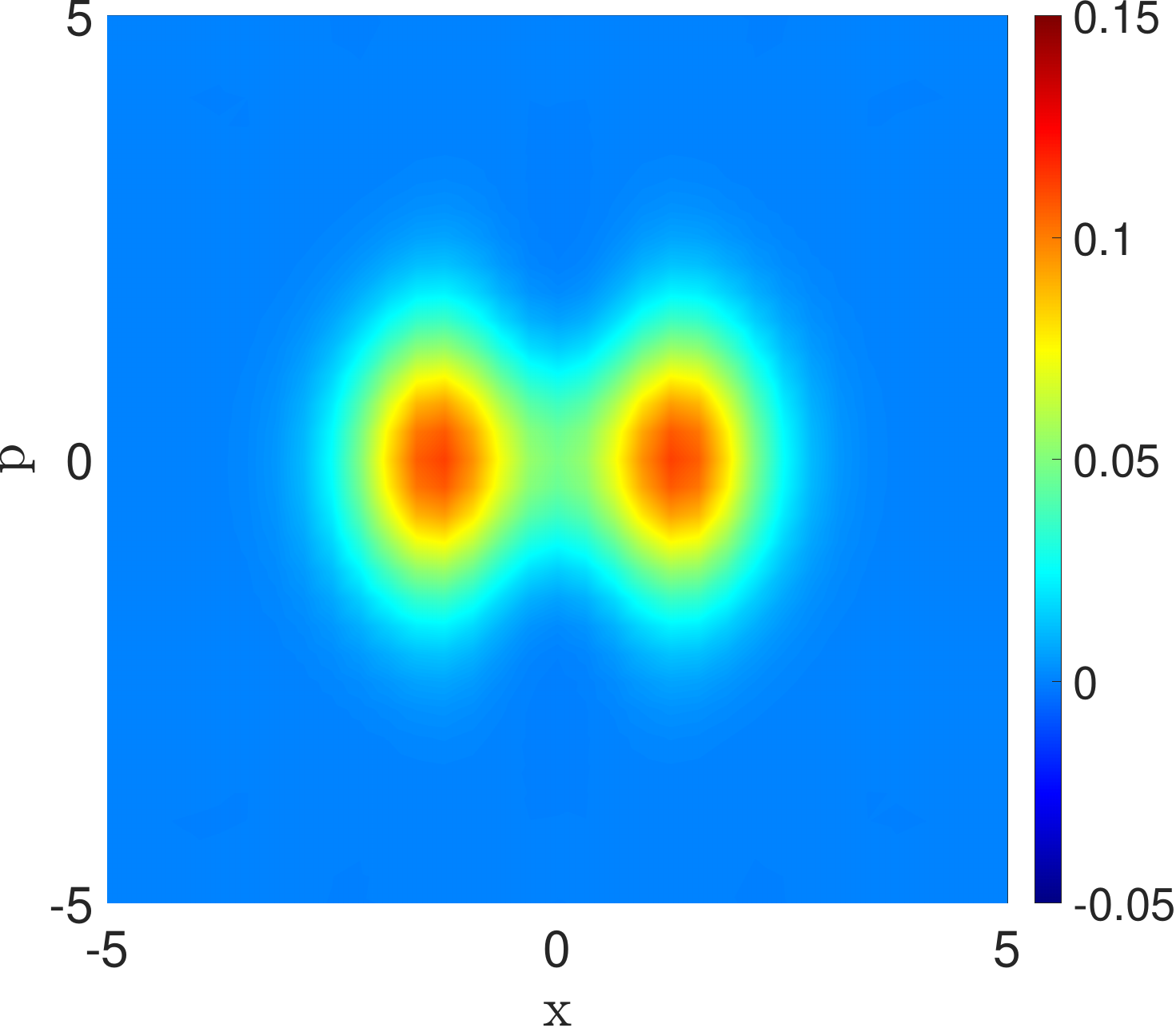}
 	 \caption{}\label{fig:CaldeiraAsymStateWignerG}
\end{subfigure}&
\begin{subfigure}[c]{.3\textwidth} 
	  \includegraphics[width=\textwidth]{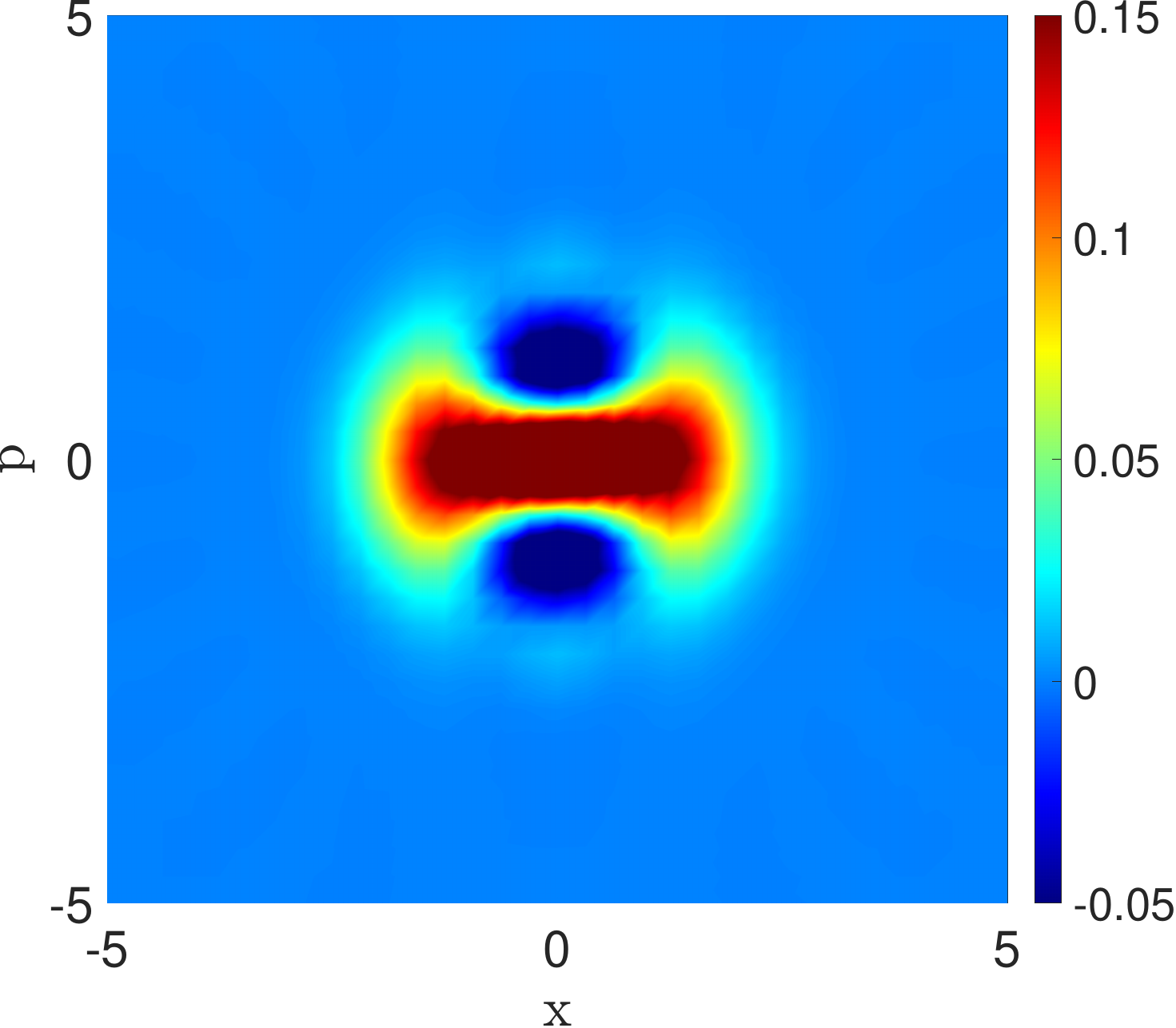}
 	 \caption{}\label{fig:ThermStateWignerG}
\end{subfigure}\\
\begin{subfigure}[c]{.3\textwidth} 
	  \includegraphics[width=\textwidth]{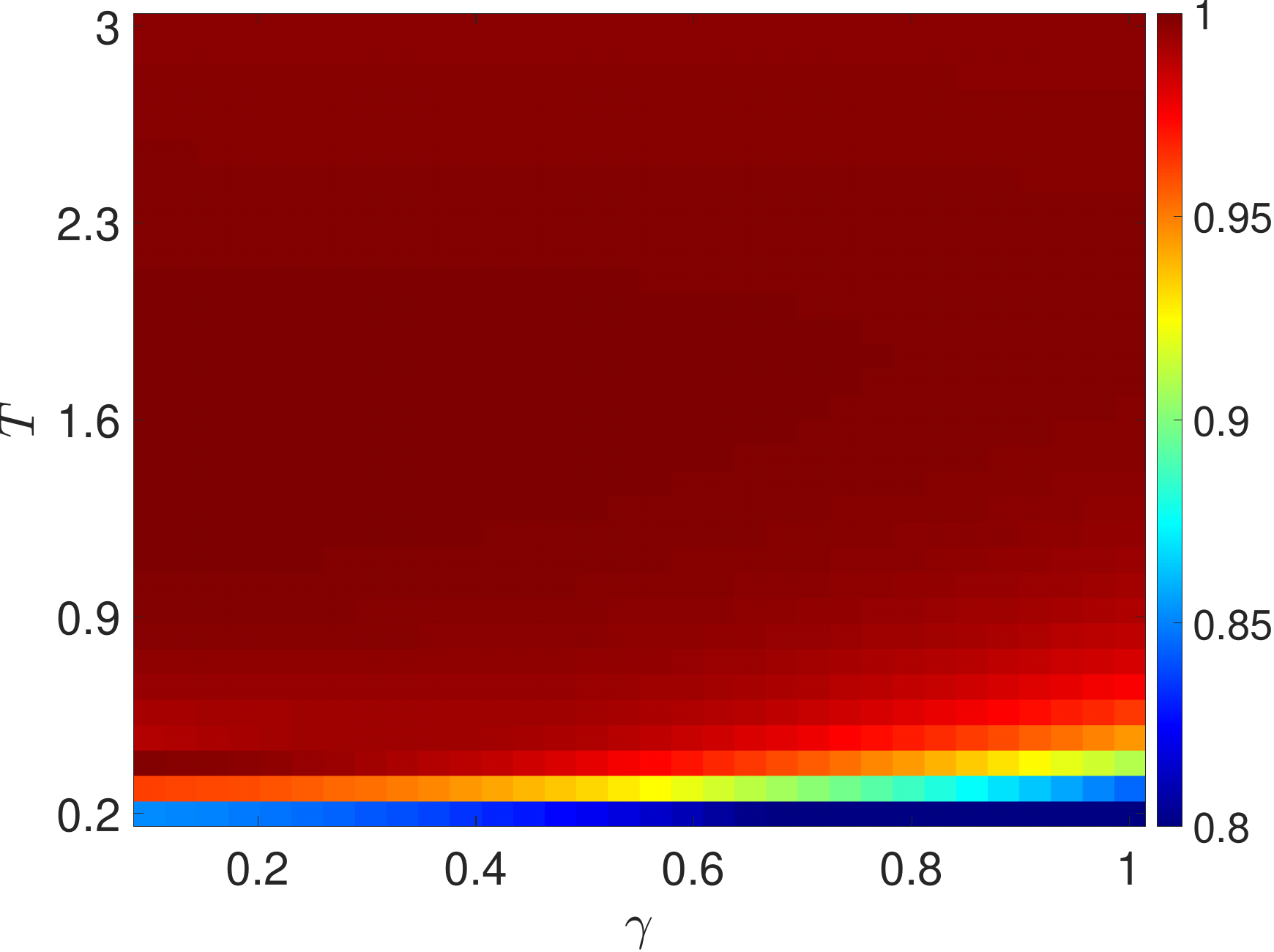}
 	 \caption{}\label{fig:G_Deep_FidelityFreeCombined}
\end{subfigure}&
\begin{subfigure}[c]{.3\textwidth} 
	  \includegraphics[width=\textwidth]{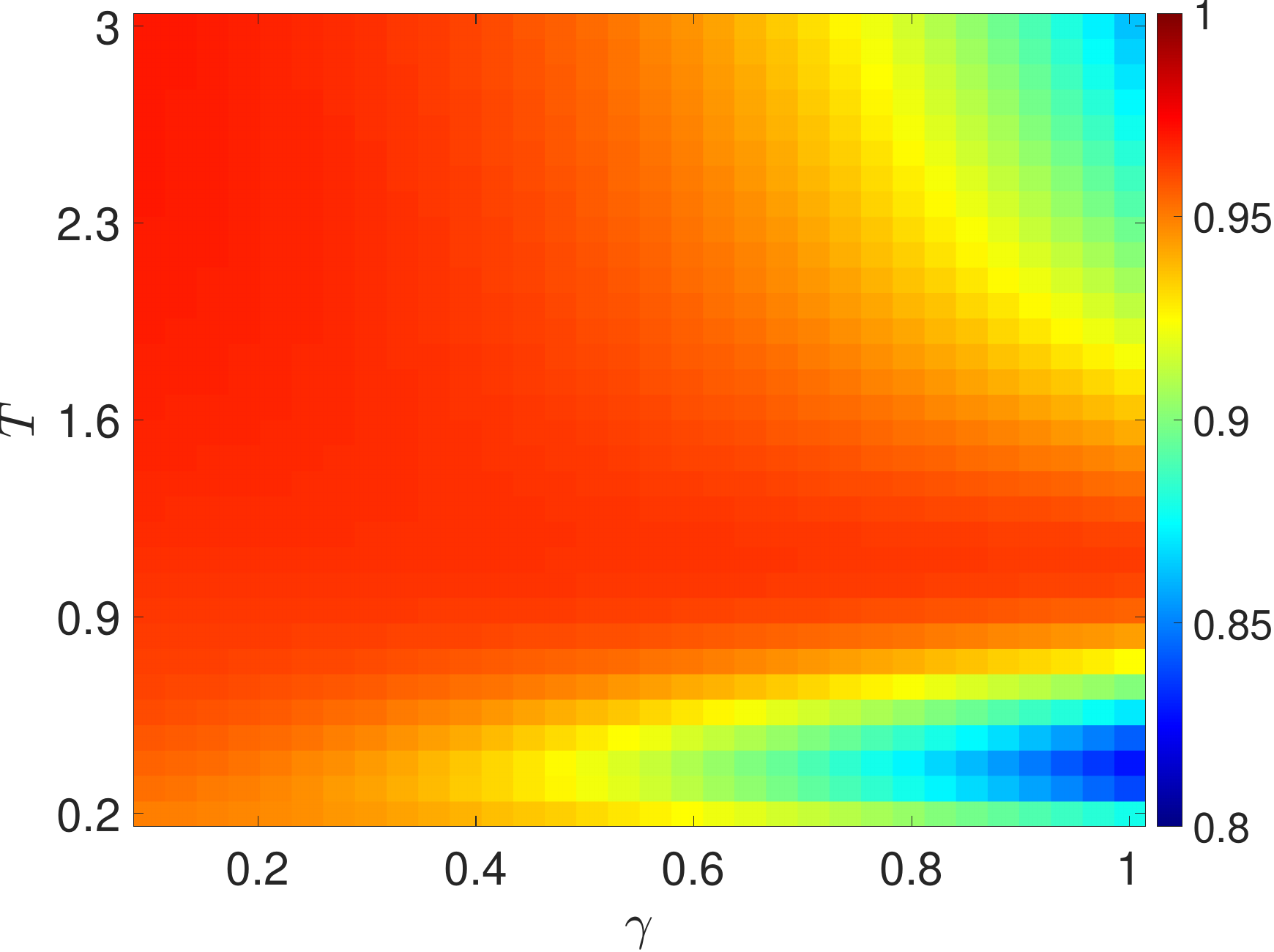}
 	 \caption{}\label{fig:G_Deep_FidelityQHOCombined}
\end{subfigure}&
\begin{subfigure}[c]{.3\textwidth} 
	  \includegraphics[width=\textwidth]{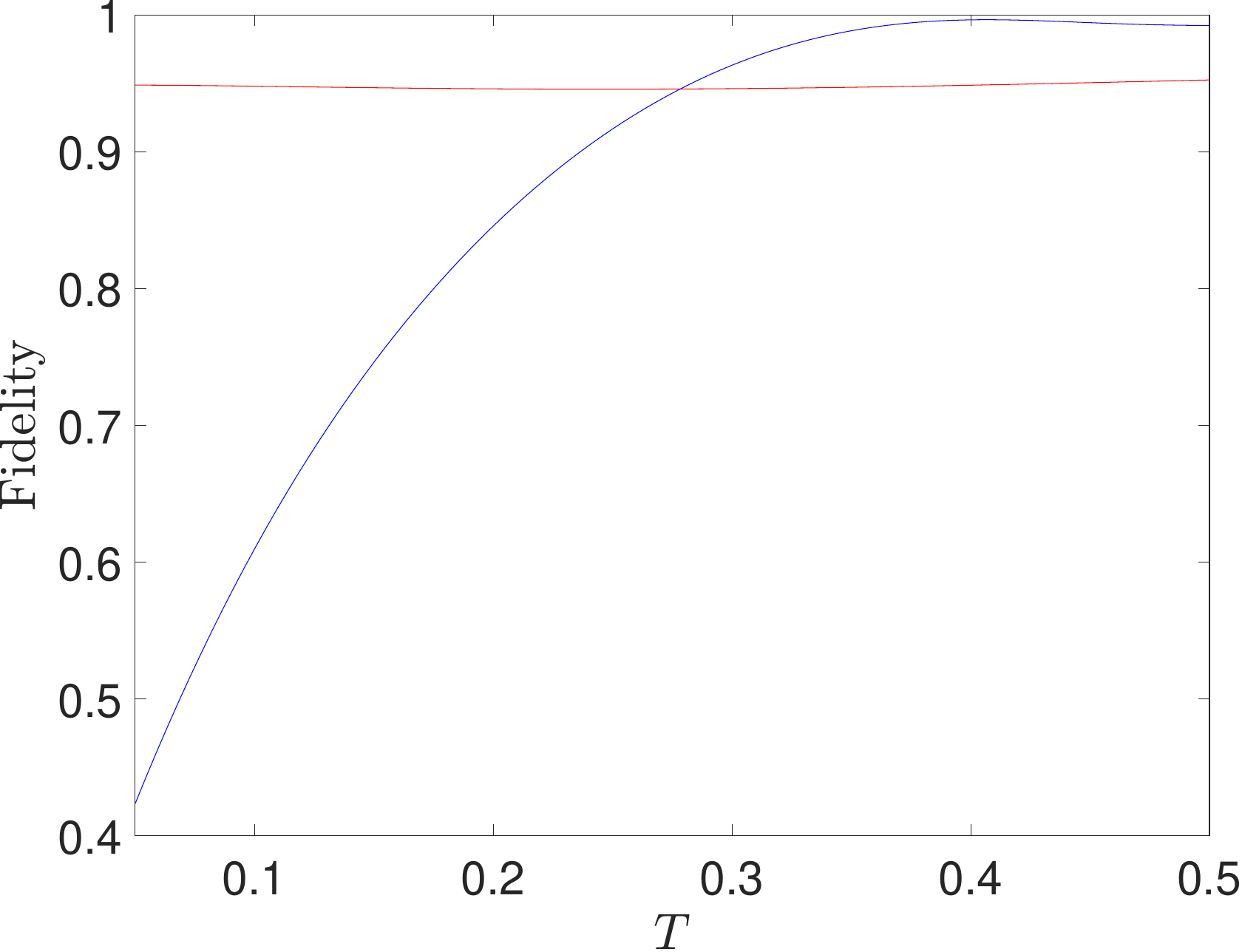}
 	 \caption{}\label{fig:G_Deep_FidelityVsTemp}
\end{subfigure}
\end{tabular}
\caption{Plots for the deep well with $A=8$, and $\sigma=1$. Top row: asymptotic state Wigner functions at $T=0.01$ (a) shows the minimally invasive asymptotic state, (b) the harmonic approximation asymptotic state (c) the Gibbs state in phase space. Bottom Row: Fidelity of the asymptotic state of Lindblad dynamics with the thermal state (d) corresponds to the minimally invasive model, (e) shows the thermal state preserving model and (f) shows a comparison of both models (harmonic approximation red, minimally invasive blue) with $\gamma=0.25$ and varying temperatures. \label{fig:G_Deep_Fidelity}}
\end{figure}

We further explore some features of the master equation dynamics in the deep double well in \cref{fig:G_Deep_Fidelity}. In the top row, from left to right, we exhibit the asymptotic state of the minimally invasive model, the harmonic approximation, and the thermal state \eqref{eq:WigTherm} at a very low temperature. It is observed that the minimally invasive state becomes highly delocalized and bears little resemblance to the thermal state. This outcome is anticipated since the high-temperature assumption was employed during the derivation of this model \cite{caldeira1983path}. The harmonic approximation state retains a similar overall shape; however, it lacks the fringes caused by tunnelling between wells that are apparent in the thermal state.

In the bottom row of \cref{fig:G_Deep_Fidelity}, we plot the fidelity of the asymptotic state relative to the thermal state. The left plot is associated with the minimally invasive model, while the middle plot represents the harmonic approximation concerning. We observe that the fidelity of the minimally invasive state is greater for most parameter choices, yet the harmonic approximation proves more accurate at very low temperatures. In the bottom right panel, a comparison of the fidelities in this extremely low-temperature region reveals the harmonic approximation aligning more closely than the minimally invasive model.

\Cref{fig:QHORatesCombined} shows the transition rate heatmaps with respect to temperature and coupling for the harmonic approximation model. In comparison with the analogous plots for the minimally invasive model in \cref{fig:G_Shall_RateFreeCombined,fig:G_Deep_RateFreeCombined} we observe similar qualitative features with the key difference that the very low-temperature increase in the transition rate is not present in the for the Harmonic approximation model.

\section{Summary} \label{sec:summary}
In conclusion, our investigation has provided insightful data on the behaviour of a particle in a double well potential, specifically focusing on the influence of temperature on tunnelling and transition rates. By juxtaposing classical Langevin dynamics with the stochastic Schrödinger dynamics derived from a modified Caldeira Leggett model, we observed a notable convergence between quantum and classical mechanics under defined conditions.

A significant aspect of our study involved the derivation of a thermal state-preserving model for the harmonic oscillator; this model provides a good description of the limit of low-temperature dynamics near a potential minimum for non-quadratic Hamiltonians and recovers the Pettrucione model in the free particle limit. By applying thermal state preservation as a constraint on the master equation, there's a potential to develop models that can be extended to understand quantum Brownian motion in a variety of systems at low temperatures.

\section{Acknowledgements}
 We would like to thank Eva-Maria Graefe for the helpful discussion throughout the project and feedback during the drafting phase.
 
R. Christie acknowledges support from EPSRC grant no. EP/W522673/1. J. Eastman acknowledges support from the Royal Society (Grant No. URF $\backslash$R$\backslash$ 201034) and from the European Research Council (ERC) under the European Union's Horizon 2020 research and innovation programme (Grant No. 758453).
J. Eastman is currently located at the Australian National University.

\appendix
\section{Equilibrium preserving master equation for a Harmonic Oscillator} \label{sec:QHO}

 Considering a Harmonic oscillator with Hamiltonian
\begin{equation} 
    \hat H =\frac{\hat P^2}{2m}+\frac{m \omega^2}{2} \hat X^2  
\end{equation}
The thermal state \eqref{eq:WigTherm} is well known for the harmonic oscillator \cite{davies1973harmonic} and given by
\begin{equation}
    \hat \rho_{eq}=2 \sinh(\frac{\hbar \omega}{2 k_B T})\sum_{n=0}^{\infty}\exp(-\frac{\hbar \omega}{k_B T}(n+\frac12))\ket{n}\bra{n}.
\end{equation}
Transforming the above into Wigner Weyl formalism, we obtain the Gaussian thermal Wigner function
\begin{equation} \label{eq:therm_wigner}
    W(x,p)=\frac{1}{\hbar \pi}\tanh(\frac{\hbar \omega }{2 k_B T}) \exp(-\frac{1}{\hbar  \omega}\tanh(\frac{\hbar \omega }{2 k_B T})(p^2+m \omega^2 x^2)).
\end{equation}
We may use the fact that the minimally invasive model \eqref{eq:heatbath} contains a Hamiltonian that is at most quadratic with a single linear Lindblad operator to infer that the asymptotic stationary state of this system can be expressed as a Wigner function of the form \cite{graefebradley,MyPaper}
\begin{equation} \label{eq:ansatz}
	\begin{aligned}
	W(z)=\frac{\sqrt{\det G}}{\pi \hbar}e^{-\frac{1}{\hbar}z \cdot G z}& &\text{with}& & z =\begin{pmatrix}x\\ p\end{pmatrix} 
	\end{aligned}
\end{equation}
With the stationary state covariance matrix $G$ corresponding to the unique fixed point of the dynamics
\begin{multline} \label{eq:GStationary}
        \frac{dG}{dt} =\left(H''+\text{Im}(\nabla L \nabla \bar{L}^{T})\right)\Omega G-G\Omega\left(H''-\text{Im}(\nabla L \nabla \bar{L}^{T})\right)\\+2 G \Omega \text{Re}(\nabla L \nabla \bar{L}^T) \Omega G,
\end{multline}
here $H$ and $L$ are the Weyl symbols of the Hamiltonian and Lindblad operators, respectively. By setting \eqref{eq:GStationary} to zero, we may solve the resulting algebraic equations to obtain the fixed point as
\begin{equation}
    G_F=\frac{1}{ f(T)}
\begin{pmatrix}
\frac{\hbar m \omega^2}{8  k_B T} + \frac{2 m  k_B T}{\hbar} & \frac{\hbar \gamma}{4  k_B T} \\
\frac{\hbar \gamma}{4  k_B T} & \frac{\hbar^2 (4 \gamma^2 + \omega^2) + 16  k_B^2T^2}{8 \hbar m T \omega^2 k_B}
\end{pmatrix}
\end{equation}
with 
\begin{equation}
    f(T)=\frac{64 \hbar^2 T^2 \omega^2 k_B^2}{\hbar^4 \omega^4 + 32 \hbar^2 k_B^2 T^2 (2 \gamma^2 + \omega^2)  + 256  k_B^4 T^4}.
\end{equation}
In comparison, the thermal state \eqref{eq:therm_wigner} may also be written in the same form as \eqref{eq:ansatz} with covariance matrix
\begin{equation} \label{eq:GThermal}
    G_T=\tanh(\frac{\hbar \omega}{2 k_B T})\begin{pmatrix}
        m \omega&0\\0& \frac{1}{\omega}.
    \end{pmatrix}
\end{equation}
In the high-temperature limit, we find that the fixed point and thermal state asymptotically approach the classical thermal state 
\begin{equation}
    W_C(x,p)=\frac{\exp(-\frac{H(x,p)}{k_B T})}{\int_{-\infty}^{\infty}\int_{-\infty}^{\infty}  \exp(-\frac{H(x,p)}{k_B T}) dx dp}
\end{equation}
as expected. However, At low temperatures the fixed point of the Lindblad dynamics has, off-diagonal $(\hat X \hat P)$ covariances whilst the thermal state does not. This discrepancy between the asymptotic state and thermal state at low temperatures is a consequence of the Lindblad operator \eqref{eq:heatbath} having a singular $T\to 0$ limit, modelling an effectively infinitely strong momentum measurement.

It has been shown in \cite{cleary2011phase} that it is possible to derive master equations for the Wigner function evolution by incorporating the constraint that the thermal state is a fixed point of the dynamics. We will now present a master equation derivation for the Harmonic oscillator that parallels this approach; however, in our analysis, we will take a step further by making explicit the connection with the Lindblad equation. By doing so, we can investigate the dynamics of the SSE unravelling. We begin by making the ansatz that the Lindblad operator is of the same general form as \eqref{eq:heatbath} with an additional $\hat P$ term that tames the singular $T\to 0$ limit and is negligible in the high-temperature limit, explicitly we have
\begin{equation} \label{eq:L_lp}
    \hat{L}= \sqrt{\frac{4 \gamma m k_B T}{\hbar}}\hat{X}+i\left(\sqrt{\frac{\gamma \hbar}{4mk_B T}}+l_p\right)\hat{P}.
\end{equation}
Similarly, we add a term to the effective Hamiltonian as we expect the $\hat X \hat P$ term to disappear in the $T \to 0$ limit. We have
\begin{equation} \label{eq:Heff_hxp}
    \hat H=\frac{\hat P^2}{2m}+\frac{m \omega^2}{2} \hat X^2 +\frac12(\gamma+h_{xp})\left(\hat X \hat P+\hat P \hat X\right).
\end{equation}
Since we demand that the thermal state is conserved, we insert \eqref{eq:GThermal} into \eqref{eq:GStationary} to obtain 
\begin{multline} 
        0=\left(H''+\text{Im}(\nabla L \nabla \bar{L}^{T})\right)\Omega G_T-G_T\Omega\left(H''-\text{Im}(\nabla L \nabla \bar{L}^{T})\right)\\+2 G_T \Omega \text{Re}(\nabla L \nabla \bar{L}^T) \Omega G_T.
\end{multline}
Explicitly the above equations reduce to the system
\begin{align}
4  k_B T \left(h_{xp} - 2 l_p \sqrt{\frac{m \gamma k_B T}{\hbar}}\right) &= -\omega \left(\hbar \gamma + 4 l_p \left(l_p m  k_B T + \sqrt{\hbar m \gamma k_B T}\right)\right) \tanh\left(\frac{\hbar \omega}{2  k_B T}\right), \\
 4 \gamma k_B T \tanh\left(\frac{\hbar \omega}{2  k_B T}\right)&=\hbar (h_{xp} + 2 \gamma) \omega + 2 l_p \omega \sqrt{\hbar m \gamma k_B T}. 
\end{align}
These equations have two sets of solutions
\begin{align} \label{eq:Former}
    l_p&=-\frac{1}{2} \sqrt{\frac{\hbar \gamma}{m k_B T}} - 2 \sqrt{\frac{k_B T \gamma}{\hbar m \omega^2}} \coth\left(\frac{\hbar \omega}{4 k_B T}\right),\\
    h_{xp}&=- \frac{\gamma \left(\hbar \omega + 4 k_B T \csch\left(\frac{\hbar \omega}{2 k_B T}\right) + 8 k_B T \csch\left(\frac{\hbar \omega}{k_B T}\right)\right)}{\hbar \omega}
\end{align}
and,
\begin{align} \label{eq:Latter}
    l_p&=-\frac{1}{2} \sqrt{\frac{\hbar \gamma}{m k_B T}} - 2 \sqrt{\frac{k_B T \gamma}{\hbar m \omega^2}} \tanh\left(\frac{\hbar \omega}{4 k_B T}\right),\\
    h_{xp}&=- \frac{\gamma \left(\hbar \omega - 4 k_B T \csch\left(\frac{\hbar \omega}{2 k_B T}\right) + 8 k_B T \csch\left(\frac{\hbar \omega}{k_B T}\right)\right)}{\hbar \omega}.
\end{align}
However of \eqref{eq:Former} and \eqref{eq:Latter}, only \eqref{eq:Latter} satisfies the condition $h_{xp}$, $l_p\to 0$ as $T\to \infty$ needed to recover the Caldeira-Leggett Master equation \eqref{eq:Caldeira-Master} in the high-temperature limit. Substituting \eqref{eq:Latter} into \eqref{eq:Heff_hxp} we obtain the effective Hamiltonian
\begin{equation} \label{eq:HQHO}
    \hat H=\hat H_{QHO} + \frac{2  \gamma k_B T}{\hbar \omega} \sech\left(\frac{\hbar \omega}{2 k_B T}\right) \tanh\left(\frac{\hbar \omega}{4 k_B T}\right) (\hat X \hat P+\hat P \hat X).
\end{equation}
We also substitute \eqref{eq:Latter} into \eqref{eq:L_lp} to give the Lindblad operator
\begin{equation} \label{eq:LQHO}
    \hat L=\sqrt{\frac{4 \gamma m k_B T }{\hbar}} \hat{X} + i \sqrt{\frac{4 \gamma k_B T }{\hbar m \omega^2}} \tanh\left(\frac{h \omega}{4 k_B T}\right) \hat{P}.
\end{equation}
\begin{figure}[ht]
\centering
\begin{tabular}{c c}
\begin{subfigure}[c]{.45\textwidth} 
	  \includegraphics[width=\textwidth, height=0.19\textheight]{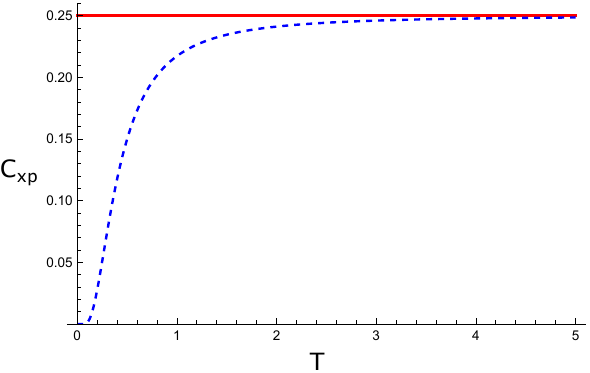}
 	 \caption{}\label{fig:HTermQHO}
\end{subfigure}&\hfill
\begin{subfigure}[c]{.45\textwidth} 
	  \includegraphics[width=\textwidth, height=0.19\textheight]{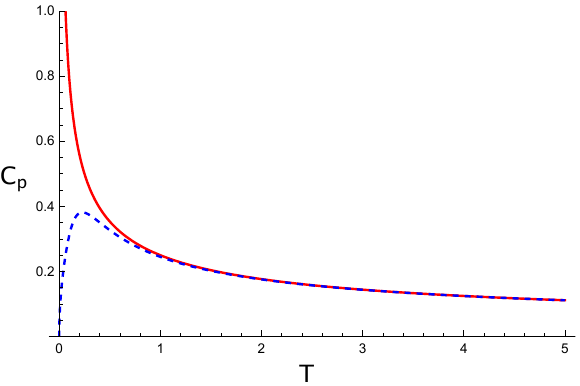}
 	 \caption{}\label{fig:LTermQHO}
\end{subfigure}
\end{tabular}
\caption{The comparison between the minimally invasive  (red-solid) and harmonic approximation (blue-dashed) Brownian motion models for $\omega=\hbar=k_B=1$ and $\gamma=0.25$. (a) shows the value of the $(\hat X \hat P+\hat P \hat X)$ term coefficent in the effective Hamiltonian against temperature. (b) shows the value of the $(\hat P)$ term coefficient in the Lindblad operator against temperature.\label{fig:QHOTerms}} 
\end{figure}
We name this combination of effective Hamiltonian and Lindblad operators the harmonic approximation model. We now compare this model with the minimally invasive model in \cite{petruccione}. \Cref{fig:HTermQHO} depicts the value of the $(\hat X \hat P+\hat P \hat X)$ term coefficient in the effective Hamiltonian in both models, given by
\begin{equation}
    C_{xp}=\gamma\quad\text{and}\quad \frac{4  \gamma k_B T}{\hbar \omega} \sech\left(\frac{\hbar \omega}{2 k_B T}\right) \tanh\left(\frac{\hbar \omega}{4 k_B T}\right)
\end{equation}
for the minimally invasive and harmonic approximation models respectively. We observe that the harmonic approximation model recovers the closed system Hamiltonian in the $T\to 0$ limit asymptotically approaching the minimally invasive effective Hamiltonian at high temperatures. \Cref{fig:LTermQHO} depicts the value of the $(\hat P)$ term coefficient in the Lindblad operator which is given by
\begin{equation}
    C_{p}=\sqrt{\frac{\gamma \hbar}{4mk_B T}}\quad\text{and}\quad  \sqrt{\frac{4 \gamma k_B T }{\hbar m \omega^2}} \tanh\left(\frac{h \omega}{4 k_B T}\right)
\end{equation}
for the minimally invasive and harmonic approximation models respectively. We observe that the harmonic approximation Lindblad operator vanishes in the $T\to 0$ compared to the minimally invasive model which is singular. At high temperatures, the harmonic approximation model Lindblad operator asymptotically approaches the minimally invasive operator.

Moreover, taking the $\omega \to 0$ limit of the harmonic approximation effective Hamiltonian \eqref{eq:HarmonicH} and Lindblad operator $\eqref{eq:HarmonicL}$ recovers the minimally invasive model \eqref{eq:heatbath}. In this sense, the minimally invasive model may be thought of as an equilibrium state-preserving model for the quantum Brownian motion of a free particle only.

We end by highlighting a notable distinction between classical and quantum models, particularly in relation to the Langevin equation \eqref{eq:Langevin_tilde} and the Stochastic Schrödinger equation \eqref{eq:SSE}. In the classical model, the ensemble dynamics maintain the classical thermal state, with the stochastic noise term remaining uninformed about the potential energy landscape in which the particle resides. In contrast, the quantum model necessitates that the stochastic term in the SSE possesses information about the potential energy landscape for the ensemble dynamics to preserve the thermal state.

\section*{References}
\bibliographystyle{IEEEtran}
\bibliography{main}{}

\end{document}